\documentclass[aps,a4paper,8pt,
superscriptaddress,
prd,onecolumn,
amssymb,amsmath,amsfonts,nofootinbib]{revtex4-2}
\usepackage{aas_macros}
\usepackage{graphicx}
\usepackage{subfig}
\graphicspath{{figs/}}
\usepackage[dvipsnames]{xcolor}
\usepackage{tikz} 
\usepackage{hyperref}
\usepackage[T1]{fontenc} 
\usepackage{tcolorbox}
\usepackage[normalem]{ulem} 
\usepackage{multirow}

\usepackage{CJKutf8} 
\newcommand{\zhsun}[1]{\begin{CJK}{UTF8}{gbsn}#1\end{CJK}} 
\newcommand{\jpmin}[1]{\begin{CJK}{UTF8}{min}#1\end{CJK}}  

\hypersetup{
    colorlinks=true,
    linkcolor=blue,
    filecolor=magenta,      
    urlcolor=blue,
    citecolor=red,
}

\newcommand{\coordinator}[1]{\noindent\emph{\small Section coordinator: #1}\vspace{0.5cm}}
\newcommand{\coordinators}[1]{\noindent\emph{\small Section coordinators: #1}\vspace{0.5cm}}
\newcommand{\contributors}[1]{\noindent\emph{\small Contributors: #1}\vspace{0.5cm}}

\newcommand{\soutPC}{\bgroup\markoverwith{\textcolor{cyan}
{\rule[0.5ex]{2pt}{1pt}}}\ULon}
\newcommand\blfootnote[1]{%
  \begingroup
  \renewcommand\thefootnote{}\footnote{#1}%
  \addtocounter{footnote}{-1}%
  \endgroup
}

\newcommand{\ba}{\begin{eqnarray}}
\newcommand{\ea}{\end{eqnarray}}
\newcommand{\be}{\begin{equation}}
\newcommand{\ee}{\end{equation}}

\newcommand{\au}{\mathrm{AU}}

\newcommand{\IN}{\mathrm{in}}
\newcommand{\OUT}{\mathrm{out}}

\tcbset{tab2/.style={fonttitle=\bfseries,fontupper=\normalsize\sffamily,
colback=yellow!10!white,colframe=red!50!black,colbacktitle=Salmon!40!white,
coltitle=black,center title}}

\begin{document}

\newcommand{\intitle}{Gravitational Wave Astronomy With TianQin}

\title[TQ Astro WhitePaper]{%
\hspace{0.85\textwidth}\vspace{-2cm} \includegraphics[width=0.15\textwidth]{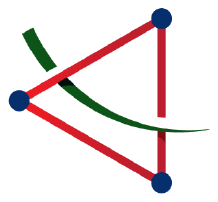}\\
\vspace{2cm} \intitle}


\newcommand{\TRC}{MOE Key Laboratory of TianQin Mission, TianQin Research Center for Gravitational Physics $\&$ Frontiers Science Center for TianQin, Gravitational Wave Research Center of CNSA, Sun Yat-sen University (Zhuhai Campus), Zhuhai 519082, China}

\newcommand{\SPA}{School of Physics and Astronomy, Sun Yat-sen University (Zhuhai Campus), Zhuhai 519082, China}

\newcommand{\HKU}{Department of Physics, The University of Hong Kong, Pokfulam Road, Hong Kong}

\newcommand{\BIMSA}{Beijing Institute of Mathematical Sciences and Applications, Beijing 101408, China}

\newcommand{\Zhaoqing}{School of Electronic and Electrical Engineering, Zhaoqing University, Zhaoqing 526061, China}

\newcommand{\KIAA}{The Kavli Institute for Astronomy and Astrophysics, Peking University, Beijing 100871, China}

\newcommand{\DLUT}{School of Physics, Dalian University of Technology, Liaoning 116024, China}

\newcommand{\UZH}{Department of Astrophysics, University of Zurich, Winterthurerstrasse 190, CH-8057 Z\"urich, Switzerland}

\newcommand{\YNO}{Yunnan Observatories, Chinese Academy of Sciences, Kunming, 650216, China}

\newcommand{\YNKL}{International Centre of Supernovae, Yunnan Key Laboratory, Kunming 650216, China}

\newcommand{\UCAS}{University of Chinese Academy of Sciences, Beijing 100049, China}

\newcommand{\Fukui}{Center for Information Science, Fukui Prefectural University, Fukui 917-0003, Japan}

\newcommand{\NJU}{School of Astronomy and Space Science, Nanjing 210093, China}

\newcommand{\tokyo}{Department of Astronomy, University of Tokyo, Tokyo 113-0033, Japan}

\newcommand{\ZJU}{School of Physics, Zhejiang University, Hangzhou 310027, China}

\newcommand{\SHAO}{Shanghai Astronomical Observatory, Shanghai 200030, China}

\newcommand{\fisk}{Department of Life and Physical Sciences, Fisk University, Nashville 37208, USA}

\newcommand{\vandy}{Department of Physics \& Astronomy, Vanderbilt University, Nashville 37235, USA}

\newcommand{\NCAC}{Nicolaus Copernicus Astronomical Center of the Polish Academy of Sciences, Warsaw 00-716, Poland}

\newcommand{\AMU}{Faculty of Mathematics and Computer Science, A. Mickiewicz University, Uniwersytetu Pozna\'nskiego 4, 61-614 Pozna\'n, Poland}

\newcommand{\NUS}{Department of Physics, National University of Singapore, Singapore 117551, Singapore}

\newcommand{\NUSM}{Department of Mathematics, National University of Singapore, Singapore 119076, Singapore}

\newcommand{\NBI}{Niels Bohr Institute, Copenhagen 2200, Denmark}

\newcommand{\FDU}{Center for Astronomy and Astrophysics, Center for Field Theory and Particle Physics, and Department of Physics, Fudan University, Shanghai 200438, China}

\newcommand{\NUU}{School of Natural Sciences and Humanities, New Uzbekistan University, Tashkent 100007, Uzbekistan}

\newcommand{\YZU}{Center for Gravitation and Cosmology, Yangzhou University, Yangzhou 225009, China}

\newcommand{\Naples}{Department of Physics, University of Naples Federico II, Naples 80131, Italy}

\newcommand{\IHEP}{Institute of High Energy Physics, Beijing 100049, China}

\newcommand{\UMIB}{Department of Physics, University of Milano-Bicocca, Milan 20126, Italy}

\newcommand{\HBU}{College of Physics Science \& Technology, Hebei University, Baoding 071002, China}

\newcommand{\THU}{Department of Astronomy, Tsinghua University, Beijing 100084, China}

\newcommand{\HUST}{School of Physics, Huazhong University of Science and Technology, Wuhan 430074, China}

\newcommand{\PMO}{Purple Mountain Observatory, Nanjing 210034, China}


\author{En-Kun Li(\zhsun{李恩坤})$^\dagger$}
\affiliation{\SPA}
\affiliation{\TRC}
\author{Shuai Liu(\zhsun{刘帅})$^\dagger$}
\affiliation{\Zhaoqing}
\author{Alejandro Torres-Orjuela$^\dagger$}
\affiliation{\BIMSA}
\affiliation{\HKU}
\author{Xian Chen(\zhsun{陈弦})$^\dagger$}
\affiliation{\KIAA}
\author{Kohei Inayoshi$^\dagger$}
\affiliation{\KIAA}
\author{Long Wang(\zhsun{王龙})$^\dagger$}
\affiliation{\SPA}
\author{Yi-Ming Hu(\zhsun{胡一鸣})$^\dagger$}
\email{huyiming@sysu.edu.cn}
\affiliation{\SPA}
\affiliation{\TRC}

\author{Pau Amaro-Seoane}
\affiliation{Institute for Multidisciplinary Mathematics, Polytechnic University of Val\`{e}ncia, Val\`{e}ncia 46022, Spain}
\author{Abbas Askar}
\affiliation{\NCAC}
\author{Cosimo Bambi}
\affiliation{\FDU}
\affiliation{\NUU}
\author{Pedro R. Capelo}
\affiliation{\UZH}
\author{Hong-Yu Chen(\zhsun{陈洪昱})}
\affiliation{\SPA}
\affiliation{\TRC}
\author{Alvin J. K. Chua(\zhsun{蔡靖康})}
\affiliation{\NUS}
\affiliation{\NUSM}
\author{Enrique Cond\'{e}s-Bre\~{n}a}
\affiliation{\KIAA}
\author{Lixin Dai(\zhsun{戴丽心})}
\affiliation{\HKU}
\author{Debtroy Das}
\affiliation{\FDU}
\author{Andrea Derdzinski}
\affiliation{\fisk}
\affiliation{\vandy}
\author{Hui-Min Fan(\zhsun{范会敏})}
\affiliation{\HBU}
\author{Michiko Fujii(\jpmin{藤井通子})}
\affiliation{\tokyo}
\author{Jie Gao(\zhsun{高洁})}
\affiliation{\SPA}
\affiliation{\TRC}
\author{Mudit Garg}
\affiliation{\UZH}
\author{Hongwei Ge(\zhsun{葛宏伟})}
\affiliation{\YNO}
\affiliation{\YNKL}
\affiliation{\UCAS}
\author{Mirek Giersz}
\affiliation{\NCAC}
\author{Shun-Jia Huang(\zhsun{黄顺佳})}
\affiliation{School of Science, Shenzhen Campus of Sun Yat-sen University, Shenzhen 518107, China}
\author{Arkadiusz Hypki}
\affiliation{\AMU}
\affiliation{\NCAC}
\author{Zheng-Cheng Liang(\zhsun{梁正程})}
\affiliation{\SPA}
\affiliation{\TRC}
\author{Bin Liu(\zhsun{刘彬})}
\affiliation{\ZJU}
\author{Dongdong Liu(\zhsun{刘栋栋})}
\affiliation{\YNO}
\author{Miaoxin Liu(\zhsun{刘淼昕})}
\affiliation{\NUS}
\author{Yunqi Liu(\zhsun{刘云旗})}
\affiliation{\YZU}
\author{Lucio Mayer}
\affiliation{\UZH}
\author{Nicola R. Napolitano}
\affiliation{\Naples}
\author{Peng Peng(\zhsun{彭朋})}
\affiliation{\KIAA}
\author{Yong Shao(\zhsun{邵勇})}
\affiliation{\NJU}
\author{Swarnim Shashank}
\affiliation{\FDU}
\author{Rongfeng Shen(\zhsun{申荣锋})}
\affiliation{\SPA}
\author{Hiromichi Tagawa(\jpmin{田川寛通})}
\affiliation{\SHAO}
\author{Ataru Tanikawa(\jpmin{谷川衝})}
\affiliation{\Fukui}
\author{Martina Toscani}
\affiliation{\UMIB}
\author{Ver\'{o}nica V\'{a}zquez-Aceves}
\affiliation{\KIAA}
\author{Hai-Tian Wang(\zhsun{王海天})}
\affiliation{\DLUT}
\author{Han Wang(\zhsun{王晗})}
\affiliation{\SPA}
\affiliation{\TRC}
\author{Shu-Xu Yi(\zhsun{易疏序})}
\affiliation{\IHEP}
\author{Jian-dong Zhang(\zhsun{张建东})}
\affiliation{\SPA}
\affiliation{\TRC}
\author{Xue-Ting Zhang(\zhsun{张雪婷})}
\affiliation{\SPA}
\affiliation{\TRC}
\author{Lianggui Zhu(\zhsun{朱良贵})}
\affiliation{\KIAA}
\author{Lorenz Zwick}
\affiliation{\NBI}

\author{Song Huang(\zhsun{黄崧})$^\ddagger$}
\affiliation{\THU}
\author{Jianwei Mei(\zhsun{梅健伟})$^\ddagger$}
\affiliation{\SPA}
\affiliation{\TRC}
\author{Yan Wang(\zhsun{王炎})$^\ddagger$}
\affiliation{\HUST}
\author{Yi Xie(\zhsun{谢懿})$^\ddagger$}
\affiliation{\PMO}
\author{Jiajun Zhang(\zhsun{张佳骏})$^\ddagger$}
\affiliation{\SHAO}

\author{Jun Luo(\zhsun{罗俊})}
\affiliation{\SPA}
\affiliation{\TRC}

\date{\today}

\newpage
\begin{abstract}
    The opening of the gravitational wave window has significantly enhanced our capacity to explore the universe's most extreme and dynamic sector.
    In the mHz frequency range, a diverse range of compact objects, from the most massive black holes at the farthest reaches of the Universe to the lightest white dwarfs in our cosmic backyard, generate a complex and dynamic symphony of gravitational wave signals.  
    Once recorded by gravitational wave detectors, these unique fingerprints have the potential to decipher the birth and growth of cosmic structures over a wide range of scales, from stellar binaries and stellar clusters to galaxies and large-scale structures. 
    The TianQin space-borne gravitational wave mission is scheduled for launch in the 2030s, with an operational lifespan of five years. 
    It will facilitate pivotal insights into the history of our universe. 
    This document presents a concise overview of the detectable sources of TianQin, outlining their characteristics, the challenges they present, and the expected impact of the TianQin observatory on our understanding of them. 
\end{abstract}

\clearpage
\thispagestyle{empty}
\begin{center}
\begin{tikzpicture}[remember picture, overlay]
    \node (O) at (current page.center) {};
    \node[inner sep=0pt] at (O) {%
    \includegraphics[width=\paperwidth,trim=60 0 60 0,clip]{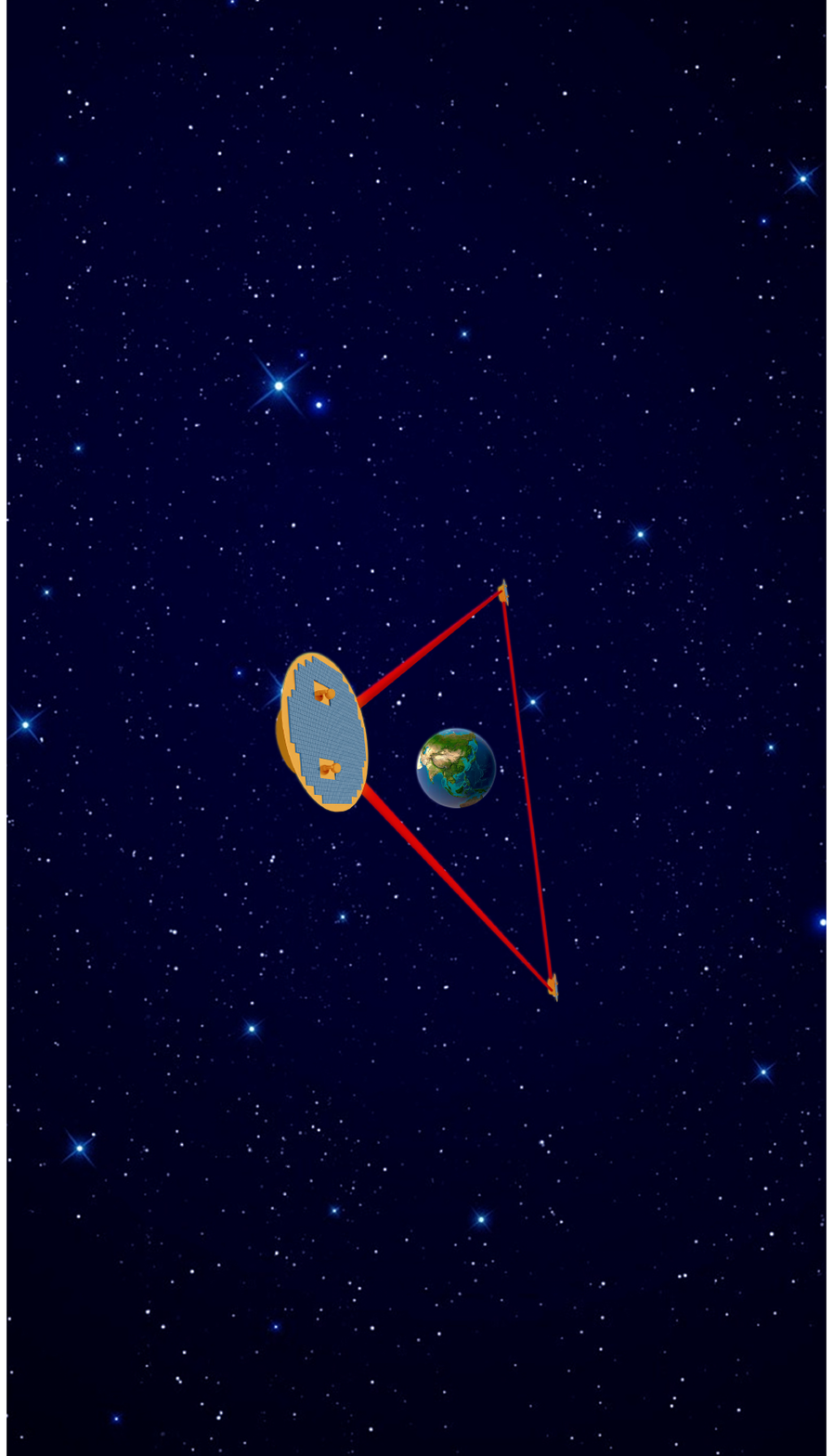}};
    \node[yshift=0.3\paperheight,
    font=\huge\color{white}]%
    at (O) {\parbox[t]{\textwidth}{\baselineskip=2em\centering\bfseries%
    \intitle}};
\end{tikzpicture}
\end{center}
\clearpage
\setcounter{page}{0}

\maketitle

\blfootnote{$^\dagger$ Section Coordinators}
\blfootnote{$^\ddagger$ Internal Reviewers}

\tableofcontents

\clearpage


\begin{tcolorbox}[tab2,title=List of acronyms,boxrule=0.5pt]
\centering
\begin{tabular}{p{0.18\linewidth}|p{0.72\linewidth}}
\textbf{Acronym} & \textbf{Definition} \\\hline\hline
AGN & Active Galactic Nuclei \\\hline
AM CVn & AM Canum Venaticorum star\\\hline
BH & Black Hole \\\hline
BBH & Binary Black Hole \\\hline
CO & Compact Object \\\hline
DWD & Double White Dwarf \\\hline
EM & Electromagnetic \\\hline
EMRI & Extreme Mass Ratio Inspiral  \\\hline
GCB & Galactic compact binary \\\hline
GR &  General Relativity \\\hline
GW &  Gravitational Wave \\\hline
IMBH & Intermediate Mass Black Hole \\\hline
IMRI & Intermediate Mass-Ratio Inspiral \\\hline
MBH &  Massive Black Hole (including IMBH and SMBH)\\\hline
NR &  Numerical Relativity\\\hline
NS &  Neutron Star \\\hline
PN &  Post-Newtonian \\\hline
Pop III & Population III \\\hline 
QPE & Quasi-Periodic Eruption \\\hline
SMBH & Supermassive Black Hole \\\hline
SN & Supernova \\ \hline
SNR & Signal-to-Noise Ratio \\ \hline
TDE & Tidal Disruption Event \\ \hline
UCXB & Ultra-compact X-ray binary \\ \hline
UV & Ultraviolet \\ \hline
VB & Verification Binary \\ \hline
VMS &  Very Massive Star \\\hline
WD &  White Dwarf \\\hline
\end{tabular}
\end{tcolorbox}

\clearpage
\section{Introduction}
\label{sec:introduction}

\coordinator{Yi-Ming Hu}

Gravitational waves (GWs), often referred to as the ``ripples of spacetime'', are a natural prediction of Einstein's general theory of relativity. 
Since their theoretical prediction over a century ago, numerous and different efforts have been made to detect these subtle disturbances through various principles~\cite{tianqin_2016, Weber:1967jye, kagra_2016, Kamionkowski:2015yta, Sazhin:1978myk}. 
Significant progress has been achieved in the hundred Hz band and the nHz band, notably by international collaborations such as the LIGO-Virgo collaboration~\cite{ligo_2016, LIGOScientific:2017vwq} and various pulsar timing array collaborations~\cite{NANOGrav:2023gor, EPTA:2023fyk, Reardon:2023gzh, Xu:2023wog}. 
We have entered a new era of GW astronomy, enabling us to uncover the secrets of the universe through this novel observational channel. 
Unlike electromagnetic (EM) waves, which interact with matter and thus get scattered and absorbed, GWs preserve and convey unaltered information from the innermost core about the most violent events in the cosmos. 
In many cases, when our interest lies in our universe's most extreme and compact objects (COs), there will be hardly any matter left to generate EM radiation in the first place, and the GW is often the only available messenger.
This makes them invaluable, particularly when studying extreme and COs where EM radiation is either minimal or non-existent. 
In this new era, we anticipate exciting discoveries that will deepen our understanding of the history and fate of stars, galaxies, and the universe, such as the evolution of massive binaries, the formation and growth of massive black holes (MBHs), and the behavior of stellar objects around galactic centers.

Similar to their EM counterparts, GWs span a wide range of frequencies, from the highest hundred-hertz detected by ground-based laser interferometers to the lowest atto-hertz imprints sought in the cosmic microwave background. 
Within this extensive frequency spectrum lies a rich array of sources in the millihertz range. 
Anticipated astronomical sources in this range include Galactic compact binaries (GCBs), mostly composed of double white dwarfs (DWDs)~\cite{Huang:2020,Ren:2023}, inspirals of nearby stellar-mass black holes (BHs)~\cite{liu_hu_2020}, extreme mass ratio inspirals (EMRIs)~\cite{fan_hu_2020}, and mergers of binary MBHs extending to the edge of our observable universe~\cite{TQ_MBH_2019}. 
The observable DWDs that are already detected through EM observation are also known as the verification binaries (VBs).
High-energy processes, such as the first-order electroweak phase transition in the early universe, may also produce observable signals in the data~\cite{liang_hu_2022}. 
The prospect of abundant observational sources in the near future holds the promise of significantly advancing and redefining the field of astrophysics.

\begin{figure}[htbp]
    \centering
    \includegraphics[width=0.8\linewidth]{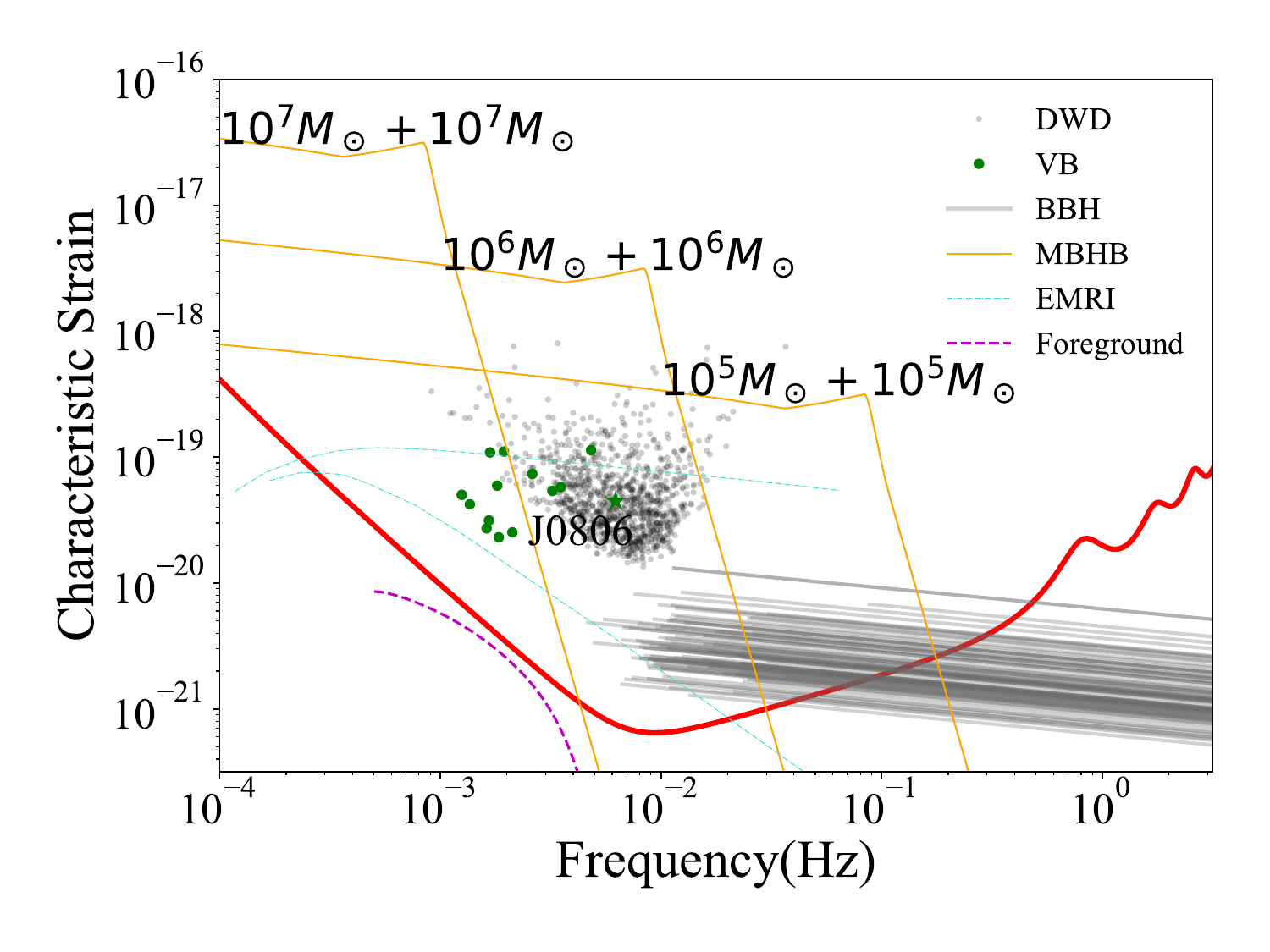}
    \caption{TianQin sensitivity (red) together with anticipated sources, showing loud DWDs and VBs, stellar mass binary black hole (BBH) inspirals, EMRIs, and mergers of binary MBHs with dots, grey lines, blue lines, and yellow lines, separately. The foreground (purple dashed line) after the five-year operation will be below the sensitivity curve.}
    \label{fig:TQ_source}
\end{figure}

TianQin, a proposed space-borne GW mission operating in the range of $10^{-4}-1$Hz, can achieve the above goals.
It is designed to have three identical satellites, operating in an orbit with a height of around $10^5$km, composing a regular triangle. 
Therefore, the arm length would be around $1.7\times 10^5$km. 
The core of the satellites are very stable test masses that are free-falling in space. 
The caging satellites monitor and follow the test masses, any non-gravitational disturbance will be compensated through the drag-free control so that ideally only the effect of passing GW will be recorded by the tiny change of arm length.
By accurately measuring the change of arm length through laser interferometers, one can depict the effect of the GW. 
Due to the geocentric configuration, the pointing of the TianQin constellation remains fixed.
In the current design, the pointing direction is HM Cancri, also known as RX J0806.3+1527, or J0806 for short.
Studies suggest that the periodic light curve in X-ray indicates that it is a binary white dwarf (WD) system with an orbital period of 5.4 minutes~\cite{2010ApJ...711L.138R}. 
The ecliptic latitude of J0806 is -4.7$^\circ$, therefore the TianQin plane is almost perpendicular to the ecliptic plane. 
During the operation, such a design is ideal for shielding the sunshine and keeping the satellites thermally stable. 
However, when the Sun's direction is too close to the orbital plane, the thermal fluctuation becomes too big and the satellite will switch off from the science mode. 
We expect two chunks of continuous three-month observation time during a whole year period. This means that the duty cycle of TianQin will be at most 50\%.
The target operation time would be in the mid-2030s.

The TianQin mission adopts a ``0-1-2-3" roadmap, with different indexes referring to the number of satellites involved in this stage. 
In the zeroth step, the TianQin team implemented the high-precision lunar laser ranging for all reflectors on the lunar surface~\cite{zhang_gao_2022}.
In the first step, the TianQin-1 technology demonstration satellite was launched, of the in-orbit experiment of all payloads indicates that their performances all met the designed requirements of TQ-1~\cite{Luo:2020bls}.
In the second step, a pair of satellites will be launched to use high-precision inter-satellite laser interferometry to map the Earth's gravity field~\cite{zhang_ming_2018}. 
Finally, in the third step, the TianQin GW observatory will be observing the gravitational universe.
The TianQin Research Center for Gravitational Physics was set up by Sun Yat-sen University in 2016 to advance the TianQin project.
By now, multiple laboratories are built to support the 200+ graduate students and 50+ research staff. 
From the perspective of GW astronomy, a first round of studies is performed to quantitatively assess TianQin's ability to different astronomical sources~\cite{Huang:2020, Ren:2023, liu_hu_2020, fan_hu_2020, TQ_MBH_2019, liang_hu_2022, torres-orjuela_huang_2023}.
Data analysis studies are also undergoing to ensure the reliable detections and parameter estimations~\cite{zhang_messenger_2022, Lu:2022ywf, cheng_li_2022, wang_harry_2024, lyu_li_2023, wu_hu_2024, chen_lyu_2024, li_et_al_2023, Gao:2024uqc, wang_chen_2024, zhang_korsakova_2024, Ye:2024bku} 

TianQin can significantly advance our understanding across a very wide scale of masses as well as distances.
Within our Galaxy, millions of compact binaries are predicted to emit GWs simultaneously.
In the nearby universe, TianQin can pick up the early inspiral signals years to decades before the final merger of the binary stellar mass BHs, while the final plunge has been identified in the hundred-hertz detectors.
Reaching out to the edge of our observable universe, in the centers of many galaxies lurks the very MBHs. Some of them will merge, while some of them can form highly unequal tight binaries.
Both these MBH mergers and EMRIs can be ideal laboratories to learn about MBHs.
The observation via GW remains the most reliable way to probe these systems, and the operation of TianQin will certainly help solve a large number of puzzles related to the birth and evolution of these sources. 

In a decade's time, the millihertz GW symphony will not be recorded by a lonely detector. 
The European-led project, LISA, which is expected to be launched in 2035, has been officially adopted by the European Space Agency and has been in the implementation phase.
Over the past decades, numerous efforts have been made to envision the millihertz GW universe associated with LISA, and have been summarized in a comprehensive review~\cite{lisa_2023}.

Since the publication of the LISA astrophysics whitepaper, new observations and simulations have been made, and our understanding of the millihertz GW sources has been updated.
In writing this whitepaper, we aim to update and reflect on the advances during the past years and refer interested readers to the LISA astrophysics whitepaper for a more complete and fundamental description.
We also hope to identify important issues that need to be solved before the TianQin operation to be better prepared for the future mHz GW study.
More importantly, TianQin has a different orbit, different frequency sweet spot, and different data downlink plan than LISA.
The operation of TianQin can bring unique benefits.
Additionally, most of the existing literature considers a single detector, while certain science goals, like highly precise sky localization, or the reliable detection of stochastic backgrounds independent of noise models, can only be implemented through a network of detectors. 
We then will highlight the contribution of TianQin on its own as well as by forming a network with LISA. 
Also, we dedicate this whitepaper to astrophysical sources so detailed discussions on the cosmological sources as well as topics like GW cosmology with TianQin will be written in separate whitepapers.

\vspace{3cm}

\begin{tcolorbox}[colback=red!5!white,colframe=red!75!black]
\textbf{\textcolor{NavyBlue}{
\begin{it}
GWs are ripples in spacetime caused by massive cosmic events. Unlike light waves, GWs do not interact significantly with matter, preserving information about events such as BH collisions.  These waves span a wide range of frequencies, from high-frequency waves detected by ground-based interferometers to low-frequency waves observed in the cosmic microwave background. The millihertz range is particularly rich, with sources including binary star systems, BH mergers, and early universe events.  TianQin, a proposed space-borne mission, aims to detect GWs in the millihertz range. It will use three satellites in an equilateral triangle formation, orbiting Earth and employing laser interferometry to measure changes in distance caused by GWs.  Expected to begin operations in the mid-2030s, TianQin will enhance our understanding of phenomena such as compact binaries within our Galaxy and MBH mergers at the universe's edge. Its unique capabilities, alongside the European-led LISA project, promise to advance the field of GW astronomy, providing new insights into the evolution of stars, galaxies, and BHs.  This document summarizes recent advancements in millihertz GW research and outlines TianQin's potential contributions, both independently and in collaboration with LISA. 
\end{it}
}}
\end{tcolorbox}

\clearpage
\section{Stellar physics}
\label{sec:stellar}
\coordinators{Shuai Liu \& Long Wang}

%





In this section, we present an overview of the potential implications for stellar physics arising from the detection of stellar-mass GW sources by TianQin. 
These GW sources originate from binary systems that include objects formed through stellar evolution, such as WDs, neutron stars (NSs), and BHs.

\begin{figure}[h]
    \centering
    \includegraphics[width=1\linewidth]{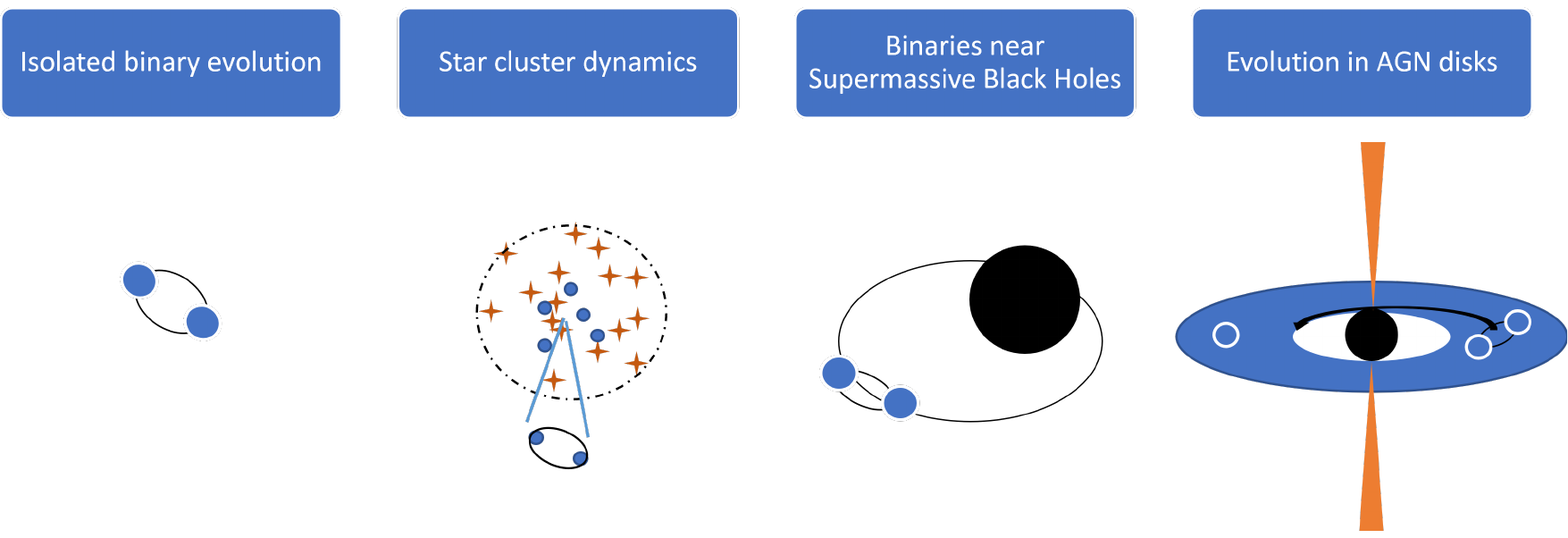}
    \caption{Stellar-mass binaries in different environments that may produce GW signals}
    \label{fig:binenv}
\end{figure}

Binary systems composed of WDs, NSs, and BHs can form in various environments, as illustrated in Figure~\ref{fig:binenv}. 
Stars may form in binary systems within giant molecular clouds, known as primordial binaries. 
Additionally, binaries can also form dynamically in dense stellar systems via three-body interactions. 
A subset of these binaries undergoes strong interactions with one another, ultimately leading to the formation of GW sources. 
These GW sources can exist within galactic fields, evolving either in isolation or within environments that influence their evolution.
Such environments include star clusters, such as low-mass open clusters, dense globular clusters and ancient Pop III 
star clusters. 
Furthermore, binaries may also exist within triple systems in nuclear star clusters at the galactic center, where an intermediate mass black hole (IMBH) or a supermassive black hole (SMBH) serves as the third component. Binaries may also form or be captured in the disk of AGN, these binary systems may interact with surrounding gas and emit EM signals together with GW.

In the following sections, we will provide a detailed discussion of binaries in various environments. 
Subsequently, we will explore the expected GW detections from these systems and discuss how TianQin can help distinguish binary systems formed under mechanisms mentioned above (e.g., Figure \ref{fig:ecc at frequency}).

\subsection{Binary evolution}
\contributors{Hongwei Ge, Ataru Tanikawa, Yong Shao, Dongdong Liu, Shunjia Huang, Long Wang, Shuai Liu, Yi-Ming Hu}

About half of the stars that we observe are in a binary system~\cite{kroupa_1995,sana_demink_2012,moe_distefano_2017}. Massive and compact binaries are considered major progenitors of the GW sources detected by LIGO and Virgo. 
However, ground-based detectors are limited in that they can only capture the final moments of binary mergers. 
In contrast, space-borne detectors, including TianQin, can measure the early phases of GW radiation from merging binaries, providing important constraints on their origins. 
In the following sections, we will focus on a few specific topics that TianQin may uncover significant discoveries.

\subsubsection{BBH within PISN mass gap}
Standard stellar evolution models predict that there are no BHs existing in the mass gap  $\sim65M_{\odot}-120M_{\odot}$ produced by pair-instability supernova (PISN)~\cite{heger_woosley_2002,Belczynski:2016jno,Woosley:2016hmi,Marchant:2018kun}. However, GW190521~\cite{LIGOScientific:2020ibl,abbott_et_al_2020} whose primary component fall in the PISN mass gap challenged the current stellar evolution models. In order to explain the formation of GW190521-like BBHs, various channels are proposed, which could be divided into two categories roughly. The first one is stellar evolution from isolated binaries with very low metallicity~\cite{Vink:2020nak,Tanikawa:2020abs,Costa:2020xbc,farrell_groh_2021}. The second one is that BHs within the PISN mass gap form by hierarchical mergers between BHs/stars~\cite{Fragione:2020han,martinez_fragione_2020,samsing_hotokezaka_2021,Anagnostou:2020tta,Gerosa:2021mno,wang_tang_2021,mapelli_et_al_2021,tagawa_kocsis_2021,Kimball:2020qyd,Liu:2020gif,DiCarlo:2019pmf,dicarlo_mapelli_2020,kremer_spera_2020,renzo_cantiello_2020}, or accretion onto BHs~\cite{safarzadeh_haiman_2020,natarajan_2020,liu_bromm_2020,rice_zhang_2021}, and then form BBHs with other BHs by gravitational capture process. The orbital eccentricities of sources formed via the former channel are expected to be lower than those of sources formed by the latter channel. Nevertheless, the current ground-based GW detectors, e.g., LIGO/Virgo, could not constrain the formation scenarios of GW190521-like BBHs by measuring their orbital eccentricities, because BBHs during merger phase tend to have negligible orbital eccentricities under GW circularization effect. 

During early inspiral phase, the orbital eccentricities of GW190521-like BBHs have not been fully circularized yet, carrying the imprint of formation channels. The inspiral GWs would last from months to years at the space-borne detectors TianQin and LISA band. Therefore, TianQin/LISA~\cite{liu_zhu_2022,toubiana_et_al_2021,sberna_babak_2022} have the potential to detect GW190521-like BBHs and constrain their formation channels by measuring orbital eccentricities.

\subsubsection{Binaries with BHs from Pop~III stars}\label{subsec:binary}

We can suppose that space-borne detectors including TianQin and LISA will detect two types of merging BBHs formed through binary evolution. For the first type, both two BHs have $\lesssim 100 M_\odot$, which are already discovered many times by LIGO and Virgo. For the second type, either of the two BHs have $\gtrsim 100M_\odot$, which has not yet been discovered by GW observations as of 2024. 
TianQin may detect much more merging BBHs due to the difference of frequency sensitivity~\cite{seto_kyutoku_2022}.
Using GW from eccentric BBH, it is also possible to constrain the modified gravity~\cite{Li2024} and potential candidates of dark matter~\cite{guo_zhong_2024}.

In this section, we describe how the second type mergers originating from Pop III 
stars may be detected by TianQin.
 
Pop~III stars are believed to be formed as massive stars ($10$--$1000M_\odot$) because of their metal-free nature~\cite{1998ApJ...508..141O, Abel_2002, 2004ARA&A..42...79B, 2008Sci...321..669Y, Hosokawa_2011, 2011MNRAS.413..543S, 2012MNRAS.422..290S, 2013RPPh...76k2901B, 2013ApJ...773..185S, 2014ApJ...792...32S, 2014ApJ...781...60H, Hirano_2015}. Moreover, several numerical simulations showed that they are born in multiple star systems as well as isolated binary stars~\cite{2010MNRAS.403...45S, 2011Sci...331.1040C, 2011ApJ...737...75G, 2011MNRAS.413..543S, 2012MNRAS.422..290S, 2013ApJ...773..185S, 2013ApJ...768..131V, 2014ApJ...792...32S, 2015MNRAS.448.1405M, Hirano_2015, Stacy2016, 2017MNRAS.470..898H, 2019ApJ...877...99S, 2020ApJ...892L..14S, sharda_federrath_2020}. These isolated binary stars may leave behind merging BHs discovered by GW observations~\cite{2004ApJ...608L..45B, 2014MNRAS.442.2963K, 2020MNRAS.498.3946K, 2021MNRAS.501L..49K, 2021PTEP.2021b1E01K, 2016MNRAS.460L..74H, 2017MNRAS.468.5020I, Wang2022, Tanikawa:2020abs, 2023MNRAS.524..307S, 2023MNRAS.525.2891C}. \citet{2021MNRAS.505L..69H} simulated Pop~III binary evolution up to $3000M_\odot$ in total, and showed that TianQin can detect their BBH mergers $1$--$10$ per year even if both the BH masses are limited to mass more than pair instability mass gap. Since the detection rate and mass distribution depend on Pop~III star formation rate and initial mass function, TianQin may reveal these features of Pop~III stars. They also found that TianQin would in principle have an advantage over LISA, finding Pop~III BBHs beyond the pair-instability mass gap. This is because TianQin is more sensitive to LISA at a high GW frequency.

\subsubsection{Probing the mass gap between NSs and BHs} 

Stellar evolution predicts that stars with masses larger than about  $8M_\odot$ undergo core collapse at the end of their lives and leave behind compact objects of either BHs or NSs. It is expected that the mass distribution of BHs and NSs can shed light on the physics of core-collapse supernovae. In the Milky Way, more than twenty BHs have been confirmed in X-ray binaries due to the measurements of their dynamical masses~\cite{Remillard2006, Casares2014}. 
The mass distribution of these BHs shows that they have a minimal mass of about $5M_\odot$~\cite{Ozel2010, Farr2011}. On the other hand, observations of radio pulsar binaries and X-ray binaries indicate a maximal mass of about $2.5M_\odot$ for NSs~\cite{Clark2002, Antoniadis2013, Cromartie2020, Romani2022}. Thus, it is suggested to exist a mass gap ($\sim 2.5-5M_\odot$) between the lightest BHs and the heaviest NSs. To explain the existence of this mass gap, a range of supernova-explosion mechanisms have been proposed \cite[e.g.,][]{Fryer2012, Ugliano2012, Kochanek2014, Liu2021}.

Recently, there is growing evidence that the mass gap is being populated \cite[see][for a compilation]{Shao2022}. For example, GRO J0422+32 is a low-mass X-ray binary with an accreting BH. With a new method to derive the orbital inclination of the binary, the dynamical mass of the BH was constrained to be $2.7^{+07}_{-0.5}M_\odot$, placing it within the mass gap~\cite{Casares2022}. In addition, the GW transient GW190814 contains a $2.59^{+0.08}_{-0.09}M_\odot$ compact object that definitely lies in the mass gap~\cite{Abbott2020}. In the globular cluster NGC 1851, PSR J0514-4002E was demonstrated to be an eccentric binary with a millisecond pulsar, in which the companion is a compact object with mass (between 2.09 and 2.71 $M_\odot$, 95\% confidence interval) being in the gap~\cite{Barr2024}. 

It is possible that mass-gap objects are the direct remnants of massive stars that experienced supernova explosions. However, mass increase of NSs via stable accretion or binary mergers may also lead to the formation of mass-gap objects \cite[e.g.,][]{Gao2022, Barr2024}. Through TianQin's observation of Galactic compact binaries, it is possible to assess the mass of the compacts. 
This survey can provide a portrait of the mass distribution of the Galactic compact objects, and has the potential to confirm the nature of the mass gap, and furthermore, constrain the formation of mass-gap objects and the mechanism of supernova explosions~\cite{Shao2021}.

\subsubsection{Binaries including WDs} 

\begin{figure}[htbp]
	\centering
	\includegraphics[width=1.0\linewidth]{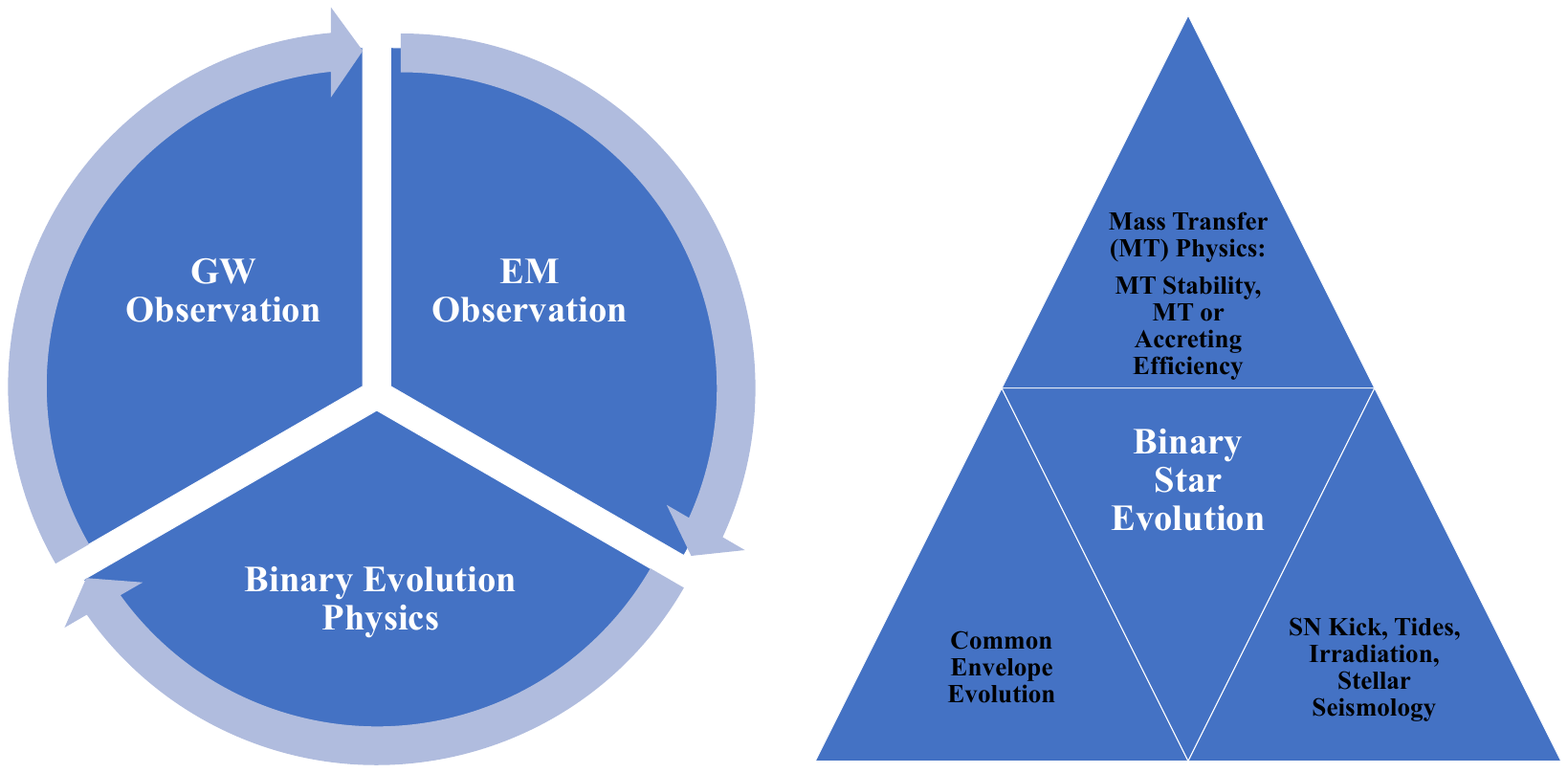}
	\caption{GW observations will verify and improve critical knowledge on binary star evolution.}
	\label{ge-binary}
\end{figure}

TianQin presents a significant opportunity to enhance our understanding of various astrophysical phenomena through the observation of GWs and associated EM wave observation with their counterparts. 
The progression of astrophysics and astronomy globally will benefit from advances in theoretical physics, the analysis of extensive data sets, the application of numerical simulations, and the observation of GWs and EM radiation. 
TianQin's observation of tens of thousands of DWD sources will aid in verifying and refining essential knowledge on binary star evolution. 
This includes mass transfer physics, common envelope evolution, supernova kick, tides, and irradiation (Figure\,\ref{ge-binary}).

From the binary evolution point of view, binary interactions, such as stable mass transfer, common envelope evolution, and tides, determine the formation and evolutionary fate of binary systems. In principle, all compact stellar binaries, including WD binaries, experienced binary interactions as the primary star evolved off the main sequence due to its evolution expansion or the system's angular momentum loss. The mass transfer process can be stable or unstable (entering a common envelope phase) against the dynamical timescale. 
It is generally believed that double compact or ultra-compact binaries suffered at least one common envelope to become the current short orbital period. 
In the theoretical modelling of the common envelope phase, it is often assumed that either the energy or the angular momentum is conserved.
However, these simplifications are likely to fail to grasp the complex nature of the common envelope phase, a large uncertainty in the physics during this stage, and more observations are needed to help understand the process. 
The TianQin observatory is expected to observe tens of thousands of compact binaries, and by looking at the population, to help decipher the common envelope process~\cite{webbink_1984, livio_soker_1988, 2005MNRAS.356..753N, ge_tout_2022, 2024ApJ...961..202G, zorotovic_schreiber_2022, 2023MNRAS.518.3966S}.

\begin{figure}[htbp]
	\centering
	\includegraphics[width=1.0\linewidth]{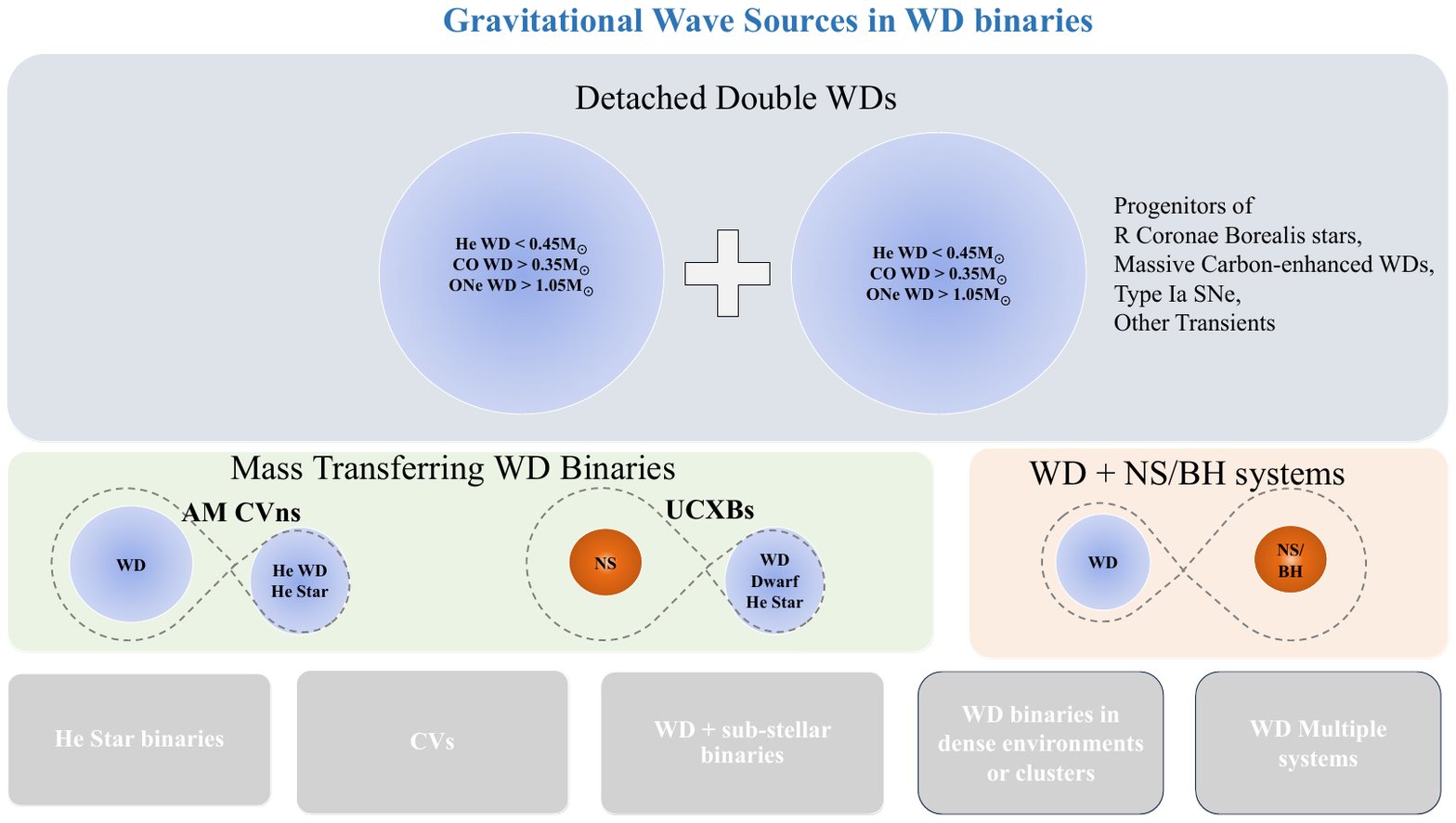}
	\caption{GW sources in WD binaries. The dominant GW sources are likely DWDs, including detached double WDs and mass-transferring WD binaries. A smaller proportion of GW sources are WD+NS/BH binaries. Other types in the grey background are possible important sources for nearby systems.}
	\label{ge-wds}
\end{figure}

For the binaries that contain a WD, from the classification point of view, the mainly TianQin sources from abundant to rare are likely DWDs, WD+NS/BH, WD+He star/evolved star/sub-stellar companion, and multiple WD systems (see Figure\,\ref{ge-wds}). Earlier studies contributing to GW sources for double WD populations cover detached WDs and mass transferring AM Canum Venaticorum systems (AM CVn) \cite[e.g.,][]{2001A&A...365..491N,2001A&A...368..939N,nelemans_yungelson_2004,2018MNRAS.480..302K,korol_toonen_2020,li_chen_2023}, WD+NS/BH including ultra-compact X-ray binaries (UCXBs) and partial low-mass X-ray binaries \cite[e.g.,][]{2001A&A...375..890N,2018PhRvL.121m1105T,2020ApJ...900L...8C,2021MNRAS.503.3540C,2021MNRAS.506.4654W, He2024}. In addition to above GW sources, nearby WD binaries are also interesting objects, such as WD+He star/evolved star/sub-stellar companion, WD binaries in multiple systems, and WD binaries in dense environments or clusters. We refer the reader to review papers and textbooks~\cite{2013A&ARv..21...59I,han_ge_2020,spera_trani_2022,2023arXiv231115778C,li_chen_2024,2023pbse.book.....T} for a more detailed background. We introduce the major scientific aspects of TianQin as follows.

TianQin is expected to detect GW signals from $\sim 10^4$ DWDs in Milky Way~\cite{Huang:2020, Jin2024}. 
In most cases, EM observation is very difficult, and therefore, the GW channel is the only practical approach.
Nevertheless, TianQin alone can detect numerous DWDs and can use these detections to constrain the underlying population. 
The deep link between the DWD systems and many observed and predicted transient signals will be put into detailed examination with TianQin observation.
For example, the mergers of DWDs possibly generate type Ia supernovae~\cite{1984ApJS...54..335I, webbink_1984, 2016MNRAS.461.3653L, 2018ApJ...865...15S, 2018MNRAS.473.5352L}, or accretion-induced collapse events \cite{1985ApJ...297..531N, 2020MNRAS.494.3422L, 2020RAA....20..135W, 2017hsn..book..317T}.
The link between DWD and SNe Ia is of particular interest to the general scientific community, as the ``normal" SNe Ia compose an important rung for the cosmic distance ladder. Many conclusions on the history and evolution of our universe are drawn based on their observations, for example, the identification of the accelerating expansion of the universe~\cite{1998AJ....116.1009R, 1999ApJ...517..565P}. 
Yet, the progenitors of normal type Ia supernovae are still under debate. Some argue that they can be DWDs~\cite{1984ApJS...54..335I, webbink_1984, 2018ApJ...865...15S, Yoshida2021}, while others believe that they are binaries with a WD and a non-degenerate star (maybe a main-sequence/subgiant, a red giant star or even a helium star)~\cite{1973ApJ...186.1007W, 1984ApJ...286..644N, 2018RAA....18...49W}. 
TianQin will contribute to resolving this long-standing problem. 
If such a nearby SN Ia explodes, we can check the archival data of TianQin so that its origin can be constrained.
GW observations with TianQin can distinguish SNe Ia progenitors by measuring the chirp masses of DWDs to determine if they exceed the Chandrasekhar limit, analyzing population statistics to confirm if their merger rates match observed SNe Ia frequencies, and identifying unique GW signatures (e.g., frequency evolution, orbital separations) to differentiate between double-degenerate (DWD) and single-degenerate (WD+non-degenerate) pathways.
Even in the absence of such a multi-messenger target, we can also constrain the origin of SNe Ia by studying the population properties of merging DWDs and comparing them to the observed SNe Ia.

A subset of the known DWDs predicted GW strains that are high enough to be individually detected, owing to their robust GW signals. These guaranteed sources, which are anticipated to be observable within weeks or a few months, are designated as VBs. At present, we know of only 18 such systems, despite theoretical predictions indicating that hundreds of them are potentially detectable within our Galaxy~\cite{Huang:2020, Ren:2023, Korol:2017}.
The challenge of finding more VBs lies in the fact that WDs are dim, and the identification of their binary nature requires long-term sensitive observations. 
Efforts have been taken to enlarge the VB samples.
For example, one can use the distance and variability of luminosity information provided by the Gaia satellite as the first screening, and use the light curve data accumulated by the Zwicky Transient Facility to check the periodicity, and the binarity by examining the actual light curve. 
In \citet{Ren:2023}, 429 candidates of short-period close binaries that contain a WD are identified.
Assuming the classical parameters, it is concluded that two (six) new VBs are found.

TianQin will also detect binaries with WDs and helium stars like TMTS J052610.43 +593445.1~\cite{2024NatAs.tmp...35L, 2018MNRAS.480..302K, 2023RAA....23a5008Q}. 
The WD$+$He star systems can also produce SNe Ia via either the single-degenerate model, the double-degenerate model, or the double detonation model~\cite{2009MNRAS.395..847W, 2013A&A...559A..94W, 2018MNRAS.473.5352L}.
SN 2012Z is a type Iax supernova. It is notable that a He star has been found at the position of SN 2012Z before the explosion occurred \cite {2014Natur.512...54M}. This strongly supports that type Iax supernovae may originate from the WD+He star systems.
Furthermore, the formation of such binaries should experience common envelope evolution. Thus, the detection of such binaries will be helpful in studying the common envelope evolution~\cite{2020A&A...636A.104B, 2021MNRAS.502.4877S}. TianQin will give a good opportunity to revisit common envelope evolution. This is similar to recent discoveries of astrometric compact binaries which indicate less efficient common envelope ejections than expected before~\cite{2023MNRAS.518.1057E, 2023MNRAS.521.4323E, 2024arXiv240206722E, 2023ApJ...946...79T, 2023AJ....166....6C, ye_chen_2024}, although a part of such astrometric compact binaries can be formed through dynamical captures~\cite{2023arXiv230613121D, 2023MNRAS.526..740R, 2024MNRAS.527.4031T}.
In addition, detached DWDs are particularly suitable for studying the physics of tides. DWDs affected by tides will yield information on the nature and origin of WD viscosity, which is still a missing piece in our understanding of WDs’ interior matter~\cite{Piro:2011,Fuller:2012,Dall'Osso:2014,mckernan:2016,Yu2020,Fiacco2024}. 

TianQin is set to play a pivotal role in mapping the intricate structure of the Milky Way. By detecting and analyzing the GW signals emitted by DWDs, TianQin has the potential to trace the spatial distribution of these celestial objects across the galaxy. This will provide an unprecedented 3D view of the galaxy's structure, offering insights into several key components: stellar disk, bulge, and dark matter halo~\cite{korol:2018a,korol:2018b,Wilhelm:2020, Korol2022}.
By combining EM observations of the motion of these DWDs in the sky, the mass of the Milky Way can be further measured via the independent GW channel.
Similarly, the DWDs detected by TianQin can also be used to study satellite galaxies in the Milky Way and the Large Magellanic Cloud and measure their mass~\cite{Korol:2020,Keim2023}.

\subsubsection{Ultra-compact X-ray binaries and AM CVn}

Ultra-compact X-ray binaries (UCXBs) are defined as NS/BH binaries that are accretion-powered X-ray usually with orbital period less than $\sim$ 1 hour \cite{2010NewAR..54...87N}. AM CVns are similar binaries but with WD accretors. The mass donors of these compact binaries are usually H-poor stars \cite{1986A&A...155...51S}. They play an important role in broad aspects of astrophysics \cite{2010NewAR..54...87N, 2012A&A...537A.104V, 2018PhRvL.121m1105T}:
(1) UCXBs and AM CVn have been thought to be strong continuous GW sources in the low-frequency region due to their short orbital periods, which can be detected by space-borne GW detectors like TianQin. 
(2) They provide important constraints on the binary evolution, such as the angular momentum loss mechanisms, the common-envelope evolution, and the mass-accretion process of compact objects, etc.
(3) They are excellent astrophysical laboratories since they are interesting X-ray sources and also the combination of ultrashort orbital periods, compact accretors, and mass donors with different chemical compositions. 
(4) UCXBs are progenitor candidates of millisecond radio pulsars.

There are several theoretical evolutionary channels for the formation of UCXBs and AM CVn.
For the formation of UCXB, it has been suggested that the NS/BH+main sequence star channel, the NS/BH+He star, the NS+WD channel and the accretion-induced collapse of ONe WD+He star/He WD channel are all possible progenitor models, e.g.\cite{1975MNRAS.172..493P, 2017MNRAS.470L...6S, 2020ApJ...900L...8C, 2021MNRAS.503.2776Y, 2021MNRAS.506.4654W, 2023MNRAS.521.6053L}.
According to these studies, there exist about several hundred detectable UCXBs by space-borne GW telescopes in our Galaxy.
For AM CVn systems, they are usually considered to be formed from CO/ONe WD+main sequence star systems, CO/ONe WD+He star systems or CO/ONe WD+He WD systems, and the predicted number is about 600 to 800, e.g.\cite{2008AstL...34..620Y, 2022ApJ...935....9C, 2021ApJ...910...22L}. The observation of space-borne GW detectors like TianQin will significantly improve our understanding of the formation and evolution of UCXBs and AM CVn~\cite{Qian2023,Wang2023,Kupfer2024,Szekerczes2023}.

\subsection{Star cluster dynamics}
\label{subsec:Nbody}

\contributors{Long Wang, Michiko Fujii, Shuai Liu, Ataru Tanikawa}

\begin{figure}[h]
    \centering
    \includegraphics[width=1\linewidth]{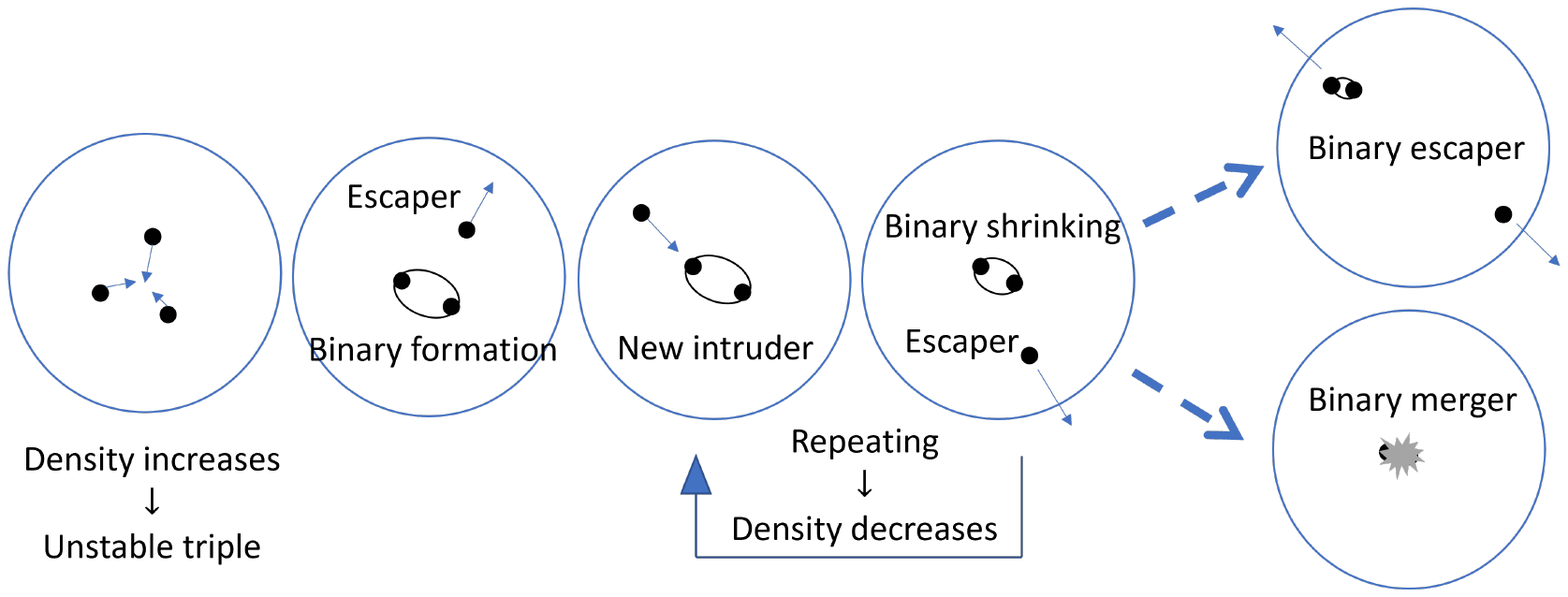}
    \caption{The binary heating scenario at the center of star clusters ejects neighboring stars to prevent core collapse while the binaries continue to shrink, ultimately leading to either mergers or ejections that produce GW sources.}
    \label{fig:binaryheating}
\end{figure}

In this section, we explore the impact of stellar dynamics within star clusters on the GW sources detected by TianQin. 
The dynamical evolution of star clusters experiences a phenomenon known as gravothermal catastrophe, which indicates that the central density of a star cluster continues to increase, ultimately driving the formation of binaries to prevent an infinite core collapse. 
As shown in Figure~\ref{fig:binaryheating}, these binaries act as a heating engine, ejecting surrounding stars outward while their orbits shrink, leading to eventual mergers into a single object or ejection from the cluster. 
This mechanism is efficient in producing GW sources, particularly when the binaries contain WDs, NSs, and BHs. 
Consequently, dense star clusters are considered ideal environments for generating GW sources. 
Additionally, the hierarchical mergers of stars may produce very massive stars, which could serve as progenitors of IMBHs or directly form IMBHs through the mergers of stellar-mass BHs. 
TianQin can detect GW signals from mergers of compact binaries (e.g., BH-BH, WD-WD) and hierarchical stellar mergers in clusters, observing their mass, spin, and redshift distributions to differentiate between IMBH formation via repeated mergers versus direct collapse.
If such IMBHs can form in ancient star clusters, such as Pop~III star clusters, they may play a crucial role in explaining the formation of SMBH seeds at high redshift. 
In this section, we will discuss how GW detection by TianQin may enhance our understanding of these topics.

\subsubsection{BBHs and IMBHs in globular clusters}
\label{subsubsec:cluster}

Globular clusters are dense stellar systems containing approximately a million stars, with a half-mass radius on the order of 10~pc. 
The high central density enables the efficient production of BBHs that may merge within a Hubble time. 
These BBHs are likely progenitors of GWs detected by ground-based observatories such as LIGO, Virgo, and KAGRA. 
However, distinguishing the progenitors of these GW events is challenging due to the limited available information. 
In particular, these GW progenitors can form either in dense star clusters through few-body interactions or in isolated binaries via binary stellar evolution. 
With a space-borne detector capable of observing low-frequency bands, it becomes possible to capture the early evolutionary phases of these BBHs, especially when their eccentricities have not yet been significantly circularized. 
This may allow us to identify high-eccentricity mergers, which are likely to originate in dense star clusters.
The formation of IMBHs in dense environments and their connection to MBH seeds is explored in Section~\ref{subsec:MBHseed}.

\begin{figure}
    \centering
    \includegraphics[width=\textwidth]{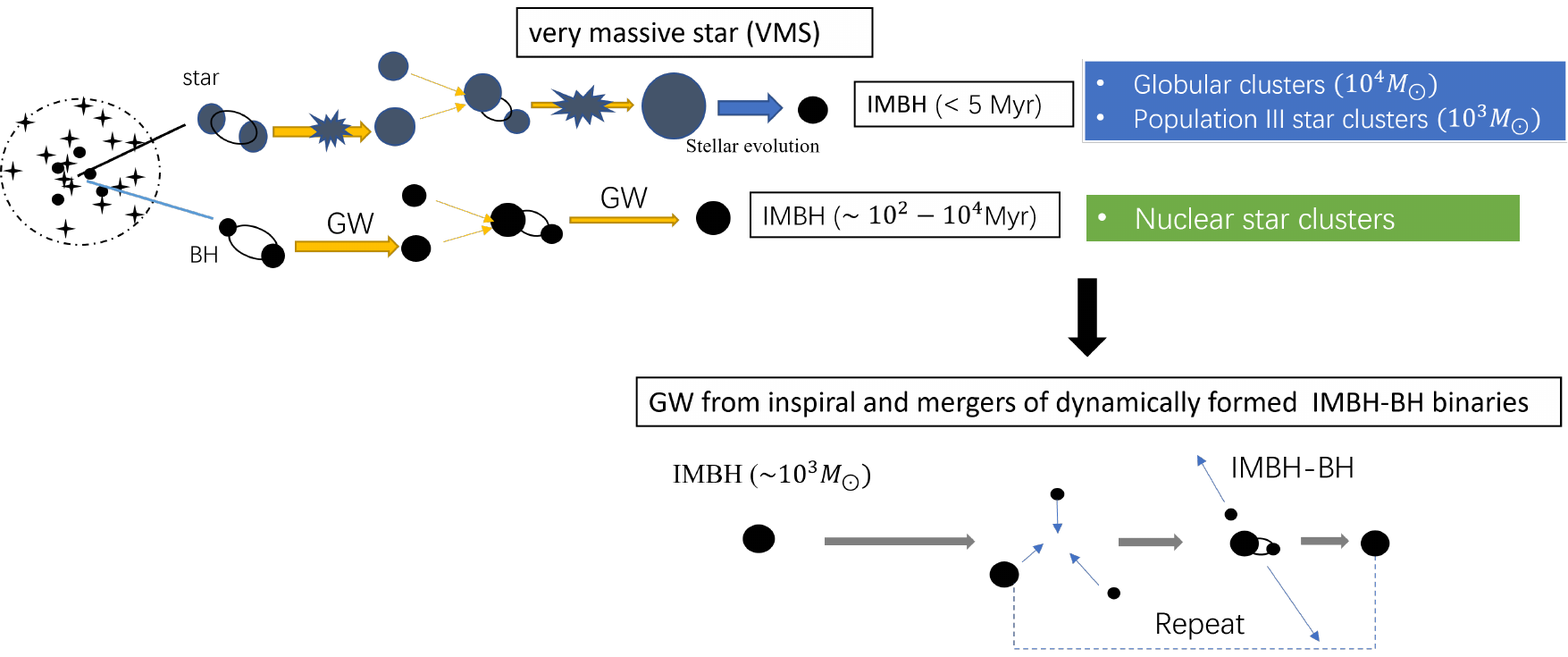}
    \caption{The two scenarios for IMBH formation in dense star clusters: 1) begins with the formation of a very massive star through runaway collisions of stars. Subsequently, this very massive star evolves into an IMBH; 2) hierarchical mergers of BBHs during the long-term evolution of star clusters. If a binary forms after the IMBH interacting with other BHs, the resulting GWs may be detectable.}
    \label{fig:imbh}
\end{figure}

In addition, globular clusters are also regarded as environments likely to host IMBHs~\cite{portegies-zwart_mcmillan_2002,portegies-zwart_baumgardt_2004,giersz_leigh_2015,Hong2020,Morawski2018,Rizzuto2021,rantala_naab_2024,fujii_wang_2024,purohit_fagrione_2024,miller_hamilton_2002, gultekin_miller_2004, freitag_gurkan_2006, Reinoso2018, antonini_gieles_2019,Fragione:2019dtr, Gonzalez2021, Fragione2022a,fragione_loeb_2022, mapelli_et_al_2021}. 
However, given that the maximum mass of a normally formed star is typically limited to around $100~M_\odot$, such stars cannot directly evolve into IMBHs with masses exceeding $100~M_\odot$. Consequently, to produce an IMBH, a massive progenitor known as a very massive star (VMS) with a mass greater than $10^3~M_\odot$ is required. 
Such VMSs may form through multiple mergers of stars. 
This scenario raises challenges though that the time between two mergers must be shorter than the timescale for wind mass loss (approximately $\le 1~\text{Myr}$) for a VMS to grow effectively. 
This rapid merger phenomenon, referred to as runaway collisions, is expected to occur in dense star clusters like globular clusters~\cite{portegies-zwart_mcmillan_2002,portegies-zwart_baumgardt_2004}. 
Figure~\ref{fig:imbh} illustrates this runaway scenario to form IMBH in dense star clusters.

Recent advancements in $N$-body simulation tools (e.g., Asura-bridge and PeTar) have enabled star cluster formation simulations that resolve individual stars and binaries without gravitational softening within giant molecular clouds~\cite{2020MNRAS.493.3398W,2020MNRAS.497..536W,2021PASJ...73.1036H}. 
The SIRIUS project~\cite{2021PASJ...73.1036H,2021PASJ...73.1074F} specifically conducted simulations of globular cluster formation to investigate the emergence of VMSs via runaway collisions of stars, revealing that globular clusters tend to host IMBHs with masses on the order of $10^4~M_\odot$~\cite{fujii_wang_2024}. 

Another potential formation pathway for IMBHs in globular clusters is through hierarchical mergers of BBHs~\cite{giersz_leigh_2015}. 
It is similar to the scenario of runaway formation of VMS but the timescale may be much longer, as demonstrated in Figure~\ref{fig:imbh}. 
However, when considering the high recoil kick velocities of BHs following mergers, the formation rate of IMBHs via this channel is influenced by the escape velocity of globular clusters, the metallicity of the stars, and the spins of the BHs~\cite{mapelli_et_al_2021}. 
This scenario may play an important role in IMBH formation in nuclear star clusters~\cite{antonini_gieles_2019}.

When globular clusters host an IMBH, a binary with IMBH and other stellar-mass BH or IMBH may form dynamically and generate GW signals during the inspiral and the merger phase, which may be possibly captured by the joint-detection of ground-based and space-borne GW detectors~\cite{Jani2020,torres-orjuela_2023}.


\subsubsection{BBHs and DWDs in young star clusters}

Young star clusters, including open clusters, are also environments where BBHs can form~\cite{2019MNRAS.486.3942K,2019MNRAS.483.1233R,DiCarlo:2019pmf,2020MNRAS.495.4268K,Banerjee2010,Banerjee2017,banerjee_2018,Banerjee2020,Banerjee2021}. 
Following the Gaia data release, thousands of open clusters have been identified in the Milky Way disk~\cite{Hunt2023}. 
These low-mass open clusters may contain fewer than 10 BHs, which can form GW sources either dynamically or through the evolution of primordial binaries. 
As smaller clusters form more frequently, the estimated total merger rate density is believed to be comparable to that of globular clusters. 
In addition to BBHs, DWDs are more common in open clusters. 
Due to their relatively low density, the evolution of DWDs is not significantly influenced by stellar dynamics, allowing them to behave similarly to those in the field where binary evolution is dominated by isolated processes such as mass transfer and GW emission.

\subsubsection{Pop III stars}

Pop III stars are first generation stars in the Universe.
The formation and evolution of Pop~III stars remain unclear.
Recently, the James Webb Space Telescope (JWST) opened a new era to explore the mystery of Pop~III star and galaxy formation at high redshift.
Observations have identified rapidly formed SMBHs at high redshift~\cite{Wu2015}.
Recent observation by the JWST have discovered an SMBH with a mass ranging from $10^7$ to $10^8 M_\odot$ at a redshift of 10.3, indicating a scenario involving heavy seeds~\cite{castellano_et_al_2022}. 
The exact process by which these seeds form remains unclear~\cite{Inayoshi_2020}. 
Pop III stars are regarded as a potential source of these seeds through the formation of VMS and IMBH.
Additionally, \citet{Xing2023} identified a potential pair-instability supernova within Pop~III VMS using data from the Large Sky Area Multi-Object Fiber Spectroscopic Telescope (LAMOST). 
Future JWST observations hold the potential to detect lensed Pop~III VMS or clusters~\cite{Bovill2024,Larkin2023}. 
The detection of potential GW signals from Pop~III IMBHs is particularly intriguing for understanding the formation of MBH seeds in the early universe.
These combined multi-messenger observations can offer valuable insights into the formation of Pop III stars and MBHs.

Pop~III stars may originate as massive stars that eventually evolve into IMBHs or SMBH seeds, potentially detectable by TianQin. 
Alternatively, as illustrated in Figure~\ref{fig:imbh}, Pop~III stars could form with lower masses, leading to the formation of star clusters within mini dark matter halos at redshifts exceeding 10.
In such scenarios, hydrodynamical studies suggest that Pop~III stars may exhibit top-heavy initial mass functions, with a majority evolving into compact objects and BHs~\cite{Stacy2016,sharda_federrath_2020,Chon2021,Latif2022}. 
This could result in the merging of massive stars to form a supermassive star, potentially giving rise to an IMBH~\cite{Sakurai2017}.

In contrast to isolated IMBHs, Pop~III star clusters have the capacity to host both IMBHs and stellar-mass BHs. 
\citet{Wang2022} suggests that if the mini dark matter halos hosting Pop~III star clusters persist until the present day, their strong gravitational potential could shield them from disruption. 
Consequently, GWs from Pop~III stars could be detected across a broad range of redshifts, from 20 to the present.
The space-borne GW observatories like TianQin can detect such events with redshifts below 2. Some of these mergers may exhibit extremely high eccentricity, allowing TianQin to observe their eccentricity evolution during detection windows.

\subsubsection{Triple BH in Pop~III cluster} 

\begin{figure}
\centering
    \includegraphics[width=0.5\textwidth]{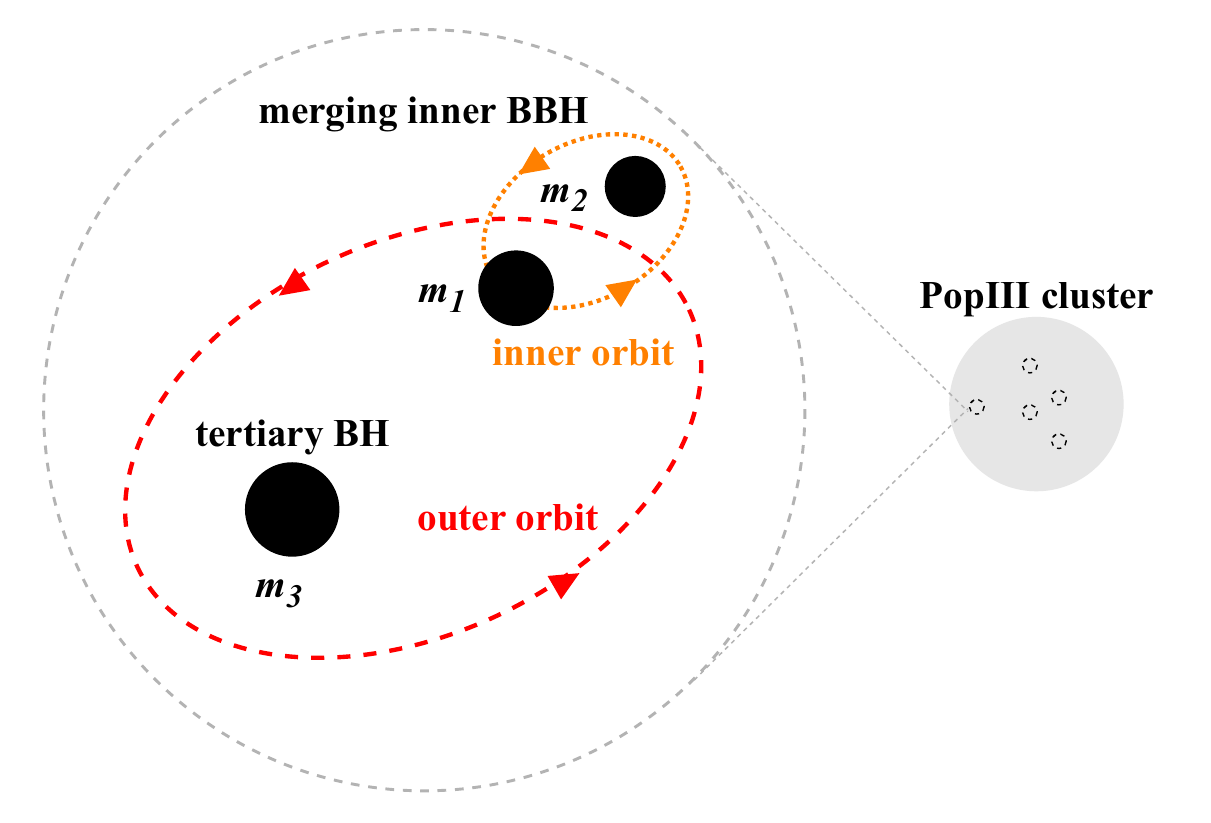}
    \caption{A sketch of merging triple BHs in Pop~III clusters. The gray ball represents a Pop~III cluster, in which small gray dashed circles are merging triple BHs. The enlarged version of a merging triple BH is in a gray large dashed circle, where a merging inner BBH with primary mass $m_{1}$ and secondary mass $m_{2}$ ($m_{2}\le m_{1}$) orbits around a tertiary BH with mass $m_{3}$. Orange dotted and red dashed circles are inner orbit with (semimajor axis $a_{1}$ and orbital eccentricity $e_{1}$) and outer orbit with (semimajor axis $a_{2}$ and orbital eccentricity $e_{2}$), respectively. The angle between inner and outer orbital planes is denoted by $i_{\rm mut}$.}
\label{fig:sketch triple bh}
\end{figure}

In dense star environments, such as globular clusters and young star clusters, BHs could form hierarchical triple BHs systems (merging triple BHs) via dynamical processes, where merging inner BBHs orbit around distant third BHs while orbiting each other \cite[e.g.,][]{2014ApJ...781...45A, Fragione:2019dtr, martinez_fragione_2020, Trani:2021tan, Britt:2021dtg}. Theoretical studies suggest that BHs could also be formed via the evolution of massive stars with very low metallicity in Pop~III clusters with thousands of stars embedded in mini dark matter haloes~\cite{Wang2022,Liu:2023zea}. BHs in Pop~III clusters also have possibility to form merging triple BHs for them via dynamical interactions~\cite{Liu:2023zea}, as shown in Figure~\ref{fig:sketch triple bh}. They could merge from the redshift of $z=20$ to the present universe, if the mini dark matter haloes protect clusters from disrupting by the potential of galaxies. The merger rate of inner BBHs within triple BHs averaged over redshift could reach up to $0.4 {\rm Gpc^{-3}yr^{-1}}$, contributing to about 3\% merger rate of all the merging BBHs in Pop~III clusters. The orbital evolution of inner BBHs within merging triple BHs will be different from that of merging isolated BBHs, due to the perturbation from the third BHs. Therefore, investigating the orbital elements of merging BBHs is important in exploring their formation channels in Pop~III clusters.

\begin{figure}[htbp]
    \centering
    \subfloat[]{
        \includegraphics[width=0.5\textwidth]{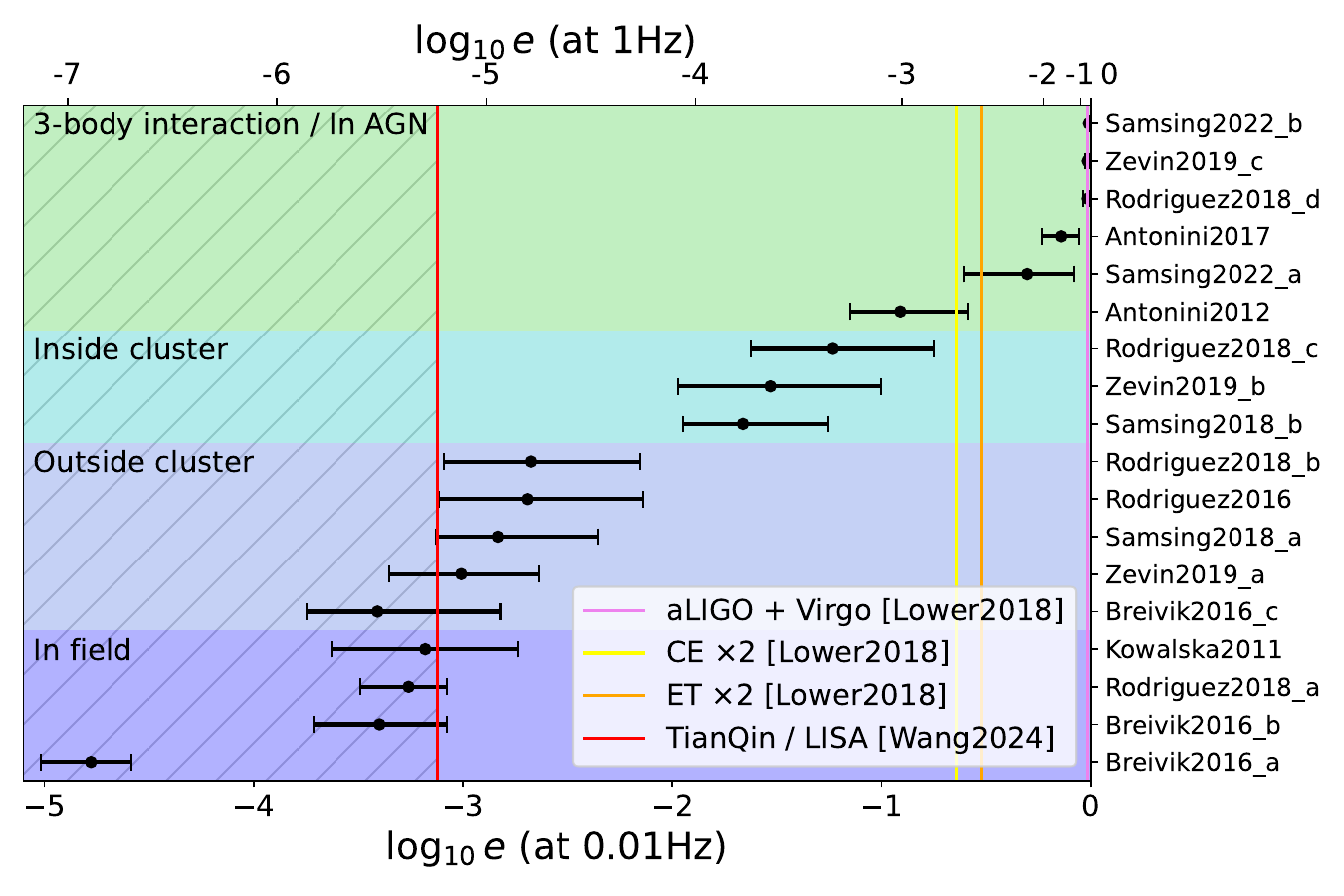}}
    \subfloat[]{
        \includegraphics[width=0.45\textwidth, height=0.3\textwidth]{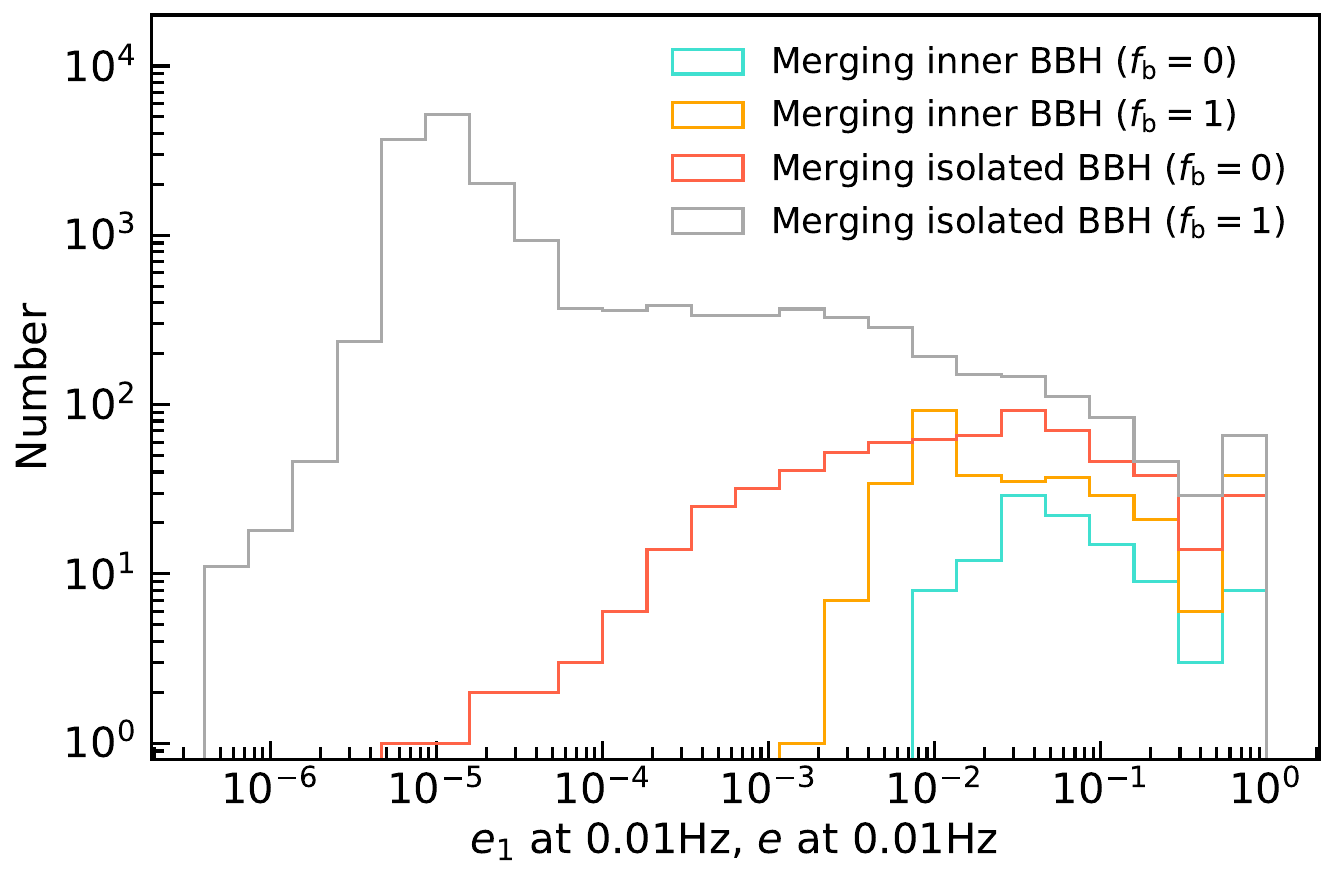}}
    \caption{(a) Predicted eccentricity distributions from different formation channels of merging BBHs at 0.01Hz where TianQin is sensitive, figure adapted from Figure 1 in \citet{wang_harry_2024}. The black dots and error bars represent the median values and 50\% credible intervals, respectively. The vertical lines show the minimum detectable eccentricities of different GW observatories.
    (b) The distribution of orbital eccentricity $e_{1}$($e$) of merging inner (isolated) BBHs at $0.01$Hz where TianQin is sensitive. In both cases, where all the primordial stars are unpaired ($f_{\rm b}=0$) and paired ($f_{\rm b}=1$), the distributions are plotted in different colors, respectively~\cite{Liu:2023zea}.}
    \label{fig:ecc at frequency}
\end{figure}

The orbital eccentricity is one of the important imprints carrying the formation channels of merging BBHs. Different formation channels would lead to merging BBHs having different orbital eccentricities, as shown on the panel (a) in Figure \ref{fig:ecc at frequency}. In Pop~III clusters, compared with merging isolated BBHs, the orbital eccentricities of merging inner BBHs $e_{1}$ are expected to be higher under the dynamical perturbation from the third BHs. The orbital eccentricities $e_{1}$ could approach unit via the interesting Zeipel-Lidov-Kozai effect~\cite{vonZeipel, Lidov, Kozai}, i.e., they oscillate approximatively periodically over time. The distribution of $e_{1}$ at 0.01 Hz where TianQin is sensitive, along with that of merging isolated BBHs $e$, are illustrated on panel (b) in Figure~\ref{fig:ecc at frequency}. Most of $e_{1}$ are larger than 0.01, with a fraction of sources having $e_{1}\sim1$. However, most merging isolated BBHs have $e<10^{-2}$. Therefore, if TianQin detected merging BBHs in Pop~III clusters, they would have the potential to distinguish their formation channels by measuring their orbital eccentricities.

%

\subsection{Dynamics of binaries near a rotating SMBHs}

\contributors{Liu Bin}\vspace{-0.5cm}

\subsubsection{Relativistic effects in triples}

In this section, we are interested in stellar-mass BBH evolution near an SMBH.
Such BBHs may exist in abundance in the nuclear cluster around the SMBH
due to various dynamical processes, such as scatterings, mass segregation and/or gas interactions in disks of AGN
\cite{OLeary2009, Leigh2018, McKernan2012, Bartos2017, Tagawa2020}.

We consider a binary with masses $m_1$, $m_2$, semimajor axis $a_{\rm IN}$ and eccentricity $e_{\rm IN}$,
moving around a rotating SMBH tertiary ($m_3$) on a wider orbit with $a_{\rm OUT}$ and $e_{\rm OUT}$.
The angular momenta of the inner, outer binaries, and SMBH spin are
denoted by $\textbf{L}_{\rm IN}$, $\textbf{L}_{\rm OUT}$, and $\textbf{S}_3$, respectively.
Gravitational perturbation from the SMBH makes the inner binary precess, and may also induce
von Zeipel-Lidov-Kozai ~\cite{vonZeipel, Lidov, Kozai} eccentricity oscillations
if the mutual inclination between two orbital planes is sufficiently high.
The first-order post-Newtonian (PN) theory introduces pericenter precession in both inner and outer binaries.

Since the tertiary mass is much larger than the masses of the inner binary,
several general relativity (GR) effects involving the SMBH can generate extra precessions
on the binary orbits, and qualitatively change the dynamics~\cite{NaozGR, C.M.WillPRD}.
By recognizing that the inner binary's orbital angular momentum
$\textbf{L}_{\rm IN}$ behaves like a ``spin",
\cite{Liu2019ApJL} identifies the impacts of several GR effects that have been little explored, including:
\begin{itemize}
\item \textit{ Effect I}: Precession of $\textbf{L}_\OUT$ around $\textbf{S}_3$ with rate\\
$~~~~~~~~~~~~~~~~~~~~~~~~~~~~~~~~~~~~~~~~~~\Omega_\mathrm{L_\OUT S_3}=\frac{GS_3(4+3m_{12}/m_3)}{2c^2a_{\rm OUT}^3(1-e_{\rm OUT}^2)^{3/2}}
~~~~~~~~$(1.5PN effect),

\item \textit{ Effect II}: Precession of $\textbf{L}_{\rm IN}$ around $\textbf{L}_\OUT$ with rate\\
$~~~~~~~~~~~~~~~~~~~~~~~~~~~~~~~~~~~~~~~~~~
\Omega_\mathrm{L_{\rm IN} L_\OUT}=\frac{3}{2}\frac{G (m_3+\mu_\OUT/3)n_\OUT}{c^2a_{\rm OUT}(1-e_{\rm OUT}^2)}~~~~~~~~$(1PN effect),

\item \textit{ Effect III}: Precession of $\textbf{L}_{\rm IN}$ around $\textbf{S}_3$ with rate\\
$~~~~~~~~~~~~~~~~~~~~~~~~~~~~~~~~~~~~~~~~~~\Omega_\mathrm{L_{\rm IN} S_3}=\frac{GS_3}{2c^2a_{\rm OUT}^3(1-e_{\rm OUT}^2)^{3/2}}~~~~~~~~$(1.5PN effect),
\end{itemize}
where $\mu_\OUT=(m_1+m_2)m_3/(m_1+m_2+m_3)$ is the reduced mass and $n_\OUT=[G(m_1+m_2+m_3)/a_{\rm OUT}^3]^{1/2}$ is the mean motion
of the outer binary. Note that the orbital eccentricity precesses in a similar way to keep $\textbf{L} \cdot \textbf{e}=0$.
Compared to the standard Zeipel-Lidov-Kozai effect, \citet{Liu2019ApJL,Liu2019ApJL} show that
including Effects I-III can dramatically widen the inclination window for $e_{\rm IN}-$excitation.
The overlapping resonances give rise to widespread chaos,
causing systems with modest inclination angles to attain extreme eccentricity growth.
The operation of space-borne GW observatories like TianQin enables the possibilities to depict the three body effects by observing these PN effects~\cite{zwick_tiede_2024}.

\subsubsection{Probing the spins of SMBHs with GWs from surrounding compact binaries}

The spins of the SMBHs at the centers of galaxies are poorly constrained; this is the
case even for Sgr A$^\ast$ in the Galactic center~\cite{Ghez1998, Ghez2008, Genzel2010}.
The spin vector of an accreting SMBH could in principle be constrained by
modeling the accretion/radiation processes~\cite{Moscibrodzka2009, Broderick2011, Shcherbakov2012} and comparing with observations, such as those of Sgr A$^\ast$ and M87
from Event Horizon Telescope~\cite{Dexter2010, Broderick2016, EHT2019}.
The Galactic center hosts a population of young massive stars~\cite{Genzel2000, Merritt2013, Alexander2017};
it has been suggested that
the relativistic frame-dragging effect on their orbits could put constraints on the Sgr A$^\ast$'s spin
\cite{Levin2003, Fragione2020,vazquez-aceves_lin_2023,vazquez-aceves_lin_2024}.

Given the fact that the S-stars around Sgr A$^\ast$ are close to the SMBH,
it is likely that binaries of compact objects could be present near SgrA$^\ast$
\cite{Antonini2012, Stephan2019}.
These compact binaries may radiate GWs in the low-frequency band ($0.001-1$ Hz), which
can be detectable by the planned space-borne GW observatory, such as TianQin~\cite{tianqin_2016}.
As the BBH precesses, the GW waveform undergoes both amplitude
and phase modulations, thereby allowing the measurement of the change in the orientation of $\hat {\textbf{L}}_{\rm IN}$~\cite{yu_chen_2021}.
\citet{liu_lai_2022} has demonstrated that the
spin of SMBH can significantly modify the orientation dynamics of the compact binary.
Therefore, in principle, detecting such compact binaries with TianQin may provide a new
probe to the spins of SMBHs, including that of SgrA$^\ast$.

\begin{figure}
\begin{centering}
\includegraphics[width=17cm]{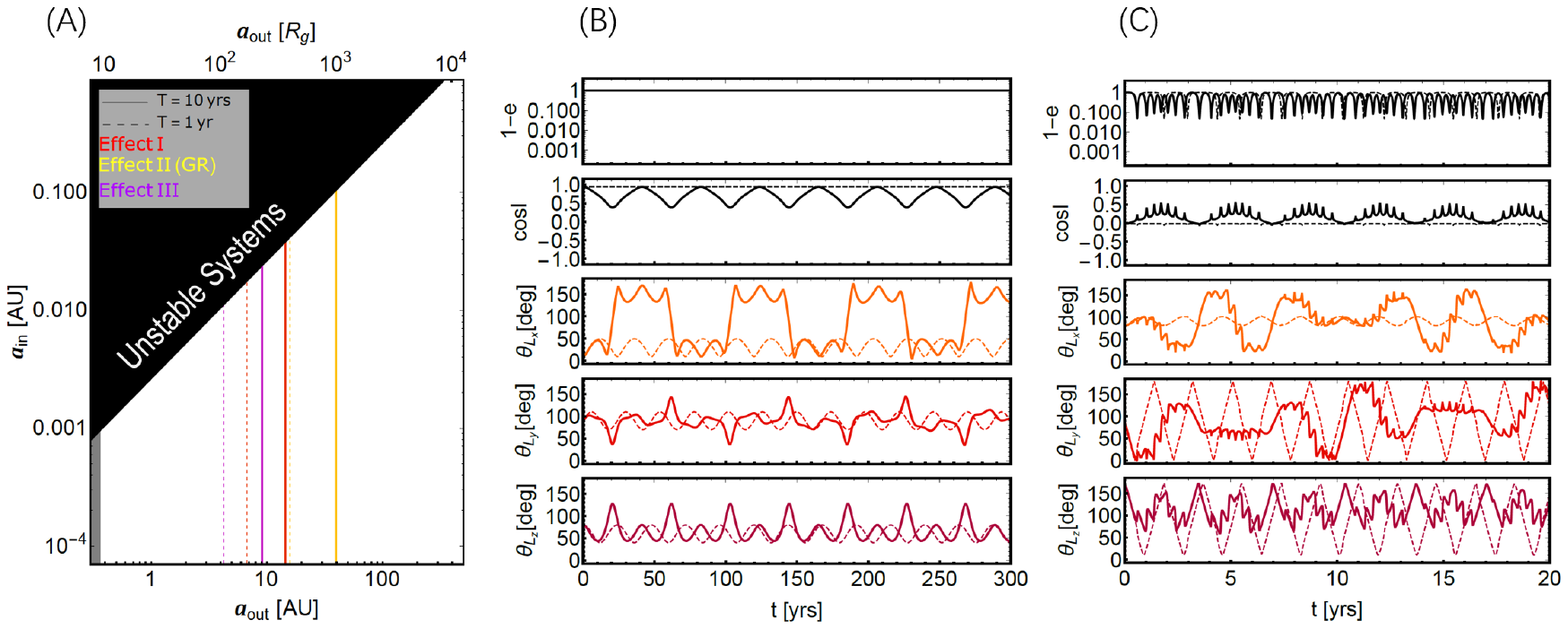}
\caption{
Panel (A) shows the parameter space in $a_{\rm IN}-a_{\rm OUT}$ plane indicating the relative importance of various GR effects.
The system parameters are $m_1=20M_\odot$, $m_2=10M_\odot$, $m_3=4\times10^6M_\odot$, $e_{\rm IN}=0$, $e_{\rm OUT}=0.5$ and $\chi_3=1$.
The black region corresponds to dynamically unstable triple systems
(the instability limit according to \citet{Kiseleva1996}), and the gray region indicates the innermost stable circular orbits 
for the outer binary, with $a_{\rm OUT}\leq9R_g=9(G m_3)/c^2$.
The solid (dashed) lines are obtained
by setting the relevant timescales.
Panels (B) and (C) show the eccentricity $e$, inclination $\text{I}$ (the angle between ${\hat{\textbf L}}_\IN$ and ${\hat{\textbf L}}_{\rm OUT}$), and
the misalignment angles ($\theta_{Lx}, \theta_{Ly}, \theta_{Lz}$) between $\hat {\textbf{L}}_\IN$ and the fixed x, y, z axes
(where the z-axis is aligned with $\hat {\textbf{S}}_3$ and the x-axis is in the initial $\hat {\textbf{S}}_3-\hat {\textbf{L}}_\OUT$ plane).
The parameters are $a_{\rm OUT}=15.8\au$, $I_0=20^\circ$, $\alpha=60^\circ$ (panel B)
and $a_{\rm OUT}=10\au$, $I_0=91^\circ$, $\alpha=80^\circ$ (panel C).
Here, $\alpha$ is the angle between $\hat {\textbf{L}}_{\rm OUT}$ and $\hat {\textbf{S}}_3$, and
$I$ is the angle between $\hat {\textbf{L}}_\IN$ and $\hat {\textbf{L}}_\OUT$.
The solid/dashed trajectories represent the evolution with/without the effects included by the SMBH spin (Effects II-III).
}
\label{fig:1}
\end{centering}
\end{figure}

Based on the analysis of relativistic effects in a BBH$+$SMBH system,
the evolution of $\textbf{L}_{\rm IN}$ is more sensitive to the SMBH spin than the orbital eccentricity.
This is because the eccentricity excitation can be completely suppressed
if the mutual inclination angle lies outside the Zeipel-Lidov-Kozai window, or the binary is relatively far away from
the SMBH. The panel (A) of Figure~\ref{fig:1} presents the parameter space indicating the relative importance of various GR effects
for BBHs around SgrA$^\ast$.
To obtain variations of the orbit on relatively short timescales ($\lesssim10$yrs),
the BBH cannot be too far away from the SMBH (i.e., $a_{\rm OUT}\lesssim100$AU).

The panels (B) and (C) of Figure~\ref{fig:1} illustrate how the GR effects induced by the spinning SMBH modify the evolution of
$\hat {\textbf{L}}_\IN$ of a BBH.
The orbital inclination $I$ undergoes significant change due to the spin effects,
and the misalignment angles $\theta_{\mathrm{L}_\mathrm{x,y,z}}$ between $\hat {\textbf{L}}_\IN$ and the fixed x, y, z axes
exhibit dramatic oscillations. 

Since $\theta_{\mathrm{L}_\mathrm{x,y,z}}$ evolves irregularly on the secular timescale that
could be longer than the mission lifetime of future GW observatories like TianQin,
the rate of change of $\hat {\textbf{L}}_\IN$ (and of $\theta_{\mathrm{L}_\mathrm{x,y,z}}$)
would be a useful observable indicator to track the evolution of the BBH.
We recognize that for typical BBHs ($m_1\sim20M_\odot$ and $m_2\sim10M_\odot$),
the maximum $|d\hat {\textbf{L}}_\IN/dt|$ can reach many tens of degrees per year for BBHs
emitting GWs in the low-frequency band ($10^{-3}-10^{-1}$Hz).
In contrast, without the spin effects of the SMBH,
the rate of change of $\hat {\textbf{L}}_\IN$ is a constant.

A joint detection of multiple compact binary systems
may be necessary to reduce the degeneracy of the GW signals on various parameters,
and provide sufficient constraints on the SMBH spin.
Future studies on detailed strategy to measure the SMBH spin using low-frequency GWs from
compact binaries would be of great value.

\subsection{Evolution in AGN disks}
\label{subsec:AGN_BBH}

\contributors{Hiromichi Tagawa}


Another environment for mergers among stellar-mass BH are AGN disks~\cite{Bartos2017,Stone17,McKernan17}. 
In an AGN disk, compact objects (COs) including stellar-mass BHs 
are embedded through capture via dynamical interactions between the nuclear star cluster and the AGN disk~\cite{Toyouchi2020,LiXinyu2020,Generozov2023,Wang2023_capture,Vaccaro2024}. 
Additionally, in-situ star formation contributes to the presence of COs in the disk 
\cite{Levin2003,Goodman2004,Milosavljevic2004,Gilbaum2022,Epstein-Martin2024}. 
COs 
in the AGN disk migrate 
and are predicted to accumulate in migration traps and gap-forming regions 
\cite{Bellovary16,Tagawa2020,Grishin2023}. 
In these regions, they efficiently encounter other COs and 
form binaries through the influence of gaseous torques~\cite{LiJiaru2022,Rowan2022,DeLaurentiis2023,Rozner2023,Whitehead2023,Rowan2023,Qian2023,Whitehead_2023b_novae} and GW emission~\cite{Boekholt2023,LiJiaru2023,Tagawa2021_Eccentricity,samsing_bartos_2022}. 
The accumulation of COs in different regions of the AGN disks, such as the outer regions (gaps~\cite{Tagawa2020} or migration traps due to thermal torques~\cite{Grishin2023}) and inner regions (migration traps due to Type I migration torques~\cite{Bellovary16,Yang2019,Yang2019_spin,Peng2021,Pan2021}), respectively, leads to the hardening of binaries mainly through binary-single interactions~\cite{Tagawa2020} and gaseous torques~\cite{Secunda2020}, before GWs trigger the mergers.

If COs frequently merge within AGN disks, it would be of great interest to observe them using space-borne GW detectors like TianQin. 
These observatories would provide valuable information about the evolution of binaries, the specific environments where the mergers occur, and the potential EM counterparts to GW signals.  
Additionally, the gas-driven migration mechanisms described here may amplify detection prospects for TianQin through multi-messenger synergies with EM counterparts (see Section~\ref{sec:IMBH_AGN} and Section~\ref{subsubsec:eimriem}).

First, the evolutionary processes of binaries can be revealed. 
Theoretical studies have shown that the evolution of binary separations~\cite{LiYanPing2021,LiYaPing2022_hotdisk,LiRixin2023_eos2,LiRixin2023_viscosity3,LiRixin2022_1,Dempsey2022,Dittmann2023}, eccentricities~\cite{LiRixin2022_1,Dempsey2022,Dittmann2023}, and spins~\cite{Tagawa2020_spin,ChenYiXian2022,LiYaPing2022_Spin_ecc,ChenYiXian2023_spin} by interacting with circum-binary disks is expected to depend on various parameters, 
though understanding of these processes is not yet complete. 
Nevertheless, gaseous torques are likely to affect GW signals and can be detected because they are usually stronger than GW torques at $\lesssim 100~{\rm mHz}$~\cite{Bartos2017}. 
Consequently, studying the influence of gaseous torques becomes crucial in order to gain a comprehensive understanding of the evolution of binaries. 

Furthermore, the evolution of binary masses is highly uncertain due to complex feedback and inflow/accretion processes~\cite{Tagawa2022,ChenKen2023,Tagawa2023_SC}. 
On the other hand, with the high precision offered by instruments like TianQin and LISA\cite{torres-orjuela_huang_2023}, 
it becomes possible to constrain accretion rates and determine the binary masses more accurately. 
If binaries form in the inner regions~\cite{Bellovary16},
such gaseous effects operate efficiently~\cite{Secunda2019,Secunda2020} 
and can be well-constrained by TianQin and LISA. 
Once gaseous hardening/softening and gas accretion are detected, it signifies that mergers are occurring in AGN disks. Additionally, this provides insights into how binaries evolve in these environments.

If BHs are accumulated in the outer regions~\cite{Tagawa2020,Grishin2023}, 
mergers are significantly facilitated through binary-single interactions. 
Due to the high frequency of the interactions and the high escape velocity from host systems, binaries are hardened to $\sim 10^{-2}~{\rm Hz}$ by the interactions~\cite{Tagawa2020}, and 
the eccentricity is distributed in $\sim 10^{-3}$--$10^{-2}$ at $10~{\rm Hz}$~\cite{Tagawa2021_Eccentricity}, 
both of which are higher compared to those predicted for mergers in globular clusters and isolated binaries. 
Furthermore, a fraction of mergers is contributed by the GW capture mechanism, resulting in binaries merging with extremely high eccentricities~\cite{samsing_bartos_2022}. 
By observing the evolution of GW frequencies and eccentricities with TianQin and LISA, the contribution of binary-single interactions to binary evolution can be constrained, leading to a better understanding of the environments conducive to mergers.

Secondly, the environments where compact object mergers occur can be directly revealed. 
If mergers take place near SMBHs, their Doppler acceleration~\cite{meiron_kocsis_2017,Inayoshi17,Wong2019,yu_chen_2021,sberna_babak_2022,Vijaykumar2023,Han_Yang2024} 
and shift~\cite{torres-orjuela_chen_2020,torres-orjuela_chen_2021} 
can be detected. 
Such detections would allow for constraints on the SMBH mass and the distance from the SMBH. 
Additionally, they would unveil the locations of gaps and/or migration traps, which are crucial for understanding the structure of the AGN disk. 
Moreover, the detection of the acceleration itself can serve as a smoking gun signature for merging environments.

The contribution of the AGN channel can also be constrained by examining their spatial distribution~\cite{Bartos17NatCom,Corley2019}. 
By directly comparing the spatial distribution of AGNs with the directions of GWs, we can estimate the contribution from AGNs. 
So far, the contribution from luminous AGNs with $\gtrsim 10^{45.5}~{\rm erg/s}$ has been constrained~\cite{Veronesi2023}. 
This strong constraint can be established by observing a large number of GW events with well-determined sky locations. 
TianQin/LISA, with its significantly improved determination of the sky location of binaries (typically within $\lesssim 1~{\rm deg}^2$\cite{torres-orjuela_huang_2023}), will be instrumental in estimating the fraction of GW events that originate from AGNs.

Thirdly, TianQin/LISA will play a significant role in determining whether EM emissions are associated with GW events. 
Thus far, several potential associations for BBH mergers have been suggested, including optical flares~\cite{Graham20,Graham2023} and gamma-ray flares~\cite{Connaughton2016,Connaughton2018,Bagoly2016}. 
However, the significance of these associations remains uncertain primarily due to the uncertainties in the locations of mergers~\cite{Ashton2020,Palmese2021,Morton2023} and the claims of detection significance~\cite{Greiner2016,Savchenko2016}. 

The detection of binaries by TianQin/LISA will significantly improve their localization, 
while also providing constraints on the timing of mergers within a range of minutes to hours~\cite{Buscicchio2021,torres-orjuela_huang_2023}. 
This capability enables the search for flares using telescopes that have high sensitivity but a narrow field of view. Hence, TianQin/LISA will contribute to identifying potential associations of EM emissions with GWs. 
If confirmed, such associations would not only unveil the astrophysical origins of GW events, but also enhance our understanding of accretion and radiation processes~\cite{Chen2023_EM,Tagawa2023_EM,RodriguezRamirez2024,Tagawa2023_SC} 
and cosmology~\cite{Gayathri2021}, and provide opportunities to test fundamental physics~\cite{Ellis2016}.

In summation, the AGN disk environment offers a fertile ground for studying stellar-mass BH mergers, posing exciting opportunities for future research facilitated by upcoming space-borne GW observatories.

\subsection{Contributions from GW observations with TianQin}
\contributors{Shuai Liu, Ataru Tanikawa, Shun-Jia Huang, Zheng-Cheng Liang, Yi-Ming Hu}


\subsubsection{GW detection of stellar-mass BBH}

\begin{figure}[h]
    \centering
    \includegraphics[width=0.8\textwidth]{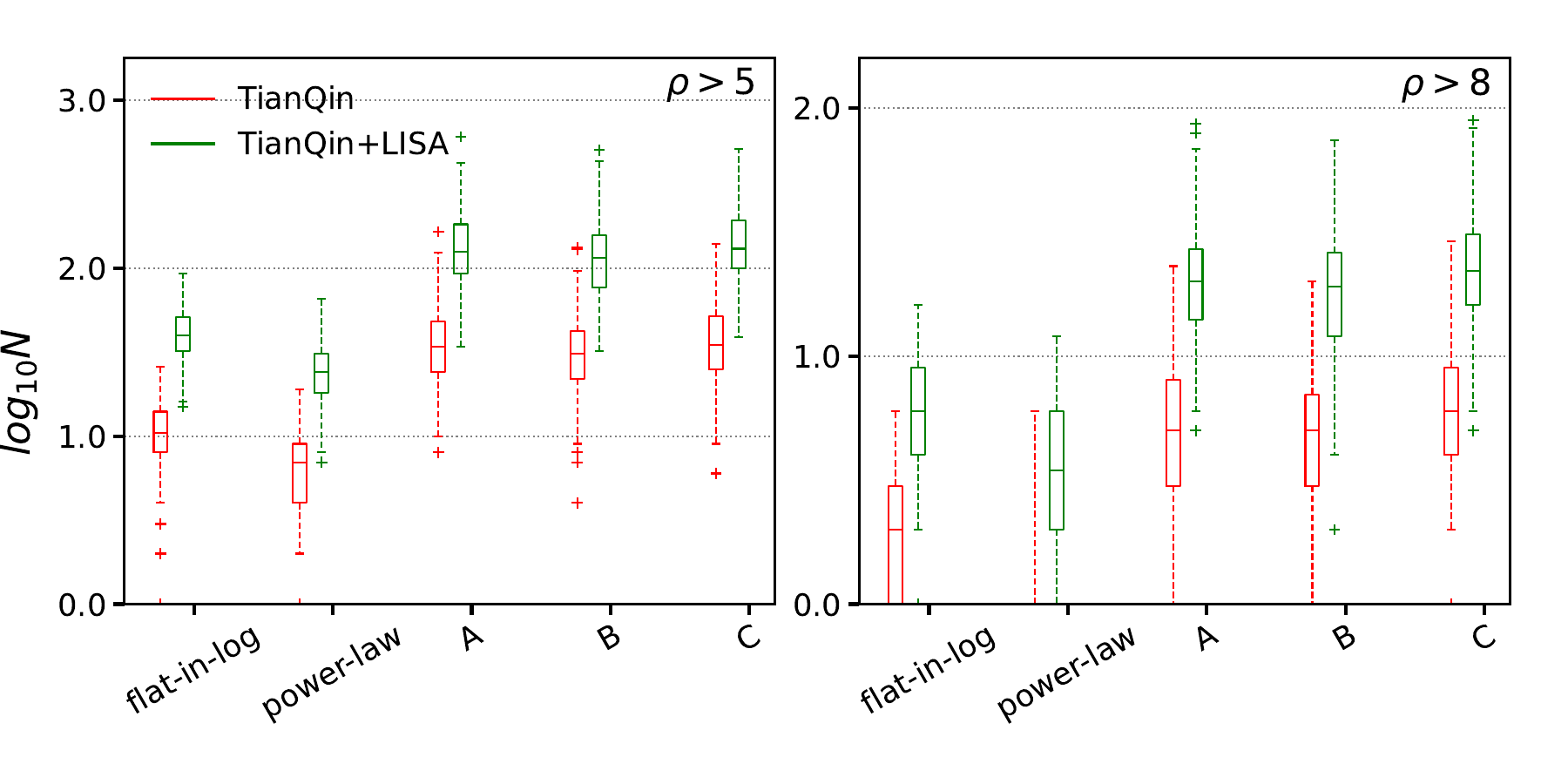}
    \caption{Detection number of stellar-mass BBHs represented by box plots under different mass distribution models (flat-in-log, power-law, A, B, and C) and detector observation scenarios (TianQin and TianQin+LISA). The left and right panels are the cases with signal-to-noise ratio $\rho>5$ and 8, respectively. Modification from Figure~3 in \citet{liu_hu_2020}.}
    \label{fig:detection number sbbh}
\end{figure}

Inspiraling stellar-mass BBHs could last from months to years within mHz space-borne GW detector bands, e.g., LISA, before they merge into ground-based GW detector bands. Stellar-mass BBHs would be detected and their source parameters would be measured accurately by LISA \cite[e.g.,][]{sesana_2016,sesana_2017,seto_2016,kyutoku_seto_2016,toubiana_et_al_2021,wong_kovetz_2018}. Compared with LISA, TianQin is more sensitive at relatively high-frequency range. \citet{liu_hu_2020,liu_zhu_2022} studied comprehensively the detection capacity of TianQin for stellar-mass BBHs. Assuming TianQin operates for 5 years with a ``3 months on + 3 months off'' scenario under different mass distribution models, TianQin could detect up to $\mathcal{O}(100)$ stellar-mass BBHs in the case of signal-to-noise ratio $\rho>5$, as shown in Fig. \ref{fig:detection number sbbh}. When $\rho>8$, it could still detect up to $\mathcal{O}(10)$ stellar-mass BBHs. About 30\% (60\%) of these sources detected would merge into ground-based detector frequency bands within 5 (10) years, which could be multiband detected, e.g., TianQin + LIGO. Furthermore, compared with TianQin or LISA alone, TianQin+LISA would improve the number of detected stellar-mass BBHs, about 3 times. Most stellar-mass BBHs detected concentrate at redshift $z<0.2$. 
    
\begin{figure}[htbp]
    \centering
    \subfloat[]{
        \includegraphics[width=0.3\textwidth]{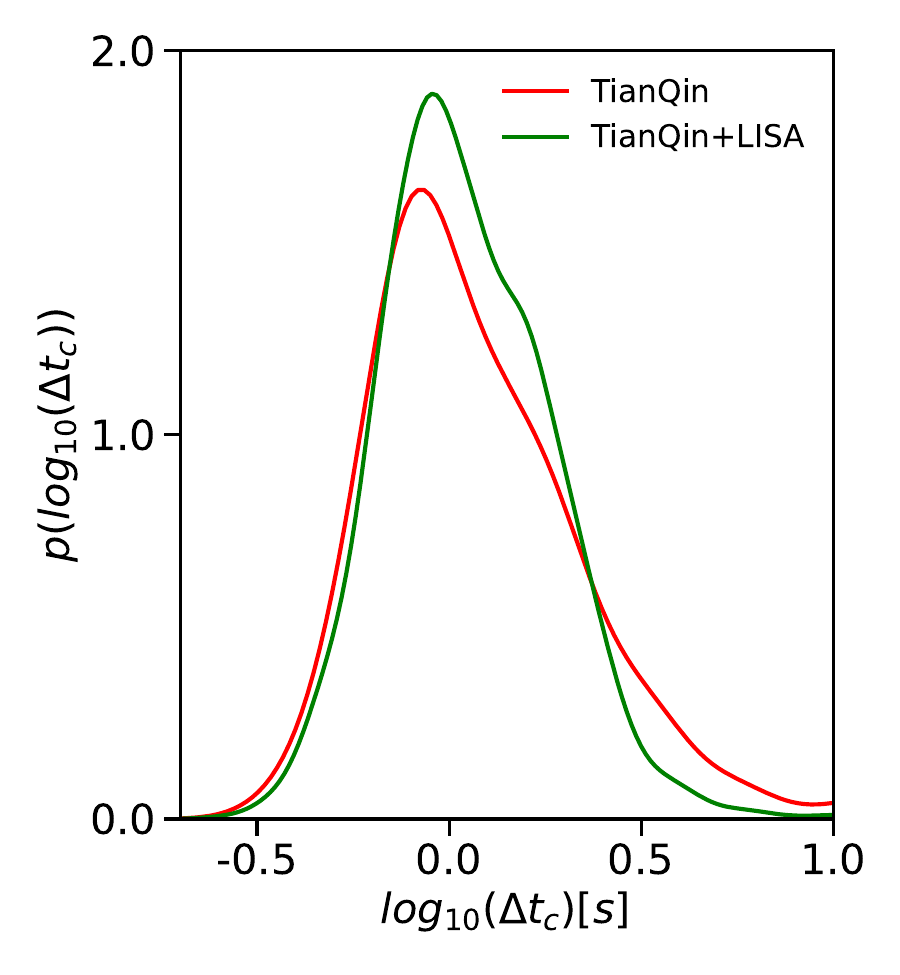}}
    \subfloat[]{
        \includegraphics[width=0.3\textwidth]{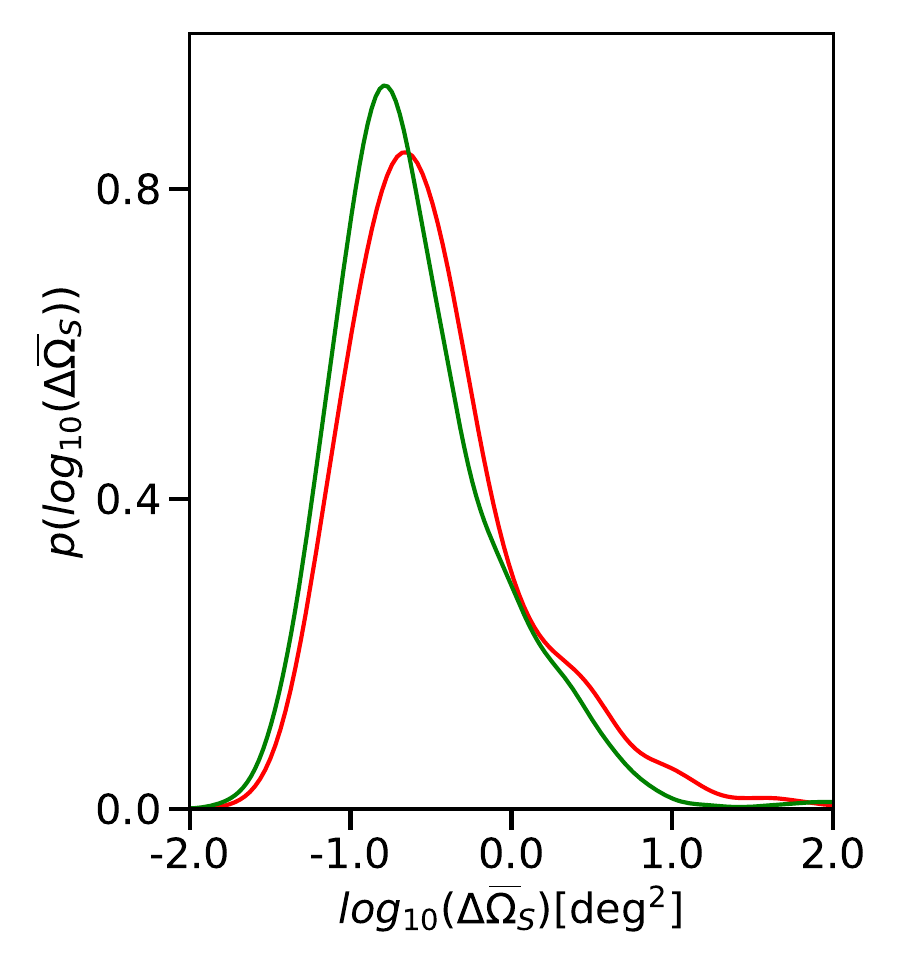}}
    \subfloat[]{
        \includegraphics[width=0.3\textwidth]{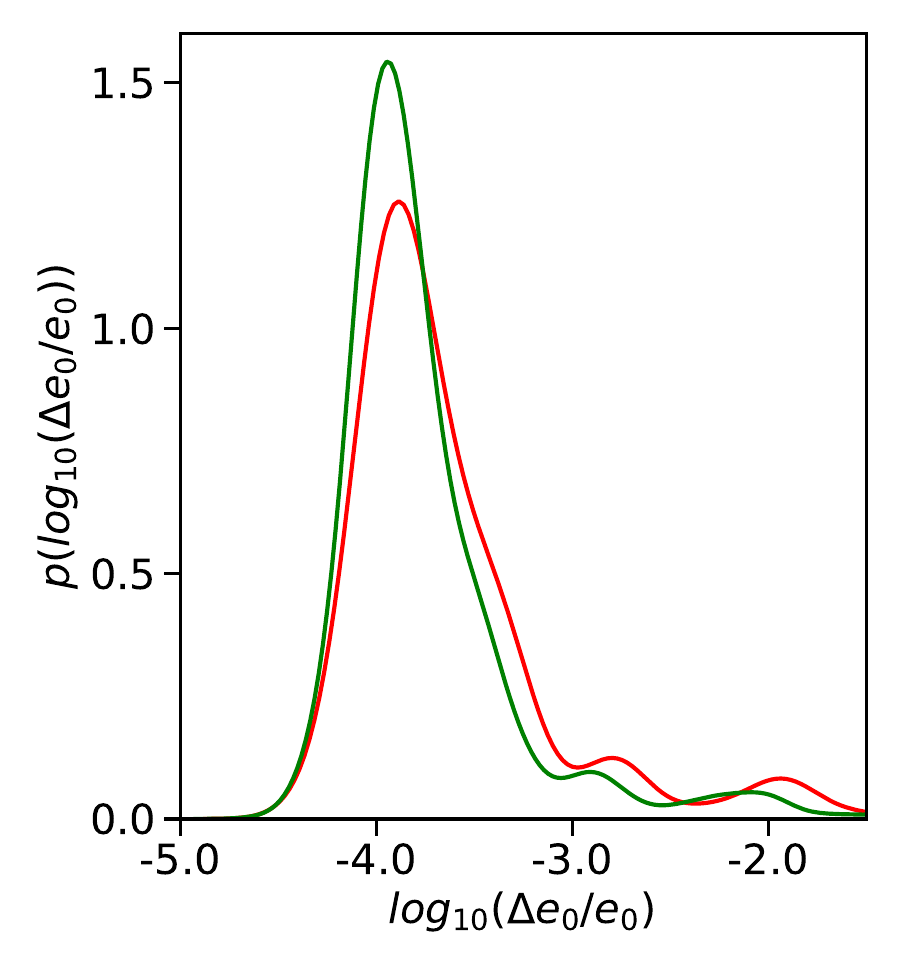}}
    \caption{Estimation precision distributions of stellar-mass BBHs merging within 5 years with $\rho>8$ performed by Fisher information matrix. (a) coalescing time $t_{c}$, (b) sky localization $\bar{\Omega}_{S}$, (c) initial orbital eccentricity $e_{0}$ at 0.01 Hz.
    Red and green lines represent TianQin and TianQin+LISA, respectively. Modification from Figure~5 in \citet{liu_hu_2020}.}
    \label{fig:PE_distri}
\end{figure}

Some parameters measurement precision of stellar-mass BBHs merging within 5 years detected are illustrated in Figure~\ref{fig:PE_distri}. Most measurement uncertainties of merger time $\Delta t_{c}$ and sky localization $\Delta\bar{\Omega}_{s}$ are within $\mathcal{O}(0.1){\rm s}-\mathcal{O}(1){\rm s}$ and  $\mathcal{O}(0.01){\rm deg}^{2}-\mathcal{O}(1){\rm deg}^{2}$, respectively. It implies that TianQin is expected to notice ground-based GW detectors as well as EM telescopes when and where events would merge in advance, so that they could be prepared for subsequent observations. Relative measurement uncertainty of luminosity distance $\Delta D_{L}/D_{L}\sim1/\rho$. It would be $\mathcal{O}(0.1)$ for stellar-mass BBHs with $\rho>8$. Combine with a $\Delta D_{L}/D_{L}\sim0.2$, for a typical source at $z=0.05$, the measurement precision of three-dimensional localization $\Delta V\sim D_{L}^{2}\Delta\bar{\Omega}_{s}\Delta D_{L}\sim50 {\rm Mpc^{3}}$. The average number density of Milky-Way-like galaxies is $0.01 {\rm Mpc^{-3}}$, so TianQin could pinpoint host galaxies of events. Orbital eccentricity $e_{0}$ at $0.01$Hz of stellar-mass BBHs could also be determined accurately, most relative errors are between $\mathcal{O}(10^{-5})-\mathcal{O}(10^{-3})$. Therefore, TianQin could use this to help distinguish that stellar-mass BBH form via isolated binary evolution or dynamical interaction. Because stellar-mass BBHs would orbit each other in huge circles during the TianQin frequency band, the mass parameters of stellar-mass BBHs could also be measured accurately. Relative error of chirp mass and symmetric mass ratio are $\Delta\mathcal{M}/\mathcal{M}\sim10^{-7}$ and $\Delta\eta/\eta\sim10^{-3}$, respectively. This means that TianQin could make contribution to the mass distribution of stellar-mass BBHs determination with high precision. Furthermore, TianQin+LISA could also improve the precision of source parameter estimation, their distribution tends to shift toward smaller values.

Therefore, TianQin detection for stellar-mass BBHs would provide early alert for ground-based GW detectors LIGO and EM telescopes, making multi-band and multi-messenger observations possible. TianQin could also make a contribution to determining the population properties of stellar-mass BBHs by measuring mass and orbital eccentricities accurately. This is useful to identify eccentric stellar-mass BBH mergers~\cite{wang_harry_2024, Fang2019,Ireland2019,Randall2019,ChenAmaroSeoane2017,ramos-buades_buonanno_2023,seto_2024} 
In addition, combined with EM observation of host galaxies of BBHs, TianQin could constrain the Hubble parameter regarding BBHs as dark sirens.

\subsubsection{GW detection of triple BH in Pop~III cluster}

In dense star environments and AGN disks, BHs could also form merging hierarchical triple BHs (merging triple BHs), where the inner merging inner BBHs orbit distant tertiary BHs \cite[e.g.,][]{2014ApJ...781...45A, Fragione:2019dtr, martinez_fragione_2020, Trani:2021tan, Britt:2021dtg,Martinez:2021tmr}, in addition to forming merging BBHs. The evolution of merging inner BBHs, from formation to merger, is accompanied by GW radiation. Under the impact from tertiary BHs, GWs from merging inner BBHs would be different from those from merging isolated BBHs, e.g., GW phase shifts. GWs from merging inner BBHs would be detected by the mHz space-borne GW detectors like TianQin and LISA, and could be used to explore the properties of tertiary BHs and environments~\cite{meiron_kocsis_2017,antonini_toonen_2017,gupta_suzuki_2020,samsing_bartos_2022,Trani:2021tan,liu_dorazio_2022,yu_chen_2021,Samsing2024,Pijnenburg:2024btj}. 

\begin{figure}[h]
\begin{centering}
\includegraphics[width=0.8\textwidth]{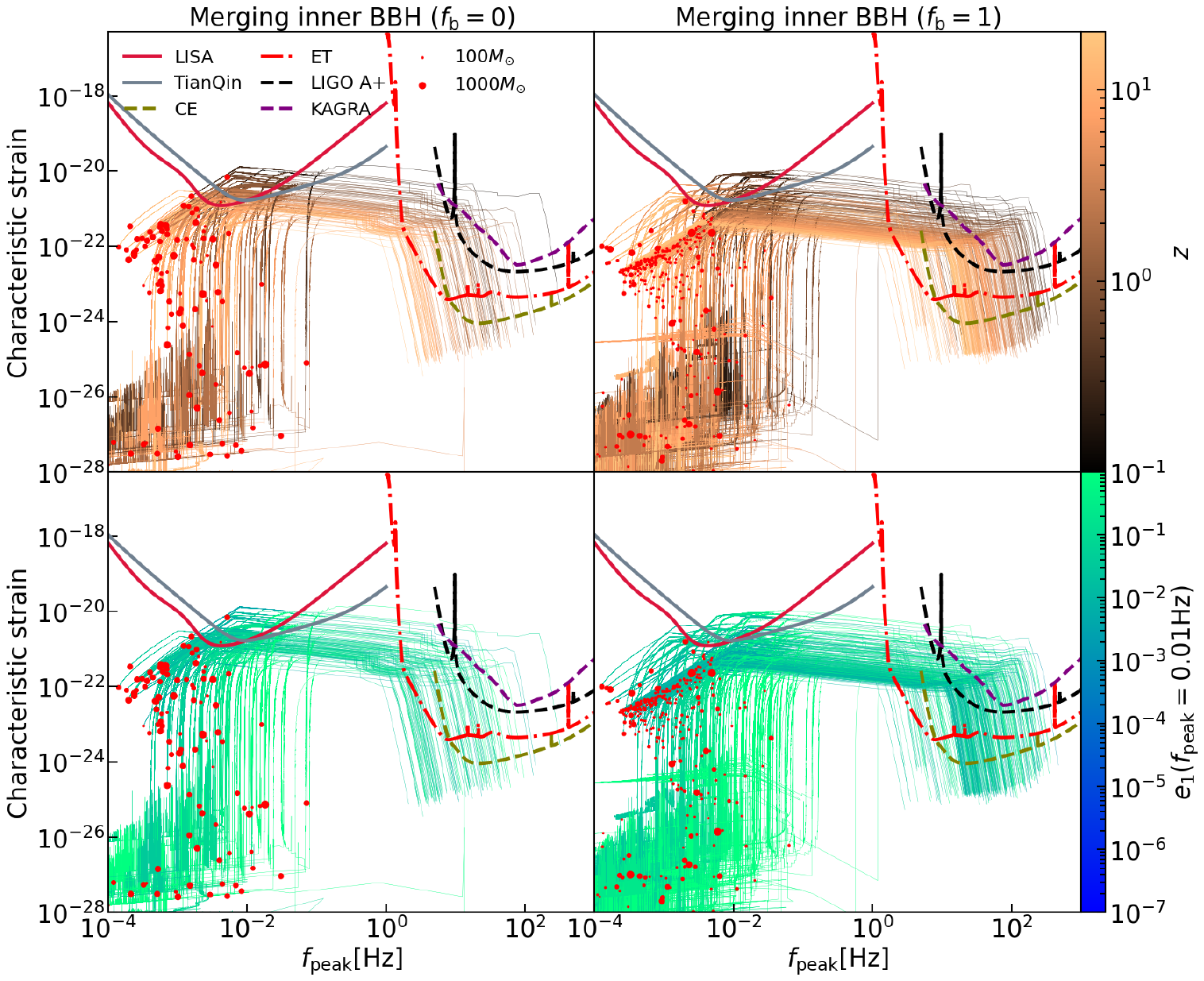}
\caption{GW detection of inner BBHs within merging triple BHs. \emph{Upper}: the characteristic strains of GWs with $f_{\rm peak}$, and
    the color bar represents the redshift of sources. \emph{Lower:} the characteristic strains of GWs with $f_{\rm peak}$, with the color bar representing the eccentricities at $f_{\rm peak}=0.01$Hz of sources. The frequency corresponding to the harmonic with maximum power of GWs from merging inner BBHs is represented by $f_{\rm peak}$. The size of dots scales with the total mass of the merging inner BBHs, the characteristic strains of noise of GW detectors~\cite{Kawamura:2011zz,Robson:2018ifk,TQ_MBH_2019,Michimura:2020xnj,LIGOScientific:2016wof,Hild:2010id} are plotted in different
    colors. The columns from left to right are cases with all the primordial stars are unpaired ($f_{\rm b}=0$) and paired ($f_{\rm b} = 1$) in Pop~III clusters, respectively. Modification from Figure~14 in \citet{Liu:2023zea}.}
\label{fig:ecc hcn fpeak inner}
\end{centering}
\end{figure}

\begin{figure}[h]
    \centering
    \includegraphics[width=0.8\textwidth]{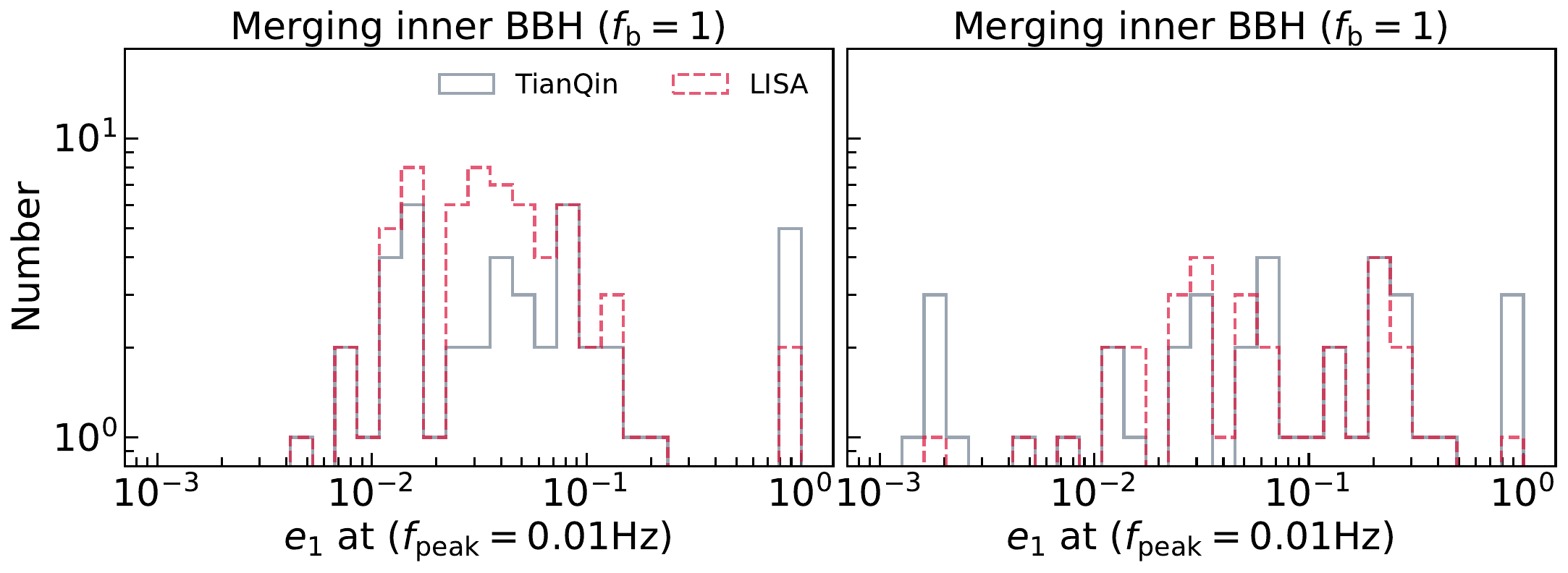}
    \caption{The distributions of orbital eccentricity at $f_{\rm peak}=0.01$Hz detectable merging inner BBHs across TianQin/LISA are illustrated using different colors and line styles. The left and right columns correspond to cases where $f_{\rm b}=0$ and 1, respectively. The detectable sources are those for which the characteristic strains of GWs exceed those of detector noise within a certain frequency range. The columns from left to right are cases with all the primordial stars are unpaired ($f_{\rm b}=0$) and paired ($f_{\rm b} = 1$) in Pop~III clusters, respectively. Modification from Figure~17 in \citet{Liu:2023zea}.}
    \label{fig:e1 at fpeak detectable inner}
\end{figure}

 
\citet{Liu:2023zea} studied GW detection of merging inner BBHs within triple BHs by TianQin. The characteristic strains of GWs from merging inner BBHs are shown in Figure~\ref{fig:ecc hcn fpeak inner}. The characteristic strains of GWs emitted by the merging inner BBHs decoupled from the third BHs are similar to those from isolated merging BBHs in Figure~19 in \citet{Liu:2023zea}, whereas the GW strains of the merging inner BBHs that are affected by the tertiary BHs display peaks, oscillations and sharp turning points at lower frequency ranges, before they are decoupled from the tertiary BHs. Most detectable sources whose GW strains are higher than those of TianQin noise within certain frequency ranges are located at $z<3$, and several sources detectable could be within $10<z<15$. The orbital eccentricities at $f_{\rm peak}=0.01$Hz of detectable merging inner BBHs for TianQin concentrate between $\mathcal{O}(10^{-3})$ and $\mathcal{O}(10^{-1})$, along with a significant fraction of detectable sources with $e_{1}\sim1$, as shown on the upper panel in Figure~\ref{fig:e1 at fpeak detectable inner}. Some merging inner BBHs with very large orbital eccentricities could not be detectable by TianQin but were detectable by at least one ground-based detector like Cosmic Explorer (CE)/ Einstein Telescope (ET) or LIGO/Virgo/KAGRA. This can be explained by the fact that high eccentricities suppress the strains of GWs so that they are lower than those of TianQin noise. The eccentricities of such sources at the mHz band could not be identified by TianQin alone, but they would be constrained via multi-band observation, e.g., TianQin+ET. This could be performed by tracking sources in ground-detected catalogs back to data streams from TianQin. If however no significant trigger is identified with TianQin, then we can conclude that the merging BBH system has a large eccentricity at the mHz band. Furthermore, TianQin could detect merging isolated BBHs in Pop~III clusters. The redshift distribution of detectable merging isolated BBHs is similar to that of merging inner BBHs, but their eccentricity distribution shifts toward smaller values.

Therefore, TianQin could detect merging BBHs located at $z<15$ in Pop~III clusters. Combined with EM observations (e.g., JWST) for these merging BBHs with high redshifts, TianQin is excepted to make contribution to studying the evolution of Pop~III stars. TianQin would have the potential to measure the orbital eccentricities of merging BBHs. Even if for the BBHs with very high orbital eccentricities which are not detectable for TianQin alone, multiband observation performed by TianQin and ground-based detectors, e.g., TianQin+ET, could also identify them. By measuring the orbital eccentricities, TianQin could distinguish the formation channels of merging BBHs, e.g., isolated evolution or dynamical interaction.

\subsubsection{Joint observations of stellar-mass BH mergers with ground-based GW observatories}

A large number of stellar-mass BH mergers have been discovered by ground-based GW observatories~\cite{2023PhRvX..13d1039A}. However, their origin remains under debate. 
Some propose that they are formed through isolated binary evolution~\cite{1998ApJ...506..780B, 2016Natur.534..512B, 2016MNRAS.458.2634M, 2016A&A...588A..50M, 2017NatCo...814906S, 2017PASA...34...58E, 2017MNRAS.472.2422M, vandenheuvel_portegies-zwart_2017, 2018MNRAS.481.1908K, 2019MNRAS.490.3740N, olejak_fishbach_2020, 2021A&A...647A.153B, kinugawa_nakamura_2021, 2022ApJ...926...83T}. Joint observations between ground-based and space-borne GW observatories could resolve this issue by determining the eccentricity. 
Ground-based GW observatories detect stellar-mass BH mergers a few seconds before the event, while space-borne observatories are expected to detect these mergers days or even years in advance~\cite{sesana_2016}, when the orbit of such systems has not circularized.
Figure \ref{fig:ecc at frequency} shows a summary of several eccentricity distributions predicted by different evolution models, with all eccentricities converted to a common reference frequency of 0.01Hz ($e_{0.01}$, for TianQin and LISA) and 1Hz (for CE and ET). 
Stellar-mass BBHs formed through isolated binary evolution can have negligible eccentricities~\cite{Breivik2016,Rodriguez2018,Kowalska2011}. 
As for stellar-mass BBHs formed in globular clusters, they can retain higher eccentricity with $e_{0.01}\sim10^{-2}$, but if they are ejected outside the cluster before merger, the eccentricity will be smaller compared with the inside case~\cite{Breivik2016,Zevin2019,Samsing2018b,rodriguez_chatterjee_2016,Rodriguez2018}. Eccentricity can be extreme when considering systems involved in complex environments like triplets or AGN disks~\cite{antonini_toonen_2017,Antonini2012,samsing_bartos_2022,Zevin2019,Rodriguez2018}. 
Figure \ref{fig:ecc at frequency} also presents the minimum detectable eccentricities of different GW observatories, and even the next-generation ground-based observatories can only distinguish extreme models~\cite{Lower2018}. One can achieve multi-band search by relying on ground-based observatories to find the systems and searching the archival data in space-borne observatories~\cite{wong_kovetz_2018}. By using the archival search, we can focus more on parameters like a system's eccentricity and make space searches computationally feasible, so that we can distinguish more evolution models~\cite{wang_harry_2024}. 
Therefore, joint observations can determine whether BBHs have eccentricities at a GW frequency of 0.01-0.1 Hz, 
and the fraction of eccentric BBHs relative to circular BBHs can provide significant constraints on the origin of stellar-mass BH mergers.
The multi-band observation can also have interesting applications like observing the non-axisymmetrically deformed neutron stars within the binary systems \cite{2018PhRvL.121m1105T,Feng:2024ulg}.

\subsubsection{GW detection of DWD}
TianQin is capable of detecting a significant number of DWD binaries up to $10^{4}$ within the Milky Way \cite{Huang:2020}.
The majority of these DWDs are expected to have a relatively low signal-to-noise ratio (SNR), while a substantial subset exhibits high SNR values, reaching up to $10^{3}$ \cite{wolz_yagi_2021,seto_2022, seto_2023c,seto_2023a,seto_2023b,guo_jin_2024,staelens_nelemans_2024}. 
These high-SNR DWDs are well-localized, making them suitable for EM follow-up and multi-messenger studies. 
For 90\% of DWD detections, the uncertainties on various parameters are as follows: the orbital period \( \Delta P/P \) is between (0.15-4.63) \( \times 10^{-7} \), the GW amplitude \( \Delta \mathcal{A}/\mathcal{A} \) is within 0.04-5.02, the inclination angle \( \Delta \cos\iota \) is within 0.02-4.95, the polarization angle \( \Delta \psi_S \) is within 0.03-4.01 rad, and the sky position \( \Delta \Omega_{\rm S} \) is within 0.02-21.36 deg$^2$. 
The median uncertainties are \( \Delta P/P = 1.41 \times 10^{-7} \), \( \Delta \mathcal{A}/\mathcal{A} = 0.26 \), \( \Delta \cos\iota = 0.20 \), \( \Delta \psi_S = 0.39 \) rad, and \( \Delta \Omega_{\rm S} = 1.85 \) deg$^2$. 
TianQin can accurately locate 39\% of DWDs within 1 deg$^2$.

When TianQin operates in conjunction with LISA, forming TianQin + LISA networks, the total number of detected DWDs is projected to double, and the parameter estimation precision better than TianQin alone \cite{Huang:2020}.
The detection capabilities of TianQin, LISA, and their joint network for DWDs are summarized as follows \cite{torres-orjuela_huang_2023}:
\begin{itemize}
    \item Below 4 mHz, LISA is superior to TianQin, with joint detection closely matching LISA's range.
    \item Between 4 mHz and 20 mHz, joint detection significantly outperforms individual networks, extending detection distances substantially.
    \item Above 20 mHz, TianQin excels over LISA, with joint detection closely tracking TianQin's capabilities.
\end{itemize}
More detail on the GW detection of DWD can be found in \citet{Huang:2020,torres-orjuela_huang_2023}.

\subsubsection{Stochastic GW backgrounds}

In addition to the individual detections, the GW signals from the stellar mass binaries can also compose the stochastic GW backgrounds, which resemble the random noise, but only have an astronomical origin~\cite{liang_hu_2022, Zhao2021}.
For the case of DWDs, it is expected that incoherent summation of the numerous DWDs adds up so high that it can exceed the anticipated noise level, forming a foreground that can cover up other signals for LISA.
Fortunately, as has been indicated in Figure~\ref{fig:TQ_source}, it is expected that after five years of operation, many of the marginal signals accumulate a strong enough signal-to-noise ratio (SNR) and can be resolved individually.
By removing these signals from the foreground, it will become lower in amplitude, and especially for TianQin, the 5-year foreground will be lower than the noise curve, making it less of an issue for other sources~\cite{Huang:2020,wu_li_2023}. 

Sometimes, the stochastic background can also be of interest, as it reflects the group properties of the underlying population.
Since its statistical properties will be identical to noise, the key to distinguishing the stochastic background from the noise is to utilize the fact that it is astronomical in origin.
Cross-correlation methods have been adopted for ground-based GW detectors and pulsar timing arrays.
The noise in different detectors/pulsars can be considered independent, while the signals are not.
With the accumulation of more data, the correlation becomes more apparent and one can thus conclude the existence of the background.
For the space-borne missions, the earlier pioneering works have to work under the assumption that there will only be one operating space-borne GW detector at a time, so the cross-correlation can not be adopted.
In this case, the null-channel method has been proposed.
In this method, one uses the fact that GW has two degrees of freedom under the GR, while the triangular-shaped space-borne GW detectors provide three interferometer readouts. 
Especially in the low-frequency limit, one can mathematically construct a data stream that by design contains no astronomical signal at all, which is referred to as the ``null channel".
If the relation between the noise properties of the null channel and the data channels is precisely known, the analysis of the null channel can be used to constrain the signal channels, and furthermore, one can identify the stochastic background and study its properties~\cite{liang_hu_2022, cheng_li_2022}. 

We can observe that in this way, a very strong assumption has to be made for the noise property, and all the following conclusions tie strongly with the validity of this assumption.
Having multiple space-borne GW missions operate simultaneously would be very attractive since it brings the model-independent cross-correlation method back to the menu. 
So the TianQin observatory functioning at the same time as LISA would present a priceless opportunity to perform cross-correlation to search and depict the nature of the stochastic background. 
Furthermore, the usage of multiple detectors can significantly enhance our sensitivity to certain spherical harmonic components of the anisotropic stochastic backgrounds.
With the case of TianQin+LISA, the improvement on certain components can reach the level of 6 orders of magnitudes~\cite{liang_li_2023}.

\vspace{3cm}
\begin{tcolorbox}[colback=red!5!white,colframe=red!75!black]
\textbf{\textcolor{NavyBlue}{
\begin{it}
The space-borne detector, with its sensitive frequency range, can track the early inspiral of binaries emitting GWs. 
The combined detection capabilities of TianQin, LISA, and ground-based instruments enhance our ability to identify binaries, allowing us to follow their multi-band GW. 
This approach is particularly advantageous for detecting the early evolution of high-eccentricity BBHs, aiding in distinguishing their progenitor origins—whether they evolve through isolated binary processes or form dynamically in dense stellar environments, near SMBHs, or within AGN disks.
It may also enable the reliable measurement of the stochastic GW background.
If binaries are confirmed to exist in AGN disks, this could provide valuable insights into the physics of these environments, which have been difficult to study through traditional EM detection methods. 
Additionally, the space-borne detector excels at identifying DWDs, advancing our understanding of their stellar evolution—particularly in ultra-compact X-ray binaries and AM CVn systems. 
It can also help measure the population of WDs in the Galaxy, contributing to our knowledge of its star formation history.
Furthermore, the space-borne detector has the potential to detect IMBHs, confirming their presence in massive star clusters and Pop~III star clusters, and enhancing our understanding of their formation scenarios. 
\end{it}
}}
\end{tcolorbox}

\clearpage
\section{Birth and growth of MBHs}
\label{sec:MBHB}
\coordinators{En-Kun Li \& Kohei Inayoshi}



The existence of SMBHs with masses exceeding $\simeq 10^9~M_\odot$ in the universe only a billion years old 
presents an intriguing astrophysical puzzle~\cite{Fan_2023}. 
Central to this enigma are questions concerning the origins of these SMBHs -- specifically, the nature of their initial ``seeds'' 
and the mechanisms by which these seed BHs were able to grow at such rapid rates, on average, $\sim 1~M_\odot~{\rm yr}^{-1}$
\cite{Inayoshi_2020,Volonteri_2021}. 
These seeds are hypothesized to be the remnants of the first generation of stars, known as Pop~III 
stars, or could alternatively originate from the direct collapse of massive gas clouds in the early universe. 
Understanding these initial conditions is crucial for elucidating the formation and subsequent evolution of SMBHs.

The discovery of a large number of high-redshift quasars, particularly the brightest ones with luminosities of 
$L_{\rm bol}\sim 10^{47}{\rm erg~s}^{-1}$ (corresponding to $M_{\rm BH}\sim 10^9~M_\odot$, when assuming Eddington luminosity),
through wide-field surveys such as the Sloan Digital Sky Survey and Subaru Hyper Suprime-Cam, has highlighted interest in these questions~\cite{Fan_2023}. 
Recently, the JWST with its unprecedented infrared sensitivity has revealed faint AGN/quasar 
populations with $L_{\rm bol}\sim 10^{45}{\rm erg~s}^{-1}$ at the epoch of cosmic reionization ($z\simeq 4$--7), 
pushing the detection threshold of BH masses down to $\sim 10^7~M_\odot$~\cite{Onoue_2023,Kocevski_2023,Harikane_2023}.
These breakthroughs allow us to address the low-mass BH populations hidden in the pre-JWST era.

\begin{figure}[b]
\includegraphics[width=130mm]{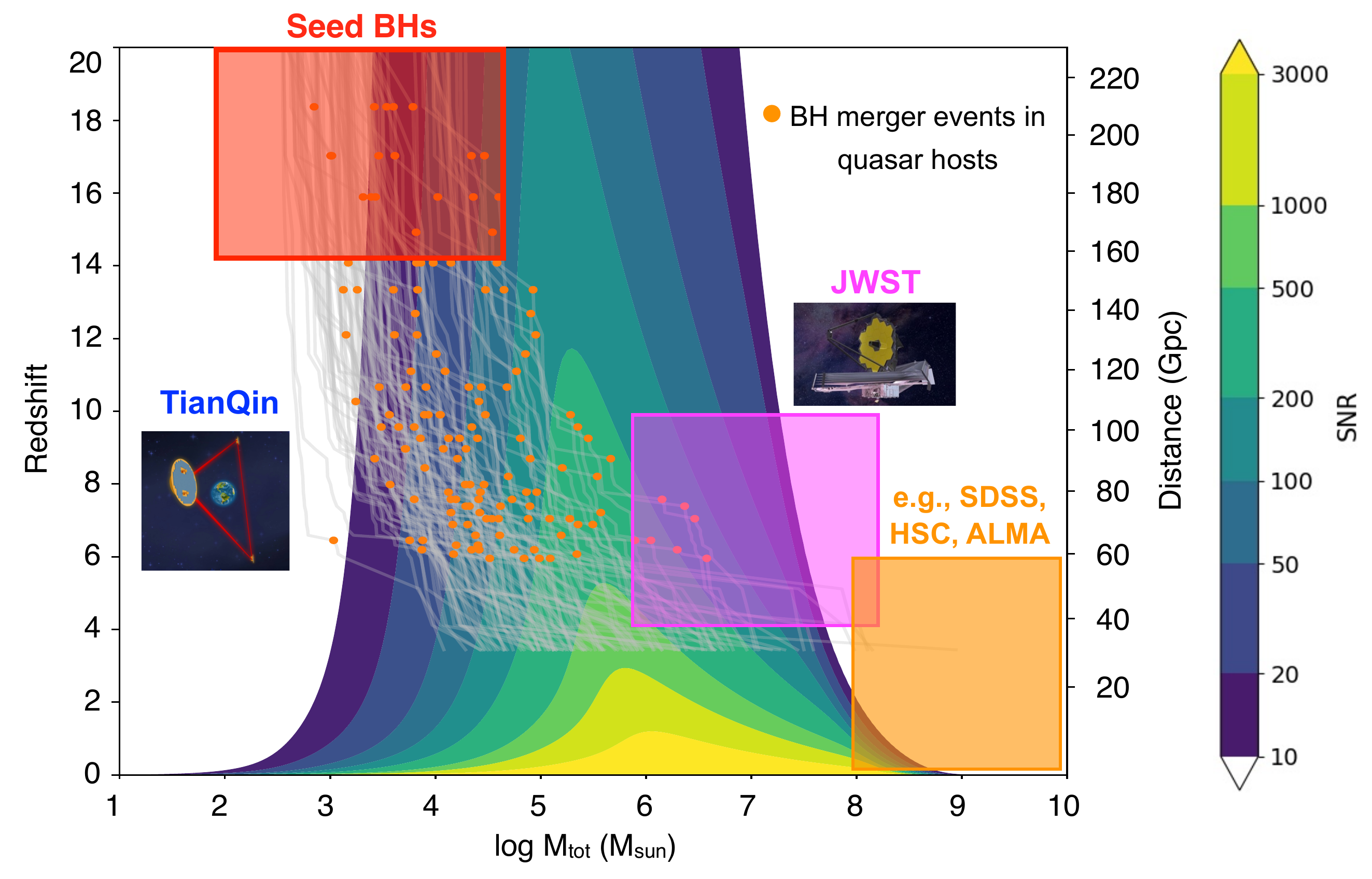}
\caption{The parameter spaces of BBH mergers in the high-redshift universe that space-borne GW interferometers such as TianQin and LISA
can explore. The color contour shows the region where GW signals from BBH mergers with total mass ($M_{\rm tot}$) are detectable
with a sufficient level of ${\rm SNR}\geq 10$.
GW observations play a crucial role in bridging the BH seeding epoch (red) and our current cosmic front ($z\sim 10$)
that ground-based quasar surveys reach (orange) and ongoing JWST observations are exploring (magenta).
Gray curves present BH growth tracks with BH merger events taken from semi-analytical calculations~\cite{Li_2023,liu2025}. 
\copyright Hanpu Liu.
}
\label{fig:roadmap}
\end{figure}

Figure~\ref{fig:roadmap} summarizes the parameter regions for BH seeding epochs (red), the areas explored by ground-based quasar surveys (orange), 
and recent JWST surveys (magenta).
To further probe lower-mass BHs, which are presumed to be closer to the massive end of seed BH populations~\cite{Inayoshi_2020,Volonteri_2021}, 
space-borne GW interferometers such as TianQin and LISA will be crucial.
The color contour in Figure~\ref{fig:roadmap} indicates the parameter spaces with a signal-to-noise ratio (SNR) $\geq 10$ 
for GW signals from binary BH mergers.
These detectors are expected to reach the mass range of $\sim 10^{4-6}~M_\odot$ at these high redshifts with a sufficiently high SNR,
providing invaluable insights into the early growth of BHs \cite{liu2025}.

As of 2024, theoretical expectations on the merger rate of MBHBs remain poorly understood and uncertain 
by more than an order of magnitude~\cite{Barausse:2012fy,Ricarte_2018,Dayal_2019,Barausse_2020}.
This uncertainty arises from the complexity and entanglement between various astrophysical processes that affect binary evolution
from large galactic separations ($\sim $ kilo-parsec scales) to small scales within $\sim $ parsec scales, where GW emission 
accelerates the coalescence. 
The most stringent bottleneck, as discussed in \citet{lisa_2023}, is at 
large/intermediate scales (the potential ``last kilo-parsec problem'') for binaries embedded in gas-rich galactic nuclei.
This is the regime expected for MBHBs in the TianQin and LISA detection windows.
While simulations and semi-analytical work continue to refine their predictions with improved physical modeling and larger statistical samples, 
the ultimate answer will come from GW experiments \cite{Berczik:2020awp, Guo:2020qkf, Volonteri:2020wkx, Avramov:2021reg, Caramete:2022ijv, Liao:2022niz, Zhong:2023pjz}. 
Having TianQin and LISA collecting data and covering the mHz GW window would be highly valuable, as differences between various models for massive binary evolution can likely be assessed only with sufficient statistical data \cite{Ding:2019tjk, Padmanabhan:2020zxj, Shinohara:2021psq}.

\subsection{MBH Seeding}
\label{subsec:MBHseed}
\contributors{Kohei Inayoshi, Pedro R. Capelo, Lucio Mayer, Pau Amaro Seoane}

\begin{figure}[t]
\includegraphics[width=160mm]{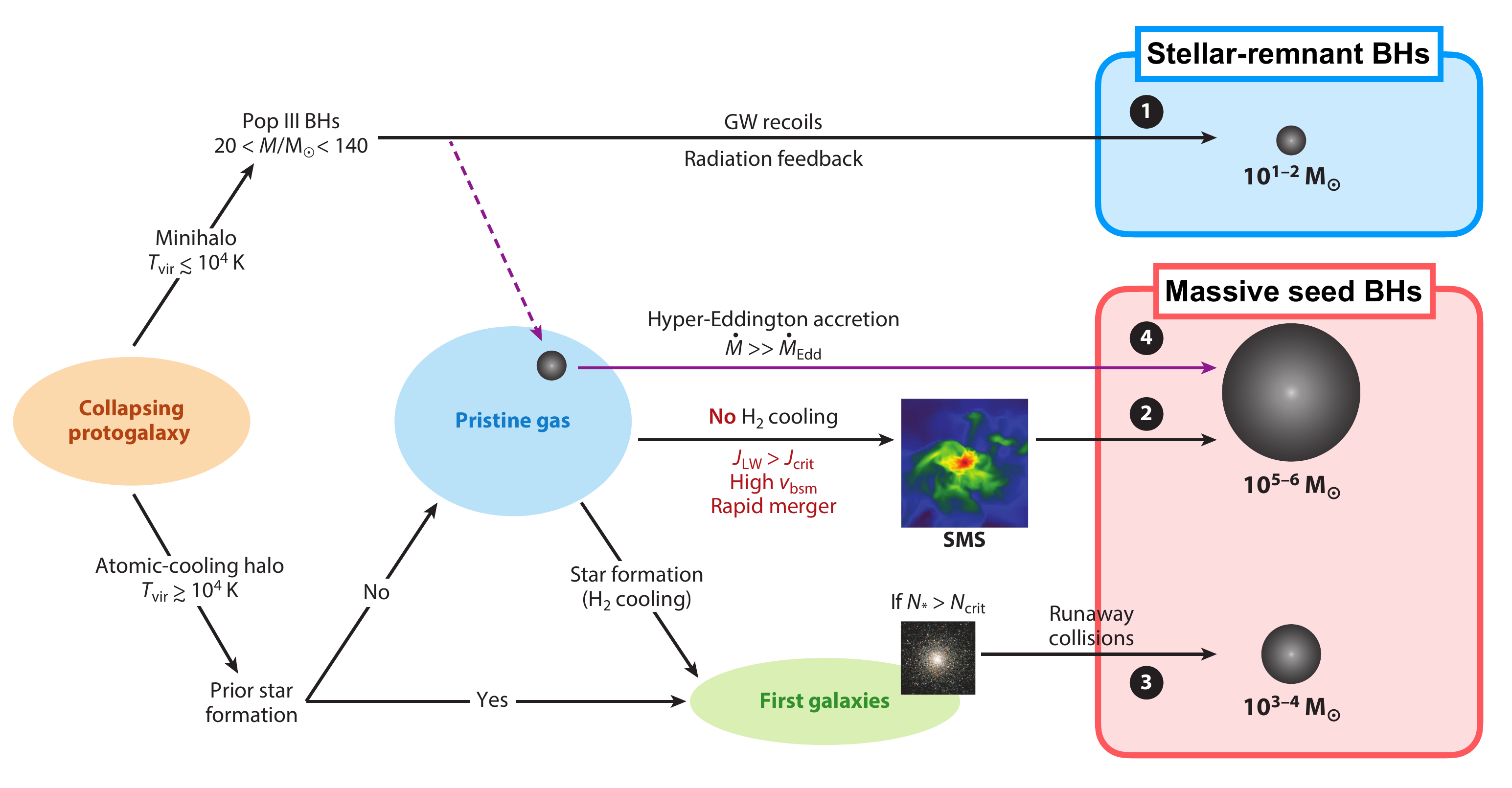}
\caption{Formation pathways of seed BHs in early protogalaxies~\cite{Inayoshi_2020}: 
Pop~III remnant BHs with a mass of $M_{\rm BH}\sim 10^{1-2}~M_\odot$,
(2) massive seed BHs with $M_{\rm BH}\sim 10^{5-6}~M_\odot$ in atomically-cooling halos under peculiar conditions such as strong Lyman-Werner
radiation, high baryon-dark matter streaming velocity, and rapid mergers of dark matter halos, 
(3) relatively massive seeds with $M_{\rm BH}\sim 10^{3-4}~M_\odot$ via runaway collisions in dense stellar clusters,
and (4) hyper-Eddington accretion onto either stellar-mass BHs or massive seed BHs.
In this paper, we classify those formation channels into two: stellar-remnant BHs (light seeds) and massive seed BHs (heavy seeds).
}
\label{fig:rees}
\end{figure}

In this section, we provide an overview of the theoretical framework for BH seeding mechanisms in early protogalaxies.
Figure~\ref{fig:rees} illustrates four potential pathways for initiating MBHs, including:
\begin{itemize}
    \item Stellar Remnant BHs ($10$--$10^2~M_\odot$): 
    The remnants of the first generation of stars (Pop~III stars) that formed in the very early universe 
    and were massive enough to leave behind BHs.
    \item Direct-collapse BHs ($10^5$--$10^6~M_\odot$): 
    MBHs formed through the direct collapse of gas clouds in early protogalaxies under specific conditions, 
    such as strong external UV radiation, frequent violent halo mergers, and low metallicity.
    \item Runaway Collisions in Dense Star Clusters ($10^3$--$10^4~M_\odot$): 
    Relatively MBHs formed through runaway stellar collisions in dense star clusters.
    \item Hyper-Eddington Accretion: Rapid accretion in protogalactic nuclei driven by intense inflows of cold gas,
    occurring at rates exceeding the Eddington accretion rate without being impeded by BH feedback.
\end{itemize}
We highlight the key physical processes and conditions required for each scenario. 
More detailed descriptions can be found in the following reviews on BH formation and growth processes~\cite{Haiman_2013, Woods_2019, Inayoshi_2020, Volonteri_2021},
as well as star formation in the earliest galaxies~\cite{Bromm_Yoshida_2011, Klessen_Glover_2023}.

\subsubsection{Stellar-remnant BHs}

One theory suggests that SMBHs originated from the remnants of the first generation of stars, known as Pop~III stars~\cite{Haiman_Loeb_2001}. 
These stars formed from primordial gas clouds in mini-dark-matter halos with masses of $\sim 10^{5-6}~M_\odot$. 
In metal-poor environments, the gas temperature is relatively warm ($\sim$ a few hundred K) compared to star-forming clouds in nearby galaxies,
due to inefficient cooling by molecular hydrogen (H$_2$) lines~\cite{Bromm_2002,Abel_2002,Yoshida_2006}. 
This warm primordial gas is expected to collapse with limited fragmentation, leading to the formation of massive Pop~III stars.

In this scenario, the growth of Pop~III stars is primarily controlled by photoheating from the stars themselves.
As Pop~III protostars accrete gas, they undergo Kelvin-Helmholtz contraction and eventually ignite stable hydrogen nuclear burning.
During their main-sequence stage, these stars reach surface temperatures of $\sim 10^5~{\rm K}$, producing intense ionizing ultraviolet (UV) radiation 
that heats the surrounding accreting material~\cite{McKee_Tan_2008,Hosokawa_2011}. 
Mass outflows driven by photo-evaporation regulate stellar accretion and ultimately determine the final stellar mass,
typically in the range of $10-100~M_\odot$~\cite{Hirano_2015}.

If Pop~III remnant BHs grew in mass through Eddington-limited gas accretion, they could potentially explain SMBHs of $\sim 10^9M_\odot$ 
(hereafter, light-seed model). 
However, this sustained, rapid growth is unlikely in more realistic scenarios within their parent mini-halos. 
Radiation feedback from the accreting remnant BH and mechanical feedback owing to supernova explosions of massive Pop~III stars would 
effectively disrupt the gas reservoir and thus regulate their mass growth~\cite{Whalen_2004,Kitayama_2005,Johnson_Bromm_2007,Alvarez_2009}. 
Therefore, while a small fraction of Pop~III BHs might grow into SMBHs in high-redshift quasars, the majority would remain at intermediate masses 
due to these feedback processes.

\subsubsection{Massive seed BHs}

The formation of massive seed BHs presents another popular scenario to explain the existence of SMBHs in the early universe. 
This theory, known as the heavy seed model, provides a more relaxed constraint on the formation timescale compared to 
the stellar-remnant BH scenario, as these massive seeds require less time to accumulate mass and become SMBHs 
\cite{Loeb_Rasio_1994,Begelman_2006,Mayer_2015}.

In many variations of the heavy-seed models, the process typically involves the rapid collapse of massive primordial gas clouds 
within atomically-cooling dark matter halos. 
These halos have virial temperatures around $T_{\rm vir}\sim 10^4~{\rm K}$ and the gravitational collapse of the cloud occurs at 
rates of approximately $0.1~M_\odot~{\rm yr}^{-1}$~\cite{Bromm_Loeb_2003,Inayoshi_2014,Becerra_2015}. 
Due to the high accretion rate, the central protostar rapidly grows in mass, reaching around $10^5~M_\odot$ within its stellar lifetime.
During the stellar growth phase, accretion shocks caused by rapid inflows inject entropy onto the stellar surface and lead to its expansion, 
while the core contracts until the central temperature reaches $10^7~{\rm K}$, the threshold of hydrogen burning~\cite{Hosokawa_2013,Woods_2017}.
Consequently, these protostars evolve similarly to red giants, characterized by lower surface temperatures around $\sim 5000~{\rm K}$, 
and thus produce very little UV radiation. 
In this case, radiative feedback mechanisms do not operate effectively, differing from the case of Pop~III star formation~\cite{Toyouchi_2023}. 
As the stellar mass exceeds $\sim 10^5~M_\odot$, the stellar interior promptly collapses into a heavy seed BH due to relativistic instabilities without 
significant mass loss~\cite{Inayoshi_2013}.

In standard scenarios, heavy seed BHs form in the first galaxies exposed to intense UV radiation from nearby bright galaxies~\cite{Omukai_2001,Shang_2010} and/or are influenced by successive galaxy mergers~\cite{Wise_2019}. 
These environmental factors prevent H$_2$, the main coolant in pristine gas, from forming through photo-dissociation, 
which delays the collapse of gas clouds and leads to higher accretion rates~\cite{Inayoshi_2014}.
However, the environmental conditions required for this process are infrequent in typical regions of the universe, 
making these heavy seed populations rare~\cite{Dijkstra_2008,Dijkstra_2014}.
A plausible solution to this rarity problem is to focus on heavy seed formation in overdense regions of the universe, 
where the progenitors of quasar hosts are likely to be located.
In such biased and intrinsically rare regions, the formation of heavy seed BHs is facilitated more efficiently due to the strong clustering of galaxies in these unique environments~\cite{Li_2021,Lupi_2021}. 

An alternative scenario for MBH seed formation relies on prominent gas inflows triggered by mergers between 
the most massive galaxies at $z\sim 10$.
In this scenario, the gas has already been metal-enriched and gas fragmentation into stars can occur
simultaneously without disrupting the gas inflow, as the mass transport proceeds faster than the typical star formation timescale~\cite{Mayer_2010,Mayer_Bonoli_2019,mayer_capelo_2024}.
This process results in a rapidly growing protostar at the nucleus, similar to the model mentioned above, and 
the star collapses into a BH through tentative nuclear burning phases~\cite{Zwick_2023,Nandal_2024}.
This direct collapse of massive clouds would induce an asymmetric structure and generate a burst of GWs at frequencies 
detectable both by TianQin and LISA~\cite{Zwick_2023}.

\begin{figure}[t]
\includegraphics[width=82mm]{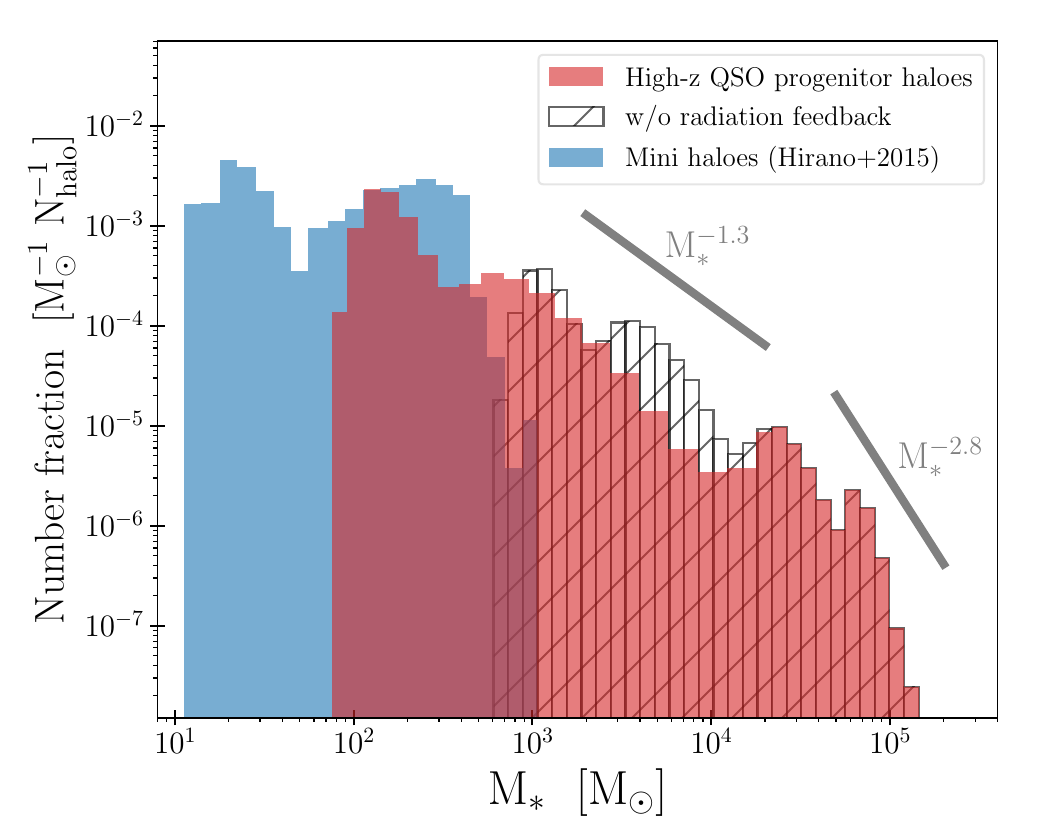}\hspace{3mm}
\includegraphics[width=86mm]{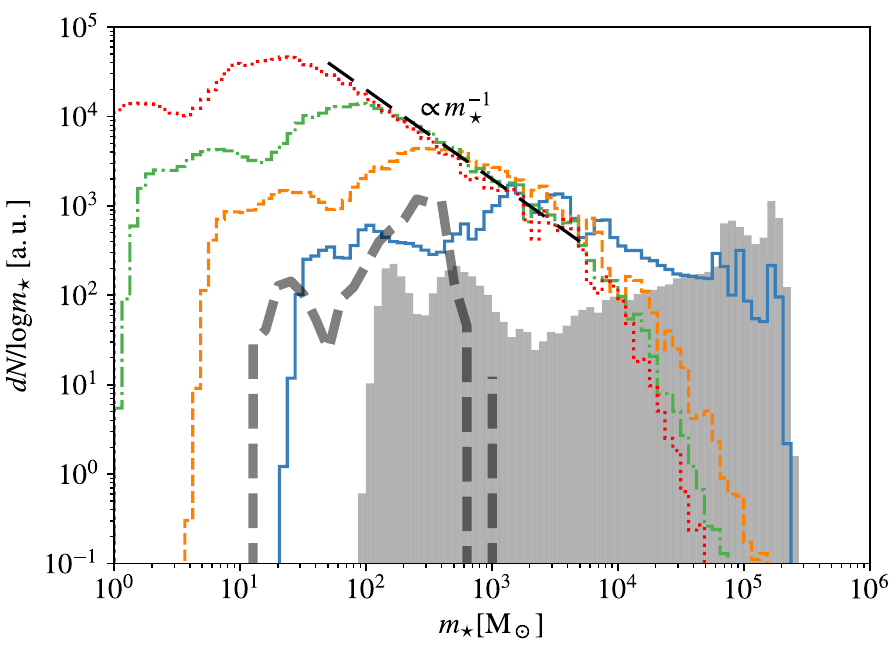}
\caption{Left panel: Mass distribution of seed BHs formed by $z\sim 20$ within progenitor dark-matter halos that end up in high-redshift quasar host galaxies 
(red; \citet{Li_2023,Toyouchi_2023}) and in typical dark-matter halos of protogalaxies (blue; \citet{Hirano_2015}). 
In the overdense region of the universe, the BH mass function for these seed populations is well approximated with $\Phi_\bullet \propto M^{-1.3}$
(c.f., $\Phi_\bullet \propto M^{-1.35}$ for a Salpeter mass distribution)
and an upper mass at $\gtrsim 10^5~M_\odot$, while the BH mass function in less-biased regions of the universe is approximated with a log-normal distribution
with a characteristic mass of $\sim 300~M_\odot$.
Right panel: Mass distribution of seed BHs taken from \citet{Liu_2024}, where the effects of fragmentation/multiplicity on the final
products during the star-formation process are considered.
Efficient cloud fragmentation maintains a mass-distribution slope of $\Phi_\bullet \propto M^{-1}$.
}
\label{fig:BHMF}
\end{figure}

\subsubsection{The Initial Mass Function of Seed BHs}

The diversity in accretion and feedback processes for Pop~III stars and heavy seed BH progenitors leads to distinct final mass outcomes.
By integrating various physical processes and accounting for the abundance of parent dark-matter halos, we can develop a theoretical model of 
the initial mass function for seed BHs (e.g., \citet{tanaka_haiman_2009}, \citet{Devecchi_2012}, and \citet{Sassano_2021}).

In Figure~\ref{fig:BHMF} (left panel), we illustrate a theoretical prediction of the initial mass function for two different seed BH populations: 
one located in overdense regions (red)~\cite{Li_2023, Toyouchi_2023}
and the other in typical regions of the universe (blue)~\cite{Hirano_2015}. 
The overall BH mass function including the two populations spans a broad range from $10~M_\odot$ to $10^5~M_\odot$. 
The mass distribution of seed BHs in the main progenitors of quasar hosts -- the most massive halos among all subbranches -- follows a Salpeter-like function 
($\Phi_{M_\bullet} \propto M_\bullet^{-1}$) and reaches an upper mass limit of $\gtrsim 10^5~M_\odot$.
Conversely, the mass spectrum of Pop~III BHs formed in typical regions is expected to be clustered around $\sim 100~M_\odot$, 
with a sharp cutoff above $10^3~M_\odot$ owing to stellar feedback in these environments.
As shown in the right panel, a similar slope of the mass distribution would be maintained down to $\sim 10~M_\odot$ when the effects of cloud fragmentation and stellar 
(presumably BH) multiplicity are taken into account~\cite{Liu_2024}. 

Importantly, the initial mass distribution of seed BHs is more continuous due to a combination of various BH seeding mechanisms. 
This stands in contrast to previously discussed scenarios where either Pop~III BHs (with masses of $10$--$100~M_\odot$) or 
heavy seed BHs (around $10^5~M_\odot$) exclusively contribute to the BH population at later epochs (e.g., \citet{Barausse2014}, \citet{Ricarte_2018}, and \citet{Barausse_2020}).
The predicted BH mass function can potentially show tight correlations with the quasars observed at lower redshifts (around $z\sim 4$--$6$).
The verification of these predictions and the examination of the proposed correlations will be a key focus of future space-borne GW observations with TianQin and LISA
through the detection of GW events from merging BHs across a wide range of masses as predicted.


\subsection{Cosmological evolution of MBHs and galaxies}
\label{subsec:Coevolution}
\contributors{Kohei Inayoshi, Hai-Tian Wang, Lucio Mayer, Pedro R. Capelo, Pau Amaro Seoane}



%
%
%
%

\subsubsection{Accretion and mergers}\label{sssec:am}

The growth of MBHs occurs via two channels: gas accretion and mergers. Gas accretion is usually associated with the presence of an accretion disk, in which gas moves inward via angular momentum transport outward~\cite{Lynden-Bell_1969}. The growth rate would be limited to the Eddington value when the incident radiation flux powered by accretion exerts the outwarding force on the inflow through electron scattering~\cite{Eddington_1916}.
Or, in some cases when the radiation energy is effectively trapped within the dense accreting matter, the rate can be even higher~\cite{begelman_1979,Abramowicz_et_al_1988,Sadowski_et_al_2016}. However, both sustained near-Eddington accretion and (episodes of) super-Eddington accretion depend on the availability of fuel and on the geometry of the flow~\cite{Inayoshi_2016, Regan_et_al_2019,Sassano_et_al_2023}. Additionally, to a minor extent, gas accretion can also occur through stellar tidal disruption events, wherein stars that venture too close to the nuclear BHs with masses of $\lesssim 10^8~M_\odot$ are disrupted~\cite{hills_1975}.

Mergers take place when two BHs coalesce with each other (in principle of any mass and mass ratio, including intermediate and extreme mass-ratio in-spirals). The merger process typically begins with the merger of the host galaxies. Following this, the BHs lose orbital energy and angular momentum through dynamical interactions with background stars and gas, and eventually become mutually gravitationally bound, thus forming a BBH~\cite{Capelo_et_al_2015}. 
As mentioned above, recent advancements in modeling the interstellar medium, star formation, and feedback processes in galaxy formation, 
along with improved galaxy merger simulations, have highlighted that the formation of a BBH is not guaranteed for BH masses in the range 
relevant to TianQin and LISA, specifically those below $M_{\rm BH}\sim 10^7~M_\odot$.
This is because such BHs, as opposed to larger ones, can be easily perturbed by the clumpiness in the ambient medium, including close encounters 
with giant molecular clouds.
Additionally, their host galaxies, being disk-dominated, can develop strong spiral patterns and bar instabilities.
These factors perturb the orbits of these BHs and cause stalling or temporary outward migration of the secondary BH. 
This phenomenon is known as the ``last kilo-parsec problem''~\cite{lisa_2023}.

Understanding the co-evolution of MBHs and their host galaxies is one fundamental factor in quantifying 
the relative importance of accretion and mergers in the last kilo-parsec problem.
Sustained gas accretion increases the mass of the BH, which can sink faster toward the center of the galaxy's potential well
due to the increased efficiency of dynamical friction.
The dense, central region provides more fuel for further accretion and a higher probability of encountering another BH. 
However, rapid gas accretion also means that more energy is radiated away and coupled to the surrounding gas.
This feedback prevents further growth and reduces the efficiency of dynamical friction~\cite{Park_Bogdanovic_2017,SouzaLima_et_al_2017, Liao:2023jci},
challenging the early growth of light seed BHs into the frequency window of TianQin and LISA.

Finally, if these barriers are overcome, a BBH can form when the two BHs are separated by less than their sphere of influence radius, 
roughly a few parsecs for a million solar-mass BH, as a result of efficient dynamical friction by the gaseous background 
in the galactic nucleus of the merger remnant (e.g. \citet{Chapon_et_al_2013} and \citet{Pfister_et_al_2017}). 
Further hardening to milli-parsec scales must occur before GW emission takes over and leads to ultimate coalescence~\cite{Begelman_et_al_1980}. 
This stage is also non-trivial, as many astrophysical processes can slow down the orbital decay of the secondary BH, 
resulting in a delay of up to a billion years in the merger of the BHs relative to that of their host galaxies, 
or even causing a reversal of orbital decay, leading to the outward migration of the secondary BH~\cite{Duffell_et_al_2024}. 
The likelihood of one outcome versus the other under realistic conditions still needs to be understood.
The mass growth via mergers can at most double the mass of the BH in each event, particularly in the case of equal-mass mergers.
Even in that case, a non-negligible fraction of the total mass is radiated away through GWs. 

\subsubsection{Importance of BH spins and other caveats}

One key quantity closely linked to the complex interplay between mergers, accretion, and feedback is the BH spin. 
The magnitude and orientation of the individual BH spins significantly impact gas accretion onto the BHs 
since the BH spin determines the size of the innermost stable circular orbit, which in turn affects 
the radiative efficiency of an accretion disk (i.e. the conversion fraction from rest-mass of the accreting gas
to photons)~\cite{Novikov_Thorne_1973, ShakuraSunyaev1973, Dong-Paez:2023qrd}. 
Additionally, the spins, mass ratio, and orbital eccentricity influence the velocity of gravitational recoils 
\cite{bonnor_rotenberg_1961}.
The recoil associated with GW emission could potentially eject the remnant BH from the centers of the host galaxy,
preventing their subsequent mergers with new incoming BHs. 
Furthermore, the spin of the remnant BH is a function of the original spins, the mass ratio, and the presence or absence of accreting gas.
Thus, an accurate understanding of BH spins -- currently accessible through X-ray reflection spectroscopy for a limited number of systems~\cite{Reynolds_2021} -- could provide valuable insights into the history of the BH assembly.

The relationship between BH mass and spin, gas accretion and feedback, and mergers, along with the influence of 
star formation and stellar feedback~\cite{vanwassenhove_capelo_2014,Pfister_et_al_2019} as well as various aspects 
of galactic dynamics -- including the effects of gas clumps~\cite{Tamburello_et_al_2017} and stellar bars~\cite{bortolas_capelo_2020,Bortolas_et_al_2022} on the sinking time-scales -- has extensively been explored 
through simulations and EM observations. 
Despite these efforts, substantial uncertainties remain in current models, with estimates of BH number densities~\cite{Habouzit_et_al_2016} and the expected numbers of GW events~\cite{lisa_2023} varying by several orders of magnitude.
Observational challenges further complicate our understanding, particularly due to the limited knowledge of obscured accretion 
and the absence of new X-ray observatories. 
Nevertheless, there is a general agreement that gas accretion is the primary mechanism driving BH growth across redshifts of 
$0\leq z \lesssim 5$~\cite{Soltan_1982,Yu_Tremaine_2002, Ueda_2014}. 
Recent findings of abundant faint AGN populations by JWST~\cite{Kocevski_2023, Harikane_2023, Matthee_2024} indicate 
higher radiative efficiency than previously accounted for in the standard So{\l}tan argument and 
suggest a highly efficient mechanism for spinning up MBHs during their early assembly~\cite{Inayoshi_Ichikawa_2024}. 
These discoveries will improve our understanding of the cosmic evolution of BH mass and spin owing to accretion and mergers, 
and their relative contributions~\cite{Volonteri_2005,Berti_2008}.

In the future, there should be a definite improvement in X-ray observations, with many missions (e.g. Athena~\cite{nandra_et_al_2013}, LynX~\cite{LynxTeam_2018}, AXIS~\cite{mushotzky_2018}, eXTP~\cite{zhang_et_al_2016}, and STROBE-X~\cite{ray_et_al_2019}) hopefully replacing or complementing eROSITA~\cite{Predehl_et_al_2010}, XRISM~\cite{XRISM_2020}, and the aging Chandra and XMM. This new deluge of X-ray data will help shed more light on the role of gas accretion onto MBHs (and on the values of their spins). Additionally, present and future surveys in other windows of the EM spectrum will also be crucial, with, e.g. JWST~\cite{Gardner_et_al_2006} providing precious high-redshift information on the host galaxies (and on the BH masses themselves, using the H$\alpha$ line emission~\cite{pacucci_nguyen_2023,Reines_et_al_2013}). EM observations, though, are not able to give us the full picture of BH mergers, since at most they can follow pairs (or, in some rare cases, binaries) of accreting BHs, the so-called dual AGN~\cite{DeRosa_et_al_2019}.
Linking dual AGN to BH mergers is extremely difficult, since the two BHs, before reaching the GW phase, must still travel enormous distances and clear several hurdles~\cite{Steinborn_et_al_2016,Capelo_et_al_2017,li_bogdanovic_2023, Liao:2023zxx}. 
Moreover, BH mergers inevitably occur after a galaxy merger, wherein tidal torques~\cite{Barnes_Hernquist_1996,Hopkins_Quataert_2010,Capelo_et_al_2015} and hydrodynamical torques~\cite{Barnes_2002,capelo_dotti_2017} funnel gas towards the central BH(s), further complicating the picture~\cite{Capelo_et_al_2023}. TianQin will provide, alone or in conjunction with other GW observatories (e.g. LISA~\cite{lisa_2023,torres-orjuela_huang_2023,Colpi_et_al_2024}), the merger history of BHs over a wide range of masses and redshifts, along with their spin distribution, thus giving us the missing piece of the ``accretion versus mergers'' puzzle.

\subsection{Multi-messenger detection}
\label{subsec:multi-det}

\contributors{Andrea Derdzinski, Rongfeng Shen}\vspace{-0.5cm}

\subsubsection{EM signatures of coalescing SMBHs}

Merging of a binary of SMBHs is believed to happen at the center of their host galaxies. As long as there exists copious gas in the vicinity of the merging site, the violent dynamics there may cause gas to generate detectable EM signatures \cite{Dong-Paez:2023qrd}. 
The list of work that have discussed this topic is long. Here we review only a very limited selection of them. For a detailed understanding, we refer the reader to reviews in \citet{2023arXiv231016896D} and \citet{2022LRR....25....3B}.

\textbf{Pre-merger to merger:} The pair of SMBHs approaching their merger likely clears out a gap inside the circum-binary gaseous disk. The gas entering the gap from the circum-binary disk will form a mini-disk around each SMBH, as shown in several hydrodynamical simulation works in both Newtonian and relativistic regimes~\cite{2014PhRvD..90j4030G,2015MNRAS.447L..80F,2023ApJ...949L..30D,2023arXiv230518538A,2022ApJ...928..137G,2021PhRvD.103j3022C,2018ApJ...865..140D,2018MNRAS.476.2249T,bowen17,bowen19,2024arXiv240110331F}. These works generally predict that modulated accretion onto the BHs manifests as a quasi-periodic variation of the disk luminosity,  
For example, the relativistic simulations of magnetized mini-disks by~\cite{bowen17,bowen19} show that the pair's interaction causes spiral density waves within each mini-disk, which in turn modulates the accretion rate toward each BH, producing variable emission at 2 - 3 times the binary orbital frequency. Several of the above works demonstrate the connection between the spectral energy distribution of the emitted radiation and the coalescence stage of the binary, predicting characteristic optical or X-ray `chirp' variability~\cite{2018MNRAS.476.2249T} as well as pre-merger X-ray and UV flux evolution~\cite{2024arXiv240110331F,2023MNRAS.526.5441K}. Similarly, relativistic Doppler modulations and lensing by the BHs can produce correlated EM variability~\cite{2017PhRvD..96b3004H}. 

In most cases, we expect the gas configuration in a galactic nucleus to resemble a disk configuration. It is interesting to note, however that when dropping this assumption, variable emission is still predicted around an SMBH binary merger. 
The relativistic, hydrodynamical simulation by \citet{bode_haas_2010} studied the late inspiral and merger of equal-mass, spinning SMBH binaries in a gas cloud. They showed that variable EM signatures correlated with GWs can arise as a consequence of shocks and accretion combined with the effect of relativistic beaming. The most striking EM variability shows an amplitude variation of a factor of 2, at the same frequency as the GW frequency. This happens for systems where spins are aligned with the orbital axis. 


\textbf{Post-merger:} Due to energy release via GW emission, the new BH after the merger has a slight, but sudden mass reduction. Many works have considered the disk's dynamical response to the sudden drop of the gravitational potential. The inner edge of the disk may rapidly retreat, causing a reduction to the disk luminosity or a change of the jet radio luminosity~\cite{o'neill09}. It may also induce oscillations in the disk which then modulate the disk internal energy and its Bremsstrahlung luminosity~\cite{megevand09}.   

More importantly, the new BH will have a recoil velocity. When this velocity has a component that is parallel to the disk plane, the disk’s dynamics are strongly impacted, giving rise to relativistic shocks that generate EM radiation. Based on hydrodynamical simulation,  \citet{corrales10} estimates that the luminosity rises steadily on the time-scale of months, and reaches a few $\times 10^{43}$ erg s$^{-1}$, corresponding to about 10 percent of the Eddington luminosity of the central binary SMBH. Moreover, depending on the orientation of the kick, the recoiled BH may capture and carry with it a significant fraction of the disk. The enhanced energy release, from both the circularization of the ``kicked'' disk material and its subsequent accretion. The luminosity can increase to $\sim 10^{43}$ erg s$^{-1}$ for a near-in-plane kick of $\sim 10^3$ km s$^{-1}$~\cite{rossi10}.\\

Ultimately, rapid cooperation between GW detectors and time-domain sky surveys is critical for enabling multimessenger discoveries. Timely localization of sources plays an important role, particularly if we want to detect pre-merger signatures of MBH binaries. These detections will provide host galaxy identification, and redshift measurements, thus turning MBH merger detections into standard sirens and probes of cosmological model parameters~\cite{2005ApJ...629...15H,haiman_xin_2023}. 
The real-time transmission and analysis of the GW data will play a key role in the multi-messenger identification and research on the MBH binary mergers.
The TianQin's geocentric orbit makes it much easier for ground receivers to collect and have all-day coverage of the TianQin data, allowing a real-time data downlink workflow.
A proof-of-principle method has been implemented, indicating that the merging MBH binaries can be identified and measured on the fly, with a refresh time as short as one hour, depicting a promising future for a multi-messenger observation with TianQin~\cite{chen_lyu_2024}. 
Joint EM-GW signatures of MBHs in AGN disks are further analyzed in Section~\ref{subsubsec:eimriem}.

\subsubsection{IMBH binaries in AGN}
\label{sec:IMBH_AGN}

TianQin has a tantalizing advantage of detecting the mergers of IMBH binaries and the early inspiral of stellar-origin binaries, given its sensitivity to higher frequencies compared to LISA. 
Both of these sources can exist in galactic nuclei or AGN environments (for the evolution of stellar-mass BBH in AGNs please refer to Section~\ref{subsec:AGN_BBH} and more information of IMBHs in AGNs refer to \ref{subsubsec:eimriem}), in which the dense stellar population and vast supply of gas facilitate BH interaction \cite{Sedda:2019btz}. 
The early inspiral stage of stellar-mass BH binaries is unlikely to coincide with bright, characteristic emission signatures. However, sources that merge in the millihertz band can produce associated flares due to the remnant's violent interaction with surrounding gas~\cite{McKernan2014}. For multimessenger prospects with TianQin, it is most interesting to focus on the IMBH binary mergers and light intermediate-mass-ratio inspiral (IMRI) mergers. These events may occur as a product of hierarchical mergers, for which the lower-mass end is detected by ground-based detectors (e.g. \citet{Morton2023} and \citet{Graham20}).

\subsection{Contributions from GW observations with TianQin}
\label{subsec:unique2}

\contributors{Pau Amaro Seoane, Hong-Yu Chen, Jie Gao, En-Kun Li, Alejandro Torres-Orjuela, Hai-Tian Wang} 

In this section, we discuss how GW observations with TianQin contribute to our understanding of 
the origin and growth processes of MBHs over cosmic time.
The sensitivity of the TianQin detector in the millihertz frequency range makes it particularly well-suited to detect sources 
of lower mass~\cite{tianqin_2016,TQ_MBH_2019}, complementing LISA, which excels in detecting higher mass sources in a different frequency range~\cite{Klein:2015hvg}. 
This complementary capability allows TianQin to play a crucial role in studying the birth and growth of MBHs, 
especially those at the earlier stages of their formation.

\subsubsection{MBHs merger rate and TianQin detection rate}
\label{subsec:MD-rate}



We begin by examining the ability of TianQin to detect the mergers of MBHBs and discuss the associated merger rates. 
While the realistic mass function of seed BHs is predicted to have a continuous distribution (see Figure~\ref{fig:BHMF}), 
for simplicity, we will frame our discussion in terms of the light-seed versus heavy-seed models, 
which suggests a bimodal distribution with peaks at distinct masses.
Our focus here is to explore how the properties of BH seeding mechanisms -- such as the mass distribution and formation epochs -- 
impact the GW event rates and detection probabilities. 
Rather than providing specific rate predictions within a realistic setup, we aim to understand how these factors influence 
the likelihood and characteristics of GW events observed by TianQin.

For this analysis, we utilize semi-analytical models that are extensively used to assess the expected LISA scientific performance 
for given experimental and detector designs~\cite{Klein:2015hvg}.
The models are based on the extended Press and Schechter formalism, focusing on the two seeding mechanisms:
light-seed and heavy-seed paradigms.

One key factor influencing the distribution of MBHs is the time delay between the merger of the MBH binary and the merger of their host galaxies.
Starting from the moment when the halos initially interact, many factors can take place and affect the binary evolution~\cite{BK:2007bvo}. 
One needs to first consider the effects of tidal disruption and heating of the satellite halo (and galaxy)~\cite{taffoni_mayer_2003}. 
Next, the dynamics at small scales need to be examined to predict the time delay between the MBH binary formation and their merger.
This timescale is highly uncertain and may exceed the Hubble time in some instances (the so-called final parsec problem \cite{Zhang:2023jrk}; see, e.g. \citet{colpi_2014} for a review).
Different timescales based on the MBH binary's environment are all possible~\cite{Klein:2015hvg,Antonini:2015sza}.
Delays can be as long as a few gigayears if the binary evolution is dominated by stellar hardening, or it can be much shorter as $10^7$--$10^8$ years if the MBH binary shrinks due to migration in a nuclear gas disk.
The impact of triple MBH interactions on these delay times can also be modeled~\cite{Antonini:2015sza}.

For the light-seed model, the numerical simulation indicates that the inclusion of time delay has little impact on the merger rate, so only the time-delay case is considered~\cite{Klein:2015hvg}. 
As a cautionary remark, these models do not account for the potentially important effect of orbital decay bottlenecks at much
larger scales, referred to as the ``last kiloparsec problem''~\cite{lisa_2023}. 
This is because the latter aspect requires more investigation before an additional delay-like prescription, or some other 
phenomenological correction, can be formulated and applied to BH-merger rate forecasts.

However, the situation is entirely different for ``heavy-seed'' models. Whether or not we consider the time delay will significantly affect the distribution of MBHs. Therefore, we have two distinct models: one that includes the time delay and one that does not.
In the heavy-seed model without a time-delay, MBH binaries merge immediately after the merger of their host galaxies. In the delayed heavy-seed model, the time interval between the merger of MBH binaries and their host galaxies can span a few gigayears.

\begin{figure}[h]
\centering
\subfloat[Light seeds\label{A.Klein distributions1}]{
    \includegraphics[width=0.5\textwidth]{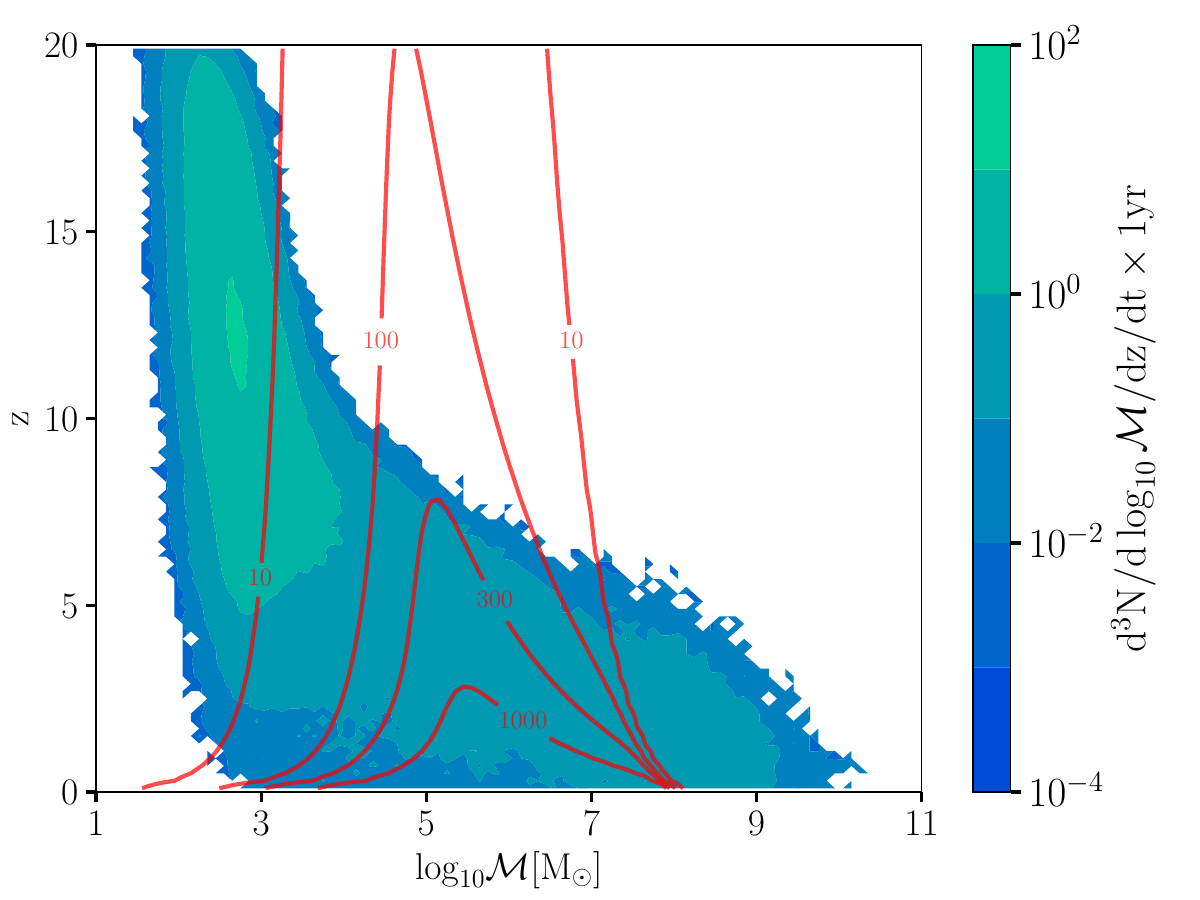}}
\subfloat[Heavy seeds\label{A.Klein distributions3}]{
    \includegraphics[width=0.5\textwidth]{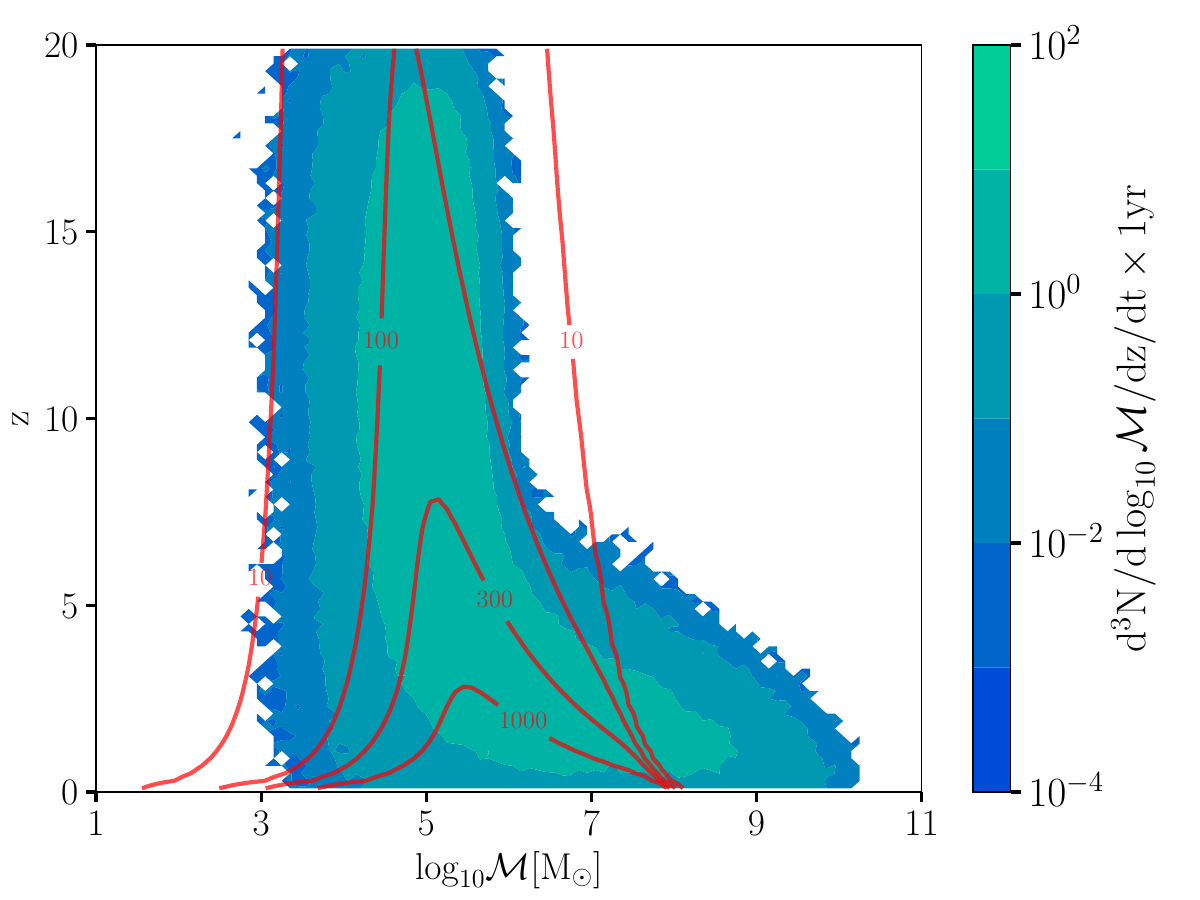}}
    
\subfloat[Heavy seeds (delay)\label{A.Klein distributions2}]{
    \includegraphics[width=0.5\textwidth]{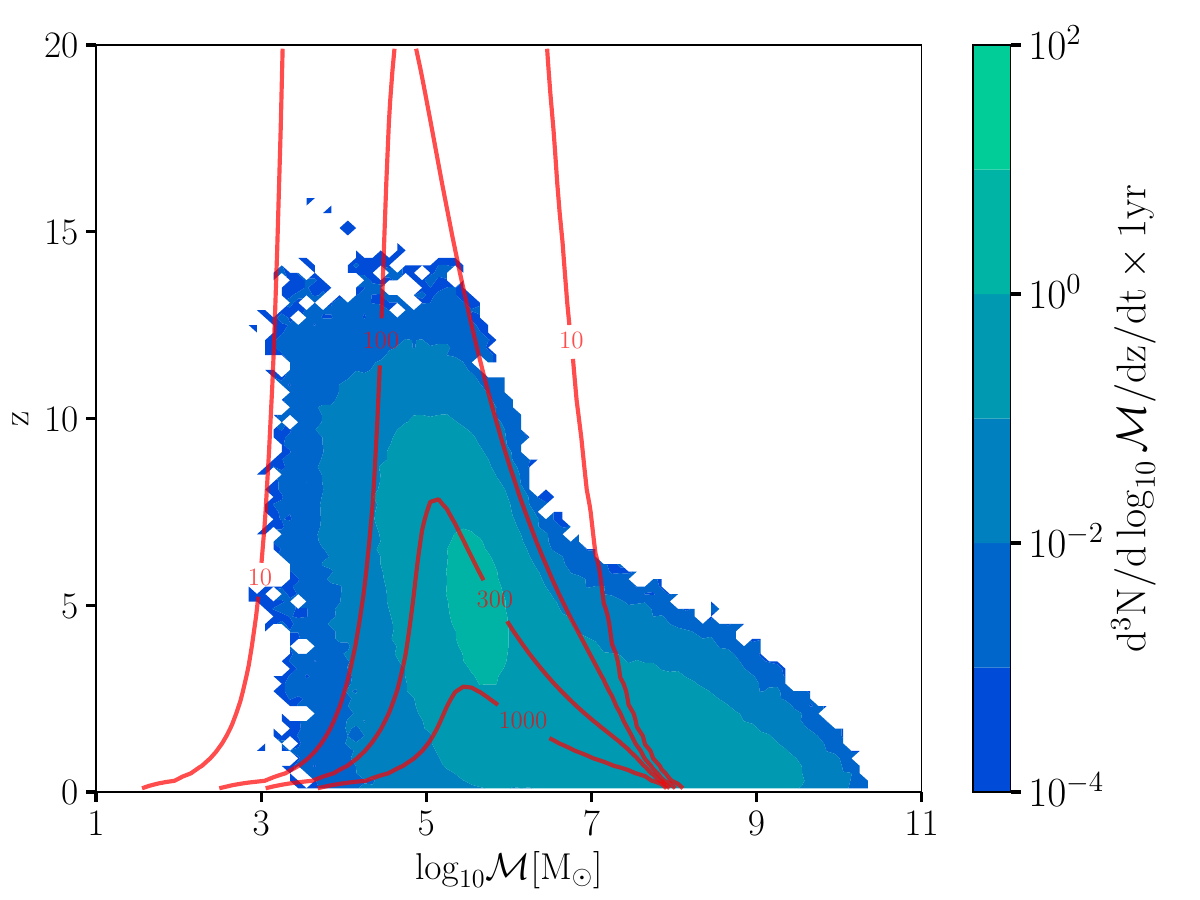}}
\caption{The distribution of mergers over chirp mass and redshift of the light-seed model (upper left panel), delayed heavy-seed model (lower panel) and heavy-seed model without a time delay (upper right panel).
The red contour line represents the average SNR in the TianQin detector, assuming equal-mass binaries~\cite{TQ_MBH_2019}.
}
\label{fig:eps_mz}
\end{figure}

In Figure~\ref{fig:eps_mz}, the binary MBHs' merger rate density is shown on the chirp mass and redshift plane.
Each model presents a distinct pattern of mass distribution over redshift.
The light-seed model suggests most detectable sources appear at $z\sim 5$--$10$, predominantly at lower masses.
In contrast, both heavy-seed models predict most mergers over a broad range of redshifts.
The impact of incorporating delays into BH mergers is demonstrated when comparing the two heavy-seed models. 
In the heavy-seed model without a time delay, mergers start as early as $z\sim 20$, whereas in the delayed heavy-seed model, they begin only at redshifts lower than $z\sim 15$.

\begin{table}
    \begin{center}
        \begin{tabular}{c|c|c|c|c|c}
            \hline
            \hline
            \multicolumn{1}{c|}{\multirow{2}{*}{model}} &
            \multicolumn{1}{c|}{\multirow{2}{*}{event rate($yr^{-1}$)}}&
            \multicolumn{2}{c|}{TianQin}&\multicolumn{2}{c}{Twin Constellations}\\ \cline{3-6}
            \multicolumn{1}{c|}{}&\multicolumn{1}{c|}{}&
            DR($yr^{-1}$)&DP&DR($yr^{-1}$)&DP \\
            \hline
            light-seed model&174.70&10.58&6.1\%&22.60&12.9\%\\
            \hline
            delayed heavy-seed model&8.18&4.42&54.0\%&8.06&98.5\%\\
            \hline
            heavy-seed model without delay&122.44&58.96&48.2\%&118.12&96.5\%\\
            \hline
            \hline
        \end{tabular}
    \end{center}
    \caption{\label{tab:event_rate} MBH binary cosmic merger rates and TianQin detection rates for the three investigated MBH population models. Detection rates (DR) are given both considering one or two TianQin detectors. We also show the detection percentage (DP), which represents the percentage of detection rate as a percentage of the event rate~\cite{TQ_MBH_2019}.}
\end{table}

Further analysis allows us to calculate the merger rates of binary MBHs from these models, as shown in Table.~\ref{tab:event_rate}, and more details can be found in \citet{TQ_MBH_2019}. 
Specifically, the expected numbers of mergers during one year are 174.7, 8.2, and 122.4 for the light-seed, the delayed heavy-seed, and the no delayed heavy-seed models, respectively. 
We then estimate the detection rates by calculating the SNRs for each event; only events with a SNR greater 
than 8 are considered detectable. 
The resulting annual detection rates for TianQin are 10.6, 4.4, and 59.0 for each model, respectively. 
These rates are approximately half of those projected for LISA, largely due to TianQin's ``3 month on + 3 month off'' observational scheme. 
If TianQin were to operate with two constellations working sequentially to cover observational gaps, it could potentially match LISA's detection rates.

Distinguishing between different seeding models is challenging for both TianQin and LISA when operated independently. 
Interestingly, TianQin performs better in the higher frequency band ($>10$ mHz), while LISA is more effective in the lower frequency band ($<10$ mHz). 
This suggests that TianQin is better suited for detecting sources of lower mass, whereas LISA excels in detecting higher mass sources. 
Thus, combining the capabilities of both TianQin and LISA could enhance our ability to differentiate between various seed models.
\vspace{1mm}

\subsubsection{IMBHs binaries in colliding galaxies}

IMBHs, with masses ranging between $10^2$ and $10^4$ solar masses, are believed to often form in the centers of dense stellar clusters. Observations suggest that star clusters frequently form in pairs or larger bound groups of colliding galaxies, making cluster mergers likely. When two clusters, each harboring an IMBH at their center, merge, the resulting IMBH binary can become a prominent source of GWs. The direct detection of these systems using traditional photon-based astronomy, which relies on monitoring the innermost stellar kinematics, seems to be a challenging task far in the future. However, the detection and characterization of IMBHs have promising prospects in lower-frequency GW astrophysics. 
To understand the dynamical processes involved in such scenarios, simulations show that within approximately 7 million years, the clusters merge, and the IMBHs form a hard binary, which then coalesces in roughly 100 million years. Interactions with surrounding stars increase the eccentricity of the IMBH binary to high values. Although the binary later circularizes due to the emission of GWs, the residual eccentricity can still be detectable through its influence on the phase of the waves, especially in the final years of inspiral~\cite{amaro-seoane_freitag_2006,amaro-seoane_miller_2009}. TianQin, with its unique sensitivity to higher frequencies, is well-suited to detect these IMBH binaries.

The dynamical evolution of IMBH binaries in stellar clusters has been a topic of increasing attention.
Studies involving simulations of equal-mass binaries of IMBHs embedded in stellar distributions with different rotational parameters indicate that eccentricities and inclinations are primarily determined by the initial conditions of the IMBHs and the influence of dynamical friction, though they are later perturbed by the scattering of field stars \cite{Arredondo:2024nsl}. 
Very large eccentricities can occur for binaries with low initial velocities, while higher initial velocities result in more circular orbits. 
Counter-rotation simulations yield significantly different results in eccentricity, suggesting a dependency on the rotational parameter~\cite{amaro-seoane_eichhorn_2010}.

For reasonable IMBH masses, there is only a mild effect on the structure of the surrounding cluster, even though the binary binding energy can exceed the cluster's binding energy.
Simulations demonstrate that the residual eccentricity should be observable in the TianQin band, inducing a measurable phase difference from circular binaries in the last year before the merger.
Despite the energy input from the binary decreasing the density of the core and slowing down interactions, the total time to coalescence is typically less than 100 million years, making these mergers unique snapshots of clustered star formation.

GW observations from TianQin will provide invaluable data on the formation and evolution of IMBH binaries. 
By detecting the early inspiral signals of stellar-mass BHs years to decades before their final merger, TianQin will be able to study the dynamics of dense stellar environments and the role of IMBHs in these systems. 
Additionally, the detection capabilities of TianQin will be enhanced by its sensitivity to the millihertz frequency range, making it particularly effective in observing the inspiral and final merger stages of IMBH binaries.

While TianQin excels in detecting IMBH binaries due to its sensitivity to higher frequencies, its collaboration with other space-borne detectors, such as LISA, will provide significant scientific advantages.
Joint observations can enhance our ability to distinguish between different models of BH formation and evolution by covering a broader range of masses and redshifts \cite{Gao:2024uqc, Shuman:2021ruh, Zhang:2021wwd, Zhang:2021kkh, Yang:2022cgm, Liu:2023hqe}. 
This synergy will help resolve uncertainties in current models and potentially uncover new phenomena. TianQin's unique capabilities, combined with those of LISA, promise to advance the field of GW astronomy, offering new insights into the formation and evolution of MBHs.


Studies have analyzed the signals from binary IMBH systems using waveform models obtained from numerical relativity simulations \cite{Wang:2024iyj} coupled with post-Newtonian calculations at the highest available order~\cite{amaro-seoane_santamaria_2010}. 
These analyses indicate that IMBH binaries with total masses between 200 and 20,000 $M_\odot$ would produce significant SNRs in Advanced LIGO, Virgo, and ET. Joint detection with TianQin would help break the degeneracies in parameter extraction.
The expected event rate of IMBH binary coalescences has been computed for different binary configurations, revealing interesting values that depend on the spin of the IMBHs. Ground-based GW observatories' prospects for detecting and characterizing IMBHs would not only provide robust tests of general relativity but also corroborate the existence of these systems. Such detections would allow astrophysicists to probe the stellar environments of IMBHs and their formation processes.

\subsubsection{Detection capabilities of TianQin}

The sensitivity of TianQin is optimal for frequencies above $10\,{\rm mHz}$, while LISA is most sensitive to frequencies below $10\,{\rm mHz}$~\cite{tianqin_2016,lisa_2017}. 
Therefore, LISA performs better at detecting GWs from heavier sources, while TianQin is more suited for identifying lighter sources~\cite{torres-orjuela_huang_2023,torres-orjuela_2024}. 
This implies that LISA will generally achieve more accurate detections of `grown' MBHs, whereas TianQin is particularly well-equipped to study the early stage of BH formation and growth. In this section, we aim to demonstrate the advantages of TianQin over LISA and show the effectiveness of joint observations using both detectors. 
To illustrate this, we study the detection distance and the averaged parameter estimation for the three detection scenarios.

\begin{figure}[htbp] 
    \centering 
    \includegraphics[width=0.6\textwidth]{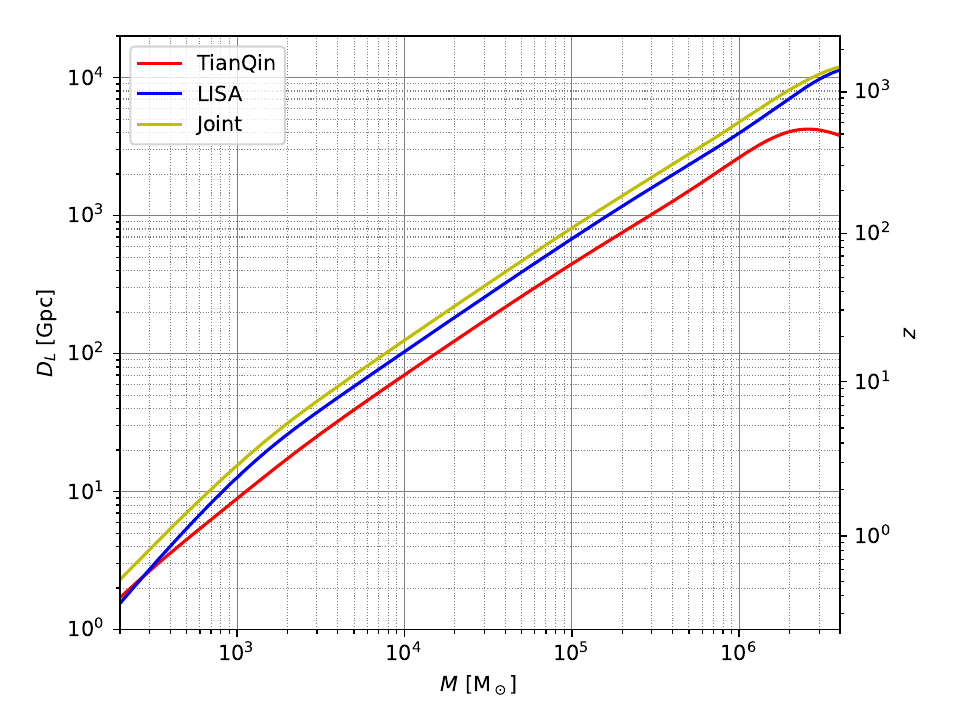}
\caption{
    The distance to which a binary MBH with a SNR of 8 can be detected by TianQin, LISA, and joint detection as a function of the total mass in the observer frame $M$. The left ordinate shows the luminosity distance $D_L$ of the source while the right ordinate shows its cosmological redshift $z$.
    }
\label{fig:mbhb_hor}
\end{figure}

First, we study the detection distance considering an equal mass binary with spins aligned to the angular momentum, $\chi_{z,1} = \chi_{z,2} = 0.5$, that is face-on and at the sky localization $\rm dec \approx +45^\circ$ and $\phi_{\rm bar} \approx 1.9^{\rm h}$. Figure~\ref{fig:mbhb_hor} shows the detection distance for MBHs with a total mass in the observer frame $M$ and a SNR $\rho=8$, as detected by TianQin (red), LISA (blue), and joint detection (yellow). For lower-mass BH mergers with $M \lesssim 300~M_\odot$, TianQin has a greater detection distance, although both detectors reach distances of a few ${\rm Gpc}$. For higher-mass BH mergers up to $M\simeq 10^3~M_\odot$, TianQin, and LISA still perform similarly. However, the detection distance by LISA expands faster than that by TianQin as the total mass increases.
Between $M=10^3~M_\odot$ and $2\times10^6~M_\odot$ in the observer frame, the detection distances of TianQin and LISA differ by a factor of $\simeq 1.5$. Despite this difference, joint detection performs better than LISA alone by a factor of $\simeq 1.2$, which translates to a detection volume larger by a factor of $\simeq 1.7$ compared to that of LISA alone. Remarkably, both TianQin and LISA will be able to detect sources with a total mass of $M\sim 10^6~M_\odot$ out to cosmological redshifts of several hundred. Given that this is the total mass in the observer frame, this means that the TianQin-LISA joint observations will be able to detect binary BH mergers with only several $10^3~M_\odot$ (in the source frame) at the earliest stages of their formation.

Next, we demonstrate the sensitivity of TianQin, LISA, and joint detection to the different parameters of an equal mass binary BHs by computing their average detection error.  We fix the distance of the binary to a cosmological redshift of $z=5$ and consider two cases for the total mass in the observer frame: $M=10^4~M_\odot$ and $M=10^5~M_\odot$. For the other parameters, we adopt ${\rm RA}\in[0,24]\,{\rm h}$, ${\rm dec}\in[-90,+90]\,{\rm deg}$, $\iota\in[0,\pi]$, and $\chi\in[0,0.98]$ and compute their average error. The average error of a parameter is obtained by computing the error for each value considered and taking the arithmetic mean of the single errors. Note that we only report one spin magnitude because we consider an equal mass binary. The errors are calculated using a Fisher matrix analysis and thus correspond to the $1\sigma$ level~\cite{coe_2009}. 

\begin{figure}[htbp]
    \centering
    \subfloat[$M=10^4\, {\rm M_\odot}$ \label{fig:radar4}]{
        \includegraphics[width=0.5\linewidth]{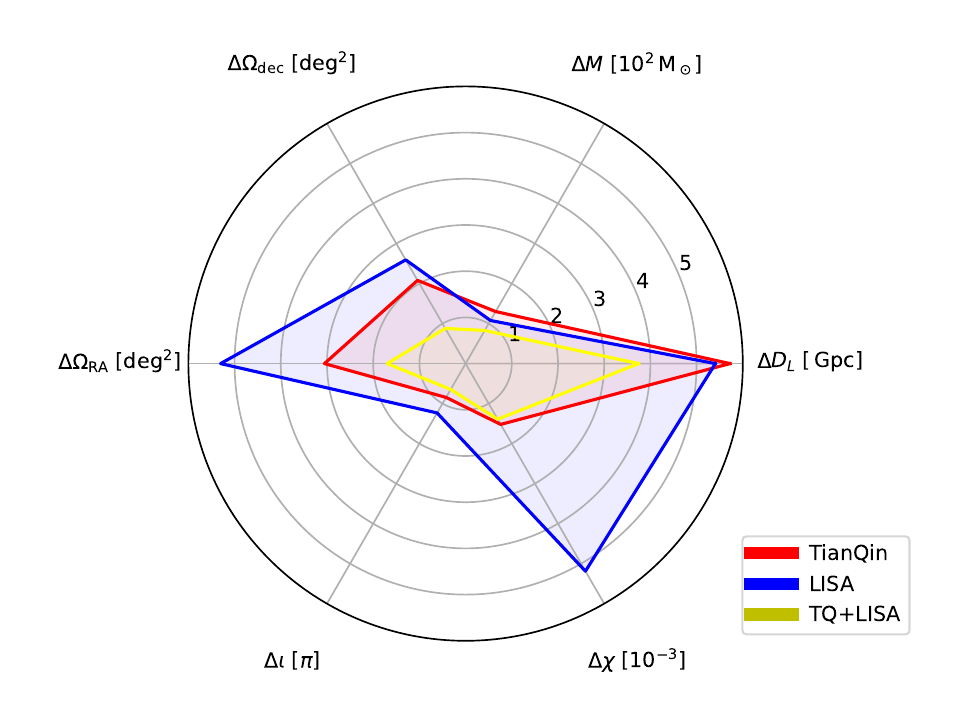}}\hspace{-5mm}
        \subfloat[$M=10^5\,{\rm M_\odot}$\label{fig:radar5}]{
        \includegraphics[width=0.5\linewidth]{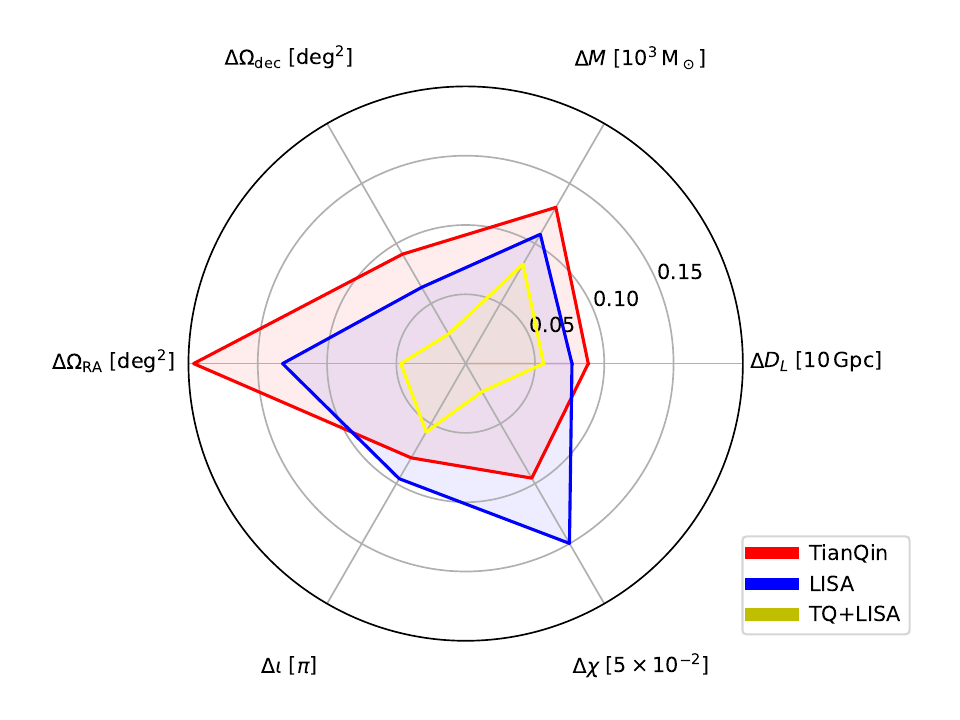}}
    \caption{The average detection error of TianQin (red), LISA (blue), and joint detection (yellow) for the luminosity distance $D_L$, the total mass in the observer frame $M$, the sky localization $\Omega$ as a function of $\theta_{\rm bar}$ and $\phi_{\rm bar}$, the inclination $\iota$, and the spin magnitude $\chi$ for an equal mass binary MBH with a total mass of $M=10^4\,{\rm M_\odot}$ (left) and $M=10^5\,{\rm M_\odot}$ (right) at a cosmological redshift of $z=5$.}
\label{fig:imri_hor}
\end{figure}

Figure~\ref{fig:radar4} presents the detection errors for a massive binary BH with a total mass $M=10^4~M_\odot$. We find that the detection error by TianQin is smaller than that by LISA for all the parameters except the total mass and the luminosity distance. Furthermore, since TianQin and LISA operate at similar sensitivity levels, joint detection always provides significant gains. Specifically, we find that TianQin measures the total mass with an error of $\Delta M \simeq 140~M_\odot$, LISA with an error of just over $100~M_\odot$, and joint detection with an error of less than $100~M_\odot$. For the luminosity distance $D_L$, both TianQin and LISA have errors of $\Delta D_L \simeq 5-6~{\rm Gpc}$, while joint detection improves this to almost $\sim 4~{\rm Gpc}$, which corresponds to an error of the order $0.1$ in the cosmological redshift. The sky localization error is estimated as $\simeq 2-3~{\rm deg^2}$ for TianQin, $2.5-5.5~{\rm deg^2}$ for LISA, and $0.8-1.7~{\rm deg^2}$ for joint detection, respectively. The inclination $\iota$ can be detected with an error of $\simeq 0.8\pi$ for TianQin and $0.6\pi$ for joint detection, respectively, but the error with LISA exceeds $\pi$, making the inclination undetectable in this case. The relatively large errors in inclination measurement are due to the reliance on detecting spherical modes beyond the quadrupolar mode, which are very weak for a light source of only $10^4~M_\odot$. The spin magnitude can be detected at a level of $10^{-3}$ in all three detection scenarios, but TianQin performs better than LISA by a factor of three.

Figure~\ref{fig:radar5} shows the detection errors for a massive binary BH with a total mass $10^5~M_\odot$. For this mass, the detection error by TianQin is smaller than that by LISA for the spin and the inclination of the source, but slightly worse for all the other parameters. The combination of TianQin and LISA significantly improves detection accuracy. Specifically, TianQin has a detection error for the total mass of $\Delta M \simeq 125~M_\odot$, LISA around $110~M_\odot$, and joint detection around $80~M_\odot$. Although the total error in mass is similar to that for a source of $10^4~M_\odot$, the relative error is approximately ten times better because heavier sources are louder. For the luminosity distance $D_L$, TianQin and LISA have errors of $\Delta D_L \simeq 0.75$ and $0.9~{\rm Gpc}$, respectively, while joint detection reduces this error to less than $0.6~{\rm Gpc}$. As a result, the error in the cosmological redshift is well below 0.1. The sky localization error is $\simeq 0.09-0.19~{\rm deg^2}$ for TianQin, $0.07-0.16~{\rm deg^2}$ for LISA, and only $0.02-0.04~{\rm deg^2}$ for joint detection. For a source with $M=10^5~M_\odot$, the inclination can always be determined with errors of $\simeq 0.08\pi$, $0.10\pi$, and $0.06\pi$ for the three cases. The spin magnitude has a detection error of $\simeq 7.5\times10^{-3}$ for LISA, $4.5\times10^{-3}$ for TianQin, and $1\times10^{-3}$ for joint detection, similar to the detection level for the lighter source of $10^4~M_\odot$.

\subsubsection{Advantage of TianQin in sky localization}
\label{sec:mbhb:pe}

TianQin is positioned in a geocentric orbit, allowing the spacecraft itself to maintain constant communication 
with Earth~\cite{tianqin_2016}.
This enables more frequent and reliable data transmission, compared to missions in heliocentric orbits, which have 
limited chances for data downlink due to the longer distance between the satellites and the Earth~\cite{Colpi_et_al_2024}.
For TianQin, real-time data transmission facilitated with real-time data analysis can lead to a real-time discovery of MBH binary mergers, which is crucial for promptly 
pinpointing the source location~\cite{chen_lyu_2024}.
Immediate analysis and localization of these events are essential as they allow astronomers to quickly initiate follow-up observations using EM telescopes \cite{chen_lyu_2024, Ruan:2024qch}.
This capability significantly enhances the scientific return of the mission.

\begin{figure*}[htbp]
  \centering
  \includegraphics[width=0.85\textwidth]{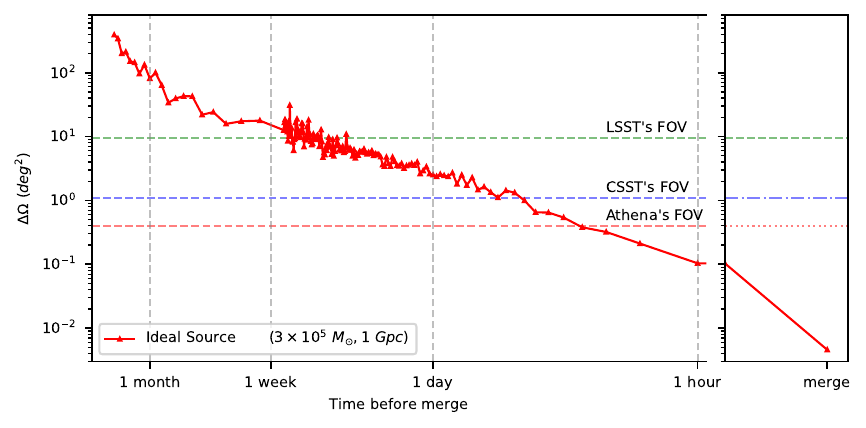}
  \caption{Time evolution of the $90\%$ confidence interval of sky localization uncertainty $\Delta \Omega$ of one source with chirp mass and luminosity distance as ($3\times 10^5 {\rm M_\odot,} 1 {\rm Gpc}$). 
  For reference, the fields of view of LSST  ($\sim$ 9.6 deg$^2$), CSST ($\sim$ 1.1 deg$^2$), and Athena ($\sim$ 0.4 deg$^2$) are denoted by dashed horizontal lines.
  Here, FOV is short for field-of-view.
  The result is obtained from \citet{chen_lyu_2024}.
  }
  \label{localizationError}
\end{figure*}

To determine whether the detectors detects a MBH binary signal, parameter inference of the source will be one of the important and challenging method \cite{Cornish:2021smq, Ruan:2023fce, Du:2023plr, Gong:2023ecg, Destounis:2023ruj, Sun:2023prd, Kong:2024ssa, Xu:2024jbo}.
Figure~\ref{localizationError} shows the temporal progression of the 90\% confidence interval for 
sky localization error $\Delta \Omega$ of one ideal source. 
The dashed horizontal lines indicate the field-of-view for three different telescopes: the Vera Rubin Observatory/LSST~\cite{LSST, 2009arXiv0912.0201L}, the Chinese Space Station Telescope (CSST)~\cite{CSST}, and the Advanced Telescope for High-Energy Astrophysics (Athena)~\cite{nandra_et_al_2013}. 
LSST is a wide-field ground-based system with a 9.6 deg$^2$ field of view, designed to study various objects in the universe with advanced technology. CSST is a space telescope with a 1.1 deg$^2$ field of view, designed and developed in China and planned to be launched and assembled in orbit as part of the Chinese space station project.
Athena is a high-energy astrophysics observatory designed by the European Space Agency with a 0.4 deg$^2$ field of view, designed to study celestial objects emitting X-rays.
All of these telescopes are expected to operate during the observation period of TianQin and can perform multi-messenger observations of binary MBHs.
The variation in localization is due to changes in accumulating SNR and the random sampling process.

As the merger approaches, the localization uncertainty significantly decreases. 
When this uncertainty falls below the horizontal lines (the field-of-view of each telescope), a single snapshot from these telescopes can potentially cover the entire region of the detected GW event, simplifying EM follow-up observation strategies, as outlined in \citet{Chan:2015bma}.
For events with high SNRs, such as the ideal source with $\rm SNR > 4000$, early warnings can be confidently issued to 
these telescopes hours or even days in advance for different telescopes. 
However, for events with lower SNRs, e.g., for sources with an $\rm SNR \sim 400$, there is a risk that EM telescopes may miss 
the target at a crucial moment, especially if data transmission is delayed~\cite{chen_lyu_2024}.

In addition to its near real-time data transmission capabilities, the operation of TianQin can also make a priceless contribution in terms of sky localization.
It has a substantial impact on resolving parameter degeneracies and enhancing the precision of parameter estimation especially for sky localizations.
The differing orbital characteristics of TianQin and LISA affect their signal response preferences.
TianQin's consistently aligned towards the specific source J0806~\cite{tianqin_2016}, 
while LISA's varying 60-degree inclination relative to the ecliptic plane~\cite{lisa_2017} create distinct sensitivities in measuring source parameters.
Having both detectors operating at the same time can achieve orders of magnitude improvement in the parameter estimation precision.

\begin{figure}[htbp]
    \centering
    \includegraphics[width=0.9\linewidth]{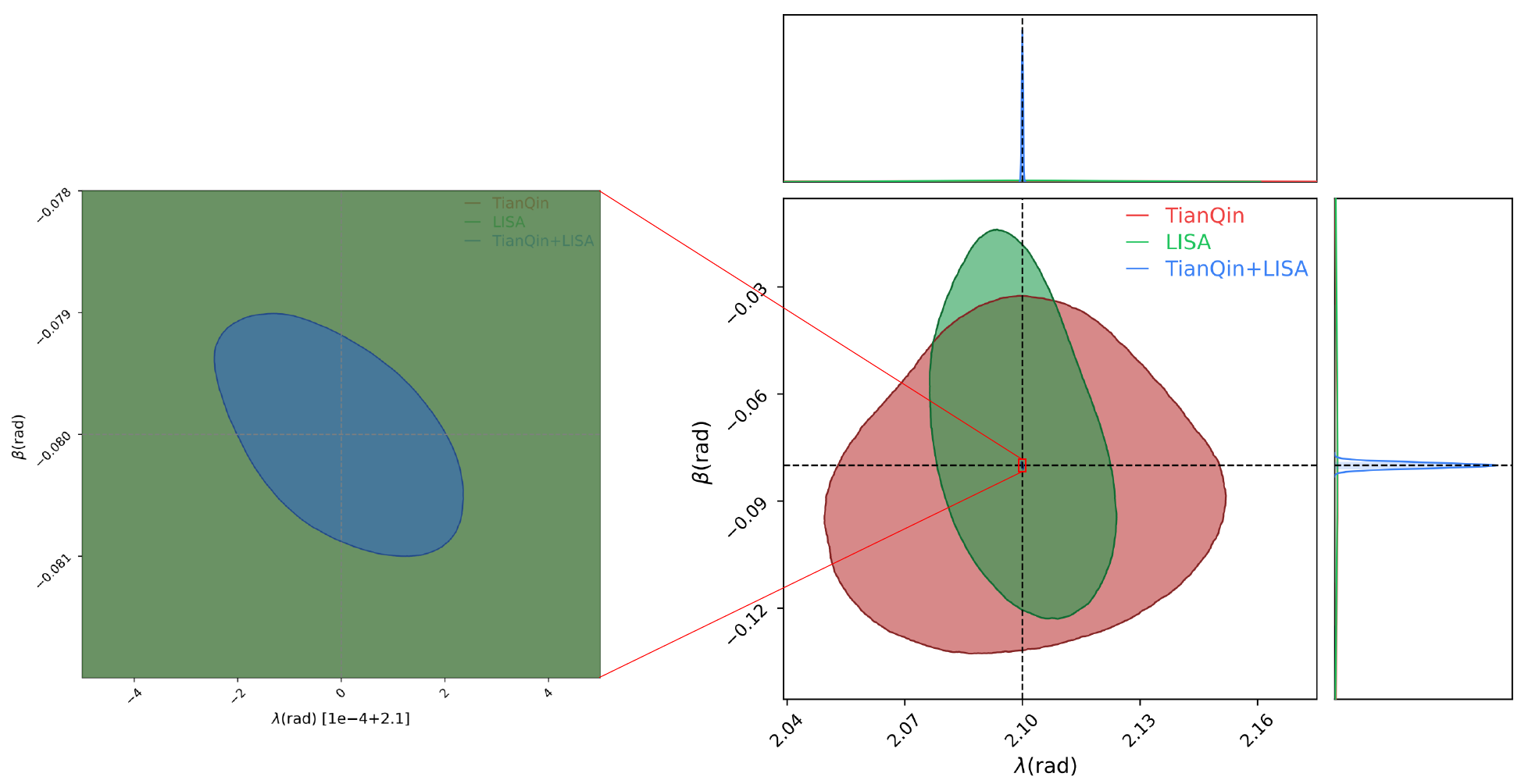}
    \caption{Constrain results of the ecliptic longitude $\lambda$ and ecliptic latitude $\beta$ with TianQin, LISA, and the joint detection of the two detectors.
    The left panel is the zoom-in of the contour plot, showing the order of magnitude improvement of the sky localization precision~\cite{Gao:2024uqc}.}
    \label{fig:PE_lambda_beta}
\end{figure}

When combined in a network, these two detectors can break some parameter degeneracies that are not possible with a single detector alone, leading to a substantial improvement in the precision of measurements~\cite{Gao:2024uqc, Shuman:2021ruh, Zhang:2021wwd, Zhang:2021kkh, Yang:2022cgm, Liu:2023hqe}.
For instance, for an ideally localized source, the constraining precisions of the source's ecliptic longitude $\lambda$ and latitude $\beta$ can be improved by orders of magnitude when a network of detectors is formed, as shown in Figure~\ref{fig:PE_lambda_beta}~\cite{Gao:2024uqc}.

Furthermore, integrating data from detectors like TianQin and LISA demonstrates the strength of collaborative efforts in GW astronomy. 
It highlights the potential for groundbreaking discoveries that arise from the synergistic combination of diverse observational methods and technologies. 
This joint detection strategy is a significant advancement, setting the stage for more accurate and detailed explorations of the universe.

\begin{tcolorbox}[colback=red!5!white,colframe=red!75!black]
\textbf{\textcolor{NavyBlue}{
\begin{it}
\textcolor{blue}{
Detecting GWs from MBHs and their predecessors, IMBHs, will provide critical insights essential for understanding how BHs gained their titanic masses. Expected to begin operations in the mid-2030s, TianQin will enhance our understanding of phenomena such as MBH mergers at the universe's edge. 
TianQin's sensitivity in the millihertz frequency range makes it particularly well-suited to detect sources of lower mass, while LISA, operating in a complementary frequency range, excels in detecting higher mass sources. 
This means that TianQin is well-equipped to study the birth and growth of MBHs, especially those at the earlier stages of their formation.  
Reaching out to the edge of our observable universe, TianQin can study the mergers of MBHs in the centers of galaxies, providing data about these extreme environments.  
The geocentric orbit of TianQin allows the implementation of real-time data transmission, thus alerts can be generated on the fly and issued to partner observatories before the merger, greatly enhancing the possibility of multi-messenger examination of MBH binary mergers.
Additionally, TianQin’s ability to operate in conjunction with LISA offers significant scientific advantages. Joint observations can enhance our ability to distinguish between different models of BH formation and evolution, by covering a broader range of masses and redshifts, and by orders of magnitude improvement in sky localization.}
\end{it}
}}
\end{tcolorbox}

\clearpage
\section{Sources around MBHs}\label{sec:aroundmbh}
\coordinators{Alejandro Torres-Orjuela \& Xian Chen}

\begin{figure}[htpb]
    \centering
    \includegraphics[width=0.8\linewidth,trim=45 300 45 300,clip]{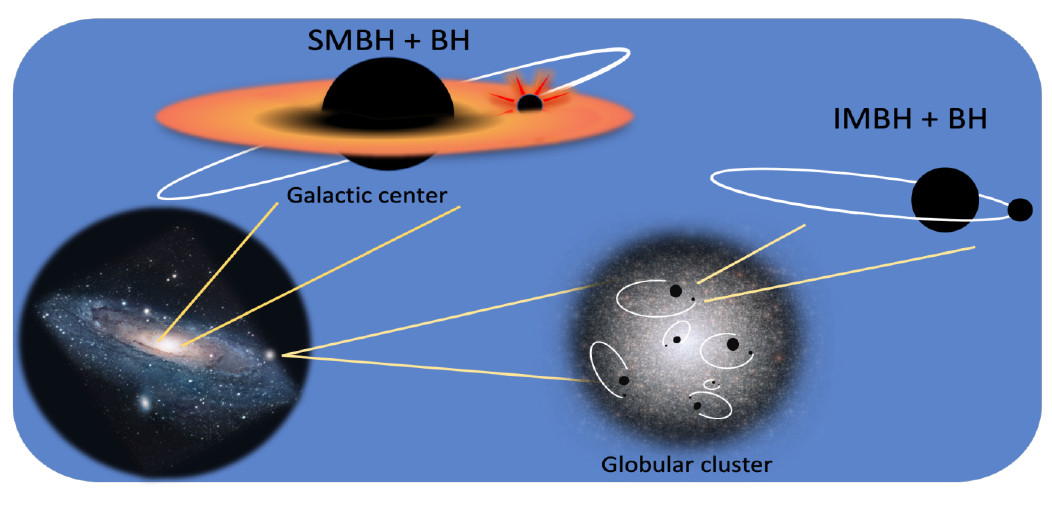}
    \caption{Typical environments for the formation of GW EMRIs and IMRIs.}    
    \label{fig:EMRI_IMRI}
\end{figure}

Unequal-mass BH mergers constitute another important type of GW source in the milli-Hz band. Such sources are normally divided into two sub-types. One is called EMRI~\cite{AmaroSeoane2022, amaro-seoane_2018b, AmaroSeoaneEtAl2015, AmaroSeoaneEtAl2007, fan_hu_2020}; normally composed of a stellar-mass BH orbiting a SMBH, with a mass ratio of about $q\sim(10^{-6}-10^{-5})$. Another type is called IMRI. The name comes from the fact that the mass ratio of the two BHs is in an intermediate range, around $10^{-4}-10^{-3}$. Given such a mass ratio, the possible configuration of an IMRI could be an IMBH orbiting a SMBH (also known as ``heavy IMRI''), or a stellar-mass BH orbiting an IMBH (``light IMRI''). Both EMRIs and IMRIs contain MBHs and such massive objects are expected to be found in the centers of either galaxies or globular star clusters. Therefore, the formation and evolution of EMRIs and IMRIs are closely related to dense stellar environments as well as gaseous accretion disks which often form around MBHs. Figure~\ref{fig:EMRI_IMRI} illustrates the typical environments of EMRIs and IMRIs.

This section is dedicated to the detectability of EMRIs and IMRIs by TianQin and the uniqueness of TianQin in elucidating some of the key astrophysical and physical questions related to these sources. Since TianQin has better sensitivity than LISA-type detectors at higher frequencies, it is more sensitive to light IMRIs than heavy IMRIs or EMRIs. Therefore, in Subsection~\ref{subsec:imri} we first review the formation and merger channels of light IMRIs, as well as the characters of the light IMRIs detectable by TianQin. Subsection~~\ref{subsec:enviro} focuses on the close interplay between EMRIs and IMRIs with their stellar and gaseous environment and the possible observational signatures in the TianQin frequency band. We also discuss the consequent impact on various tests of gravity theories given the important role of EMRIs and IMRIs in testing fundamental physics. The interplay of EMRIs and IMRIs with their astrophysical environments is expected to produce a rich spectrum of electromagnetic counterparts. Such a possibility opens up new prospects for multi-messenger observations, which is the subject of Subsection~\ref{subsec:multi}. Finally, in Subsection~\ref{subsec:unique} we highlight the science payback that can be offered by conducting joint observation using both TianQin and LISA. In this light, we also identify the parameter space of EMRIs and IMRIs where TianQin will play a unique role in their detection and parameter estimation.

\subsection{Light IMRIs}\label{subsec:imri}

\contributors{Verónica Vázquez-Aceves, Pau Amaro Seoane, Abbas Askar, Alvin Chua, Mirek Giersz, Arkadiusz Hypki, Miaoxin Liu, Alejandro Torres-Orjuela}

Light IMRIs, with mass ratios between $10$ to $10^4$, are binary systems formed by an IMBH and a stellar-mass BH. The orbit of these systems evolves only due to the energy loss by gravitational radiation, making them excellent GW sources for future space-borne GW detectors. In 2020, the LIGO and Virgo Collaboration announced the detection of a BH merger which resulted in a $142\,{\rm M_\odot}$ BH~\cite{abbott_et_al_2020} confirming the existence of IMBHs on the light mass-end. However, the detection of IMRIs will give irrefutable evidence of the existence of IMBHs with a wide range of masses, improving our understanding of the growth processes of BHs, the characteristics, and dynamics of the embedding system, details about space-time, and BH physics, among others.

In order to form light IMRIs, it is necessary to first form IMBHs ($10^{2}-10^{5}\,{\rm M_\odot}$). BHs in this mass range can form through different pathways, and may have an important role in seeding SMBHs (see~\cite{2020ARA&A..58..257G,2023arXiv231112118A} and references therein, see Section~\ref{subsec:MBHseed} for more information about MBH seeding). Dynamically active environments like globular and nuclear star clusters have been suggested as potential sites for forming and growing IMBHs (see Section~\ref{subsubsec:cluster}). The stellar density in the cores of these dense and massive star clusters can exceed $\gtrsim 10^{6}-10^{7}\,{\rm M_\odot\,pc^{-3}}$ and gravitational encounters between stars can frequently occur. These interactions drive the dynamical evolution of the star clusters and can also lead to the formation of an IMBH. For instance, runaway mergers between massive stars during the early evolution of a dense star cluster can lead to the formation of a VMS. In low-metallicity environments, such a VMS could potentially evolve into an IMBH~\cite{SPZEtAl1999,portegies-zwart_mcmillan_2002,portegies-zwart_baumgardt_2004,freitag_rasio_2006,freitag_gurkan_2006}. Another formation pathway for IMBHs is the repeated or hierarchical merger of BHs in binary systems through GW emission. If the merged BH is retained within the dense dynamical environment~\cite{miller_hamilton_2002,Konstantinidis_2013,Morawski2018,fragione_ginsburg_2018}, it can pair up with other BHs and merge again, resulting in the formation and growth of an IMBH. 

Observations of ultraluminous X-ray sources, not associated with AGN, and dynamics in globular clusters suggest the existence of IMBHs~\cite{Tremaineetal2002,Holley_Bockelmann_2008,lutzgendorf_kissler-patig_2013,Lutzgendorf_2014,Reines_2015,Tremou_2018,2020_dwarfgal}, for example, \citet{Gebhardt2002} used data from the  Hubble space telescope to suggest the existence of a $\sim2\times10^4\,{\rm M_\odot}$ IMBH in the stellar cluster G1, the most massive stellar cluster in M31. This result was controversial, but the detection of weak X-ray and radio emission from the cluster center gave some more arguments in favor of the IMBH scenario. Also, observations from Hubble space telescope and the Very Large Telescope suggest the existence of an IMBH with a mass $\approx1.7\times10^4\,{\rm M_\odot}$ in the globular cluster NGC 6388~\cite{Ltzgendorf2011} and of an IMBH of $\sim4.7\times 10^4$M$_{\odot}$ at the center of NGC 5139 ($\omega$ Centauri)~\cite{noyola_gebhardt_2010}.

Understanding the formation channels and orbital properties of IMRIs is essential to develop tools to identify and extract information from their gravitational radiation. The formation of IMRIs in dense systems has been predicted in several studies~\cite{amaro-seoane_freitag_2006,Konstantinidis_2013,arca-sedda_amaro-seoane_2021,arca-sedda_capuzzo-dolcetta_2019} and new formation channels have been recently considered such as the formation of IMRIs in merging galaxies~\cite{VVA_2023} and in AGN disks~\cite{peng_chen_2023}.  The formation mechanisms also give us information about the orbital parameters of these sources when entering the TianQin detection band. Several studies have been made for LISA and a joint detection from TianQin and LISA~\cite{amaro-seoane_2018, arca-sedda_amaro-seoane_2021, torres-orjuela_huang_2023}; however, TianQin, being more sensitive than LISA above $\sim$ 0.01 Hz, can give more details about the evolution of IMRIs.

\subsubsection{Simulations and light IMRI data}

\begin{figure}[htpb] \centering 
\includegraphics[width=0.45\textwidth]{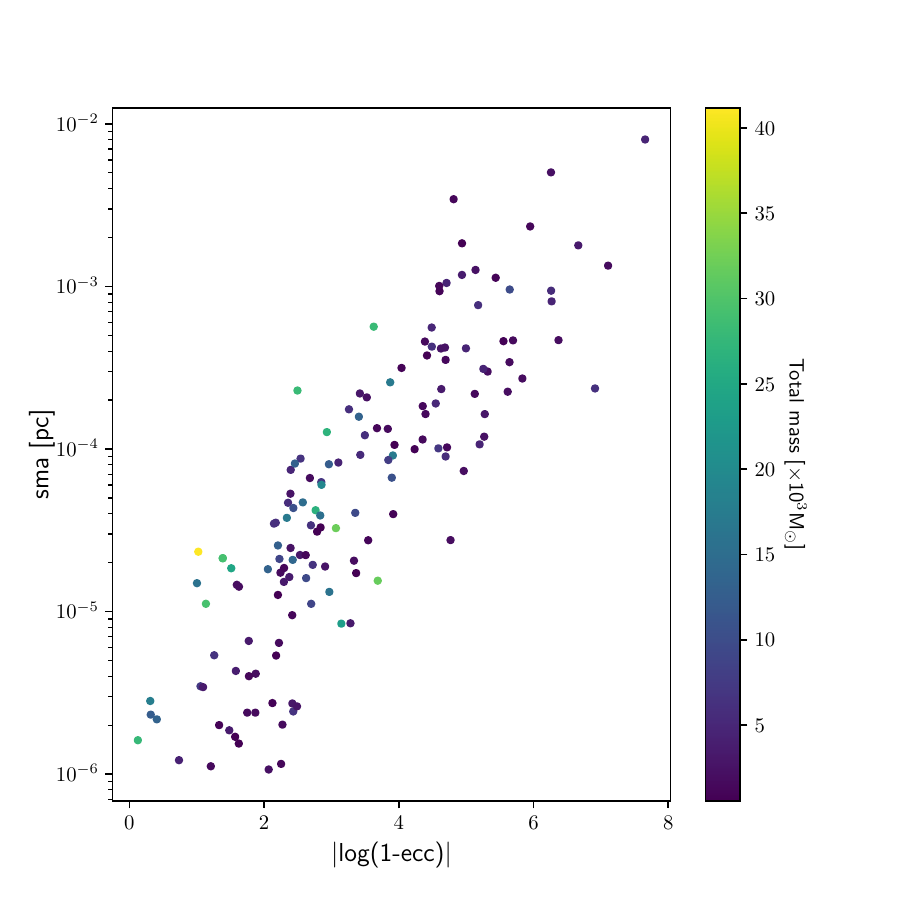}
\hspace{\fill}
\includegraphics[width=0.45\textwidth]{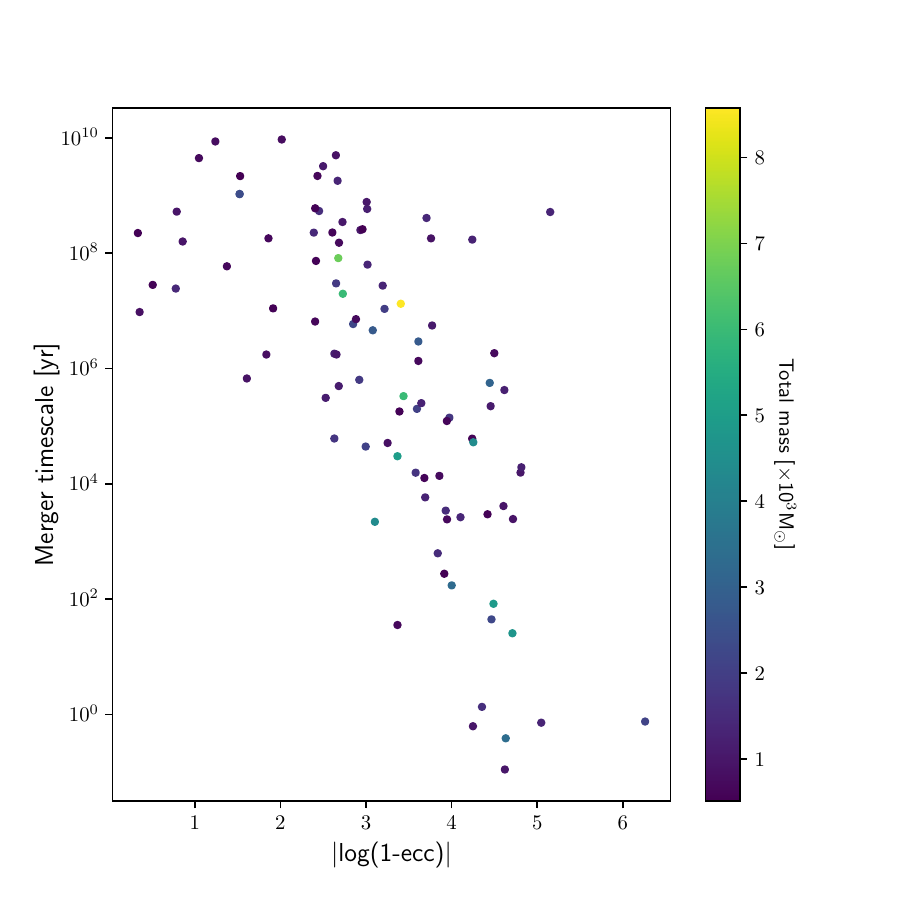}
\caption{Left: Distribution of the semimajor axis (sma), eccentricity (ecc), and total mass of light IMRIs formed dynamically in stellar dynamical simulations~\cite{askar_szkudlarek_2017}, 87$\%$ of these binaries have eccentricities larger than 0.9 before the merger. Right: Distribution of the merger timescale and eccentricity of light IMRIs formed by binary-single and binary-binary encounters, 90$\%$ of the binaries formed by this process have eccentricities larger than 0.9.}
\label{fig:imris_params}
\end{figure}

GWs emitted by light IMRIs are well within the $\rm mHz$ band, shifting to higher frequencies during their evolution, which makes them ideal sources for joint detections involving space-borne GW detectors and ground-based detectors. It is well established that the inspiral phase of light IMRIs can be detected by TianQin and LISA to large distances of the order $10^2\,{\rm Mpc}$~\cite{amaro-seoane_2018}. Depending on the orbital parameters, the merger, and ringdown will likely occur within the detection band of the ground-based detectors; thus, TianQin and LISA can generate alerts to facilitate the detection of these mergers. Nevertheless, a comprehensive analysis of their GW signal is needed to determine the range of mass ratios, distances, and orbital parameters that TianQin can detect at different stages of the IMRI's evolution.


Figure~\ref{fig:imris_params} shows the total mass, orbital parameters, and merger timescales of IMRIs formed inside globular clusters, obtained with stellar dynamical simulations~\cite{Hypki_2012I, giersz_heggie_2013,askar_szkudlarek_2017}. In these simulations, IMRIs are formed dynamically and also through binary-single and binary-binary encounters in which the final mass and orbital parameters of these systems are acquired by repeated mergers through which an IMBH can even double 
its mass, the largest IMBH formed through this mechanism is of about $8\times10^3\,{\rm M_\odot}$. Depending on the distance, mass ratio, and orbital parameters, the higher modes of these eccentric sources might be detectable up to the merger phase; however, the details of their merger detectability are poorly constrained as currently, due to their mass ratio and high eccentricity, there is no accurate way of simulating the evolution at this stage. 

\subsubsection{Light IMRIs detectability}

\begin{figure}[tpb] \centering \includegraphics[width=0.48\textwidth]{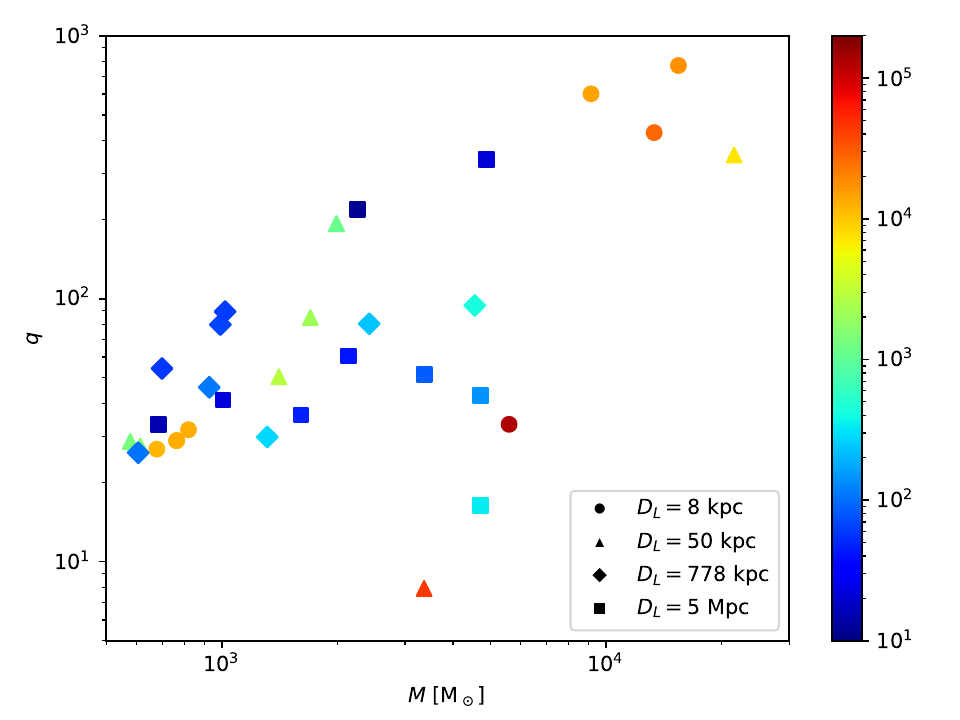}
 \caption{SNR of circular face-on IMRIs a few minutes away from merger as a function of their total mass $M$. The color represents the SNR and the shapes represent the different distances.}
 \label{fig:SNR_IMRIs0}
 \end{figure}

Using the waveform introduced in~\cite{islam_field_2022}, we compute the SNR of some representative circular sources, located within $5\,{\rm Mpc}$, short before the merger. Figure~\ref{fig:SNR_IMRIs0} shows the SNR of 34 face-on sources averaged over 100 evenly distributed sky localization located at 4 different distances representing sources located in dense systems like globular clusters or nuclear star clusters within the Milky Way ($8\,{\rm kpc}$), the Magellanic Clouds ($50\,{\rm kpc}$), the Andromeda Galaxy ($778\,{\rm kpc}$), and up to $5\,{\rm Mpc}$. The detection time corresponds to around $86\,{\rm s}$ for the lightest systems ($577\,{\rm M_\odot}$) and to around $54\,{\rm min}$ for the heaviest system ($21,500\,{\rm M_\odot}$). As expected, the heavier and nearest systems have the highest SNR, but even for systems with a total mass of $\sim10^3\,{\rm M_\odot}$ located at $5\,{\rm Mpc}$, the SNR is high enough to be detected by TianQin. Additionally, we find that for systems with total masses $\gtrsim10^4\,{\rm M_\odot}$ and distances $\gtrsim10\,{\rm kpc}$, the higher order modes emitted close to the merger fall within the $10^{-2}-10^{-1}\,{\rm Hz}$ range and can potentially be detected by TianQin.

It is important to note that the waveform model used here only considers circular orbits and contains modes up to an order of $\ell=10$. The close-to-merger systems in Figure~\ref{fig:imris_params} might emit a significant amount of their energy at higher-order modes during the merger due to residual eccentricity and their mass ratio, we take only up to $\ell=5$ as these waveforms agree particularly well with numerical relativity (NR) for the comparable mass ratio range. The resulting SNR is an initial estimate that must be improved as the waveform model still needs testing for the IMRIs eccentricity and mass ratio.

\subsubsection{IMRI waveforms}

IMRIs present significant challenges for gravitational waveform modeling, primarily due to the disparity in mass ratios which can range significantly, impacting the dynamical evolution of these systems. The fast and accurate simulation of these events is crucial for the detection and parameter estimation of TianQin. Therefore, we discuss briefly the existing methods, their limits, and how they could be applied to the IMRI case. 

NR simulations offer the most accurate descriptions of binary mergers by numerically solving Einstein's equations~\cite{AlcubierreBook2008, BonaPalenzuelaBona2009,
baumgarteShapiroBook, GourgoulhonBook2012, ShibataBook2015}; however, it cannot be applied to the IMRI case as it is restricted to mass ratios of $\lesssim 0.1$. Gravitational self-force approaches, which consider the perturbative effect of a smaller object's self-force on its motion through the spacetime of a larger BH, are especially relevant for IMRIs but require significant advancements to fully capture the complex dynamics involved in more generic orbits~\cite{Burke:2019yek, Gourgoulhon:2019iyu, Chua:2020stf, Hughes:2021exa, Fujita:2020zxe, Munna:2020iju, Isoyama:2021jjd, Lynch:2023gpu, Skoupy:2023lih, Compere:2021zfj, Lynch:2021ogr}.

The effective-one-body model can be useful for the parameter space relevant to IMRIs as it combines insights from NR and gravitational self-force, offering a semi-analytical approach that maps the two-body problem into an effective problem of a single body moving in a deformed metric~\cite{Huerta:2012zy, Antonelli:2019fmq, Barausse:2011dq, vandeMeent:2023ols, Albertini:2023aol, Yunes:2009ef, Yunes:2010zj, xin_han_2019, Zhang:2021fgy, Albanesi:2021rby, Albertini:2022dmc, Albertini:2022rfe,Ramos-Buades:2023ehm}. Moreover, surrogate models represent an alternative way of waveform modeling. These models are trained on accurate NR data and can be extended to IMRI regimes that are computationally inaccessible for direct NR simulations offering rapid waveform generation and high accuracy. Nevertheless, current surrogate models like \texttt{BHPTNRSur1dq1e4} from the `Black Hole Perturbation Toolkit', which is the waveform used in this section~\cite{islam_field_2022,field_galley_2014,BHPToolkit}, can only simulate IMRIs on quasi-circular orbits with non-spinning BHs which is very different from what we expect from realistic sources. Moreover, \texttt{BHPTNRSur1dq1e4} can only describe the late inspiral and merger face, but the early inspiral is expected to significantly contribute to the detection of light IMRIs by TianQin.

\vspace{1cm}

\begin{tcolorbox}[colback=red!5!white,colframe=red!75!black]
\textbf{\textcolor{NavyBlue}{
\begin{it}
\textcolor{blue}{
Light IMRIs, consisting of a stellar-mass BH inspiraling into an IMBH, are stand-out sources of GWs for TianQin. IMBHs can form through several pathways, particularly in dense stellar environments such as globular and nuclear star clusters. These dense environments facilitate frequent gravitational encounters, driving the dynamical evolution of the clusters and leading to the formation of IMBHs. The formation and evolution of IMRIs in dense stellar systems have been predicted through simulations, showing that IMRIs can form dynamically in environments like globular clusters, often through binary-single and binary-binary encounters, resulting in systems with high eccentricities. TianQin's sensitivity to higher frequencies allows it to detect these highly eccentric binaries, providing crucial information about their formation and evolution. TianQin will be able to detect light IMRIs out to distances of $5\,{\rm Mpc}$ even in the case of relatively light systems with total masses around $10^3\,{\rm M_\odot}$.}
\end{it}
}}
\end{tcolorbox}

\subsection{Environmental effects}\label{subsec:enviro}

\contributors{Peng Peng, Lorenz Zwick, Cosimo Bambi, Xian Chen, Alvin Chua, Debtroy Das, Mudit Garg, Yunqi Liu, Nicola R. Napolitano, Swarnim Shashank, Alejandro Torres-Orjuela, Shu-Xu Yi, Jian-dong Zhang}

The term environmental effect refers to the myriad of astrophysical processes that can influence a binary composed of compact objects, altering its evolution and perturbing its GW emission.  The field has gained considerable attraction within the last decade and now provides some very interesting connective issues between the astrophysical and fundamental science objectives of future space-borne GW observatories. The study of environmental effects is primarily motivated by the realization that a majority of sources targeted by space-borne detectors are likely to originate deep within the centers of galaxies, in particular, in the gaseous accretion disks or dense nuclear star clusters surrounding the SMBHs. These extremely dense, complex environments imprint characteristic signatures in the GW emission of the binaries they envelop.

Environmental effects may have profound implications on the science objectives of future missions. On one hand, it is crucial to remember that the analysis of GW signals requires pre-generated waveform templates which are used to filter the raw data outputs of detectors. However, current waveform generation models are exclusively based on vacuum solutions of General Relativity. If not accounted for, environmental influences can confound results based on vacuum expectations, limiting the effectiveness of parameter estimation and introducing spurious biases to important parameters such as the mass of the system and the spin of the BHs. The risk here is to severely hinder the purported scientific goals of placing extremely stringent constraints on the strong field regime of gravity. On the other hand, environmental signatures in GWs can be used to probe the astrophysical surroundings of the sources. Common types of environmental effects include the shift of the phase of GWs due to interactions with stellar and gaseous backgrounds, and if detected could be used to place unprecedented constraints on the surroundings of GW sources.


\subsubsection{Peculiar velocity and Brownian motion}\label{subsubsec:motion}

IMRIs are expected to form in dense stellar systems such as nuclear star clusters in the center of galaxies or globular clusters~\cite{arca-sedda_amaro-seoane_2021}. Therefore, we can expect IMRIs to be moving, either due to the motion of the host system or due to its interaction with other components of the system. Depending on the time scale of the velocities induced, IMRIs can be considered as moving with a constant or an accelerated velocity which induces different effects.

\paragraph{Peculiar velocity:} The motion of the host system -- often denoted as `peculiar velocity' -- can be considered constant for TianQin's observation time of five years~\cite{tianqin_2016}. For a globular cluster moving inside a galaxy like the Milky Way, the peculiar velocity ranges between $200\,{\rm km\,s^{-1}}$ and $250\,{\rm km\,s^{-1}}$~\cite{ou_eilers_2024} while the peculiar velocity of the host galaxy ranges from a few $100\,{\rm km\,s^{-1}}$ up to almost $2000\,{\rm km\,s^{-1}}$~\cite{girardi_fadda_1996,ruel_bazin_2014}. When only detecting the quadrupolar mode of a moving GW source, the effect of a constant velocity $v$ reduces to a change in its observed mass, distance, and sky localization proportional to $v/c$, where $c$ is the speed of light~\cite{yan_chen_2023,torres-orjuela_chen_2019,torres-orjuela_chen_2023}. However, if we can detect modes of the GW beyond the quadrupolar or $(2,2)$ mode~\cite{torres-orjuela_2024b} the detection of a velocity becomes detectable by a change of the spherical modes proportional to $v/c$~\cite{torres-orjuela_chen_2021,torres-orjuela_amaro-seoane_2021,boyle_2016,woodford_boyle_2019}.

\begin{figure}[tpb] \centering \includegraphics[width=0.48\textwidth]{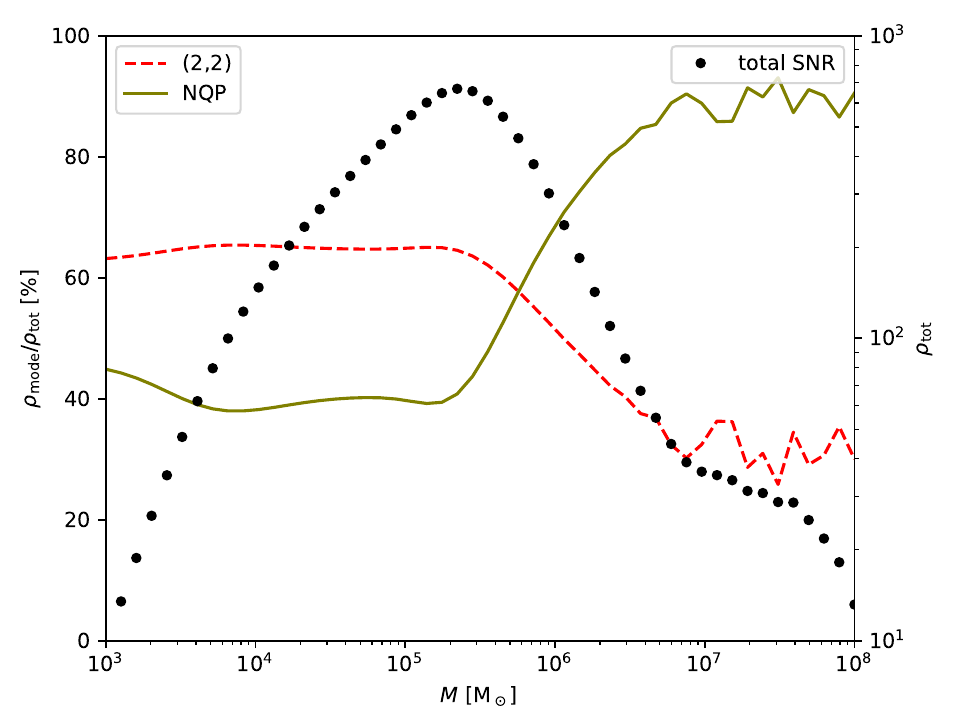}\caption{
Detectability of different GW modes emitted from an IMRI. The plot shows        
the relative contribution of $(2,2)$-mode and the sum of the non-quadrupolar (`NQP') modes of an IMRI to the SNR for a source detected by TianQin as a function of the system's total mass $M$. The right coordinate shows the total SNR of the source. Note that the total SNR as well as the sum of the non-quadrupolar modes also contain the contribution from modes with negative $m$.
    }
\label{fig:imri_modes}
\end{figure}

Figure~\ref{fig:imri_modes} shows the contribution of the $(2,2)$-mode and the sum of the contribution of all non-quadrupolar modes, marked as `NQP', to the total SNR of an IMRI detected by TianQin as a function of the source's total mass $M$ assuming a mass ratio $q=100$ and a luminosity distance $D_L=3\,{\rm Mpc}$; the waveform is generated using \texttt{BHPTNRSur1dq1e4}~\cite{islam_field_2022}. The prominence of the non-quadrupolar modes is apparent. The measurability of the velocity of a GW source lies in the measurability of the change of non-quadrupolar modes due to peculiar velocity. Because the change of the non-quadrupolar modes is proportional to the source's velocity, we get that the mismatch between the signal for a moving source is $\sim(\rho_{\rm NQP}/\rho)^2(v/c)$. Therefore, for $M\lesssim2\times10^5\,{\rm M_\odot}$ the velocity of a globular cluster will introduce a mismatch of around $10^{-4}$ while for heavy systems with $M\gtrsim3\times10^6\,{\rm M_\odot}$ and the fastest moving host galaxies the mismatch can go up to around $5\times10^{-3}$ which are at the same order of the mismatch introduced by a difference in the phase of around $0.01\,{\rm rad}$ and $0.1\,{\rm rad}$, respectively. Considering the high SNRs of several tens up to almost 1000 we expect for IMRIs (cf. Figure~\ref{fig:imri_modes}), such a small mismatch is probably detectable.

\paragraph{Brownian motion:} It has been discussed for a long time that IMBHs in star clusters may undergo a `Brownian motion' or `random walk' due to repeated interactions with the surrounding stars~\cite{lin_tremaine_1980,gurzadian_1982,milosavljevic_merritt_2003}. It is speculated that the Brownian motion plays an important role in the dynamics and
evolution of the star cluster but results remain inconclusive due to a lack of
direct observation of IMBHs. However, the acceleration of an IMRI due to Brownian motion
leads to a phase shift in the GWs emitted and hence it is a potential observable.

One can use a simple Plummer model~\cite{binney_tremaine_2008} for the star cluster to estimate the frequency of star-IMRI encounters and the magnitude of the acceleration~\cite{torres-orjuela_vazquez-aceves_2025}. For a big globular cluster with a mass $M=1.23\times10^6\,{\rm M_\odot}$, a half-mass radius $R_h=0.66\,{\rm pc}$, and $N=2.04\times10^6$ stars, the IMRI has a close encounter with a star at the order of less than a years to some hundred years while for a small cluster ($M=3.18\times10^4\,{\rm M_\odot}$, $R_h=3.81\,{\rm pc}$, and $N=8.14\times10^4$), a close encounter happens at the order of tens to hundred thousands of years. Considering TianQin's lifetime of $5\,{\rm yr}$~\cite{tianqin_2016}, we need to detect $\mathcal{O}(10^{-2} \text{--} 10^4)$ IMRIs per year to observe such a close encounter; strongly depending on the size of the host cluster. \citet{arca-sedda_amaro-seoane_2021} estimate that we will detect between 0.2 and 60 IMRIs per year in the LISA band while 6 to 3000 IMRIs could be detected in the DECIGO band every year~\cite{decigo_2021}. There is no such estimation for TianQin's detection band but considering that TianQin is more sensitive at higher frequencies than LISA, we can assume that the detection of some tens to hundreds of IMRIs per year by TianQin is realistic.

The average acceleration due to Brownian motion can be estimated assuming that the closest possible encounter does not significantly alter the intrinsic parameters of the IMRI~\cite{torres-orjuela_vazquez-aceves_2025}. The magnitude of the acceleration $a$ depends on the size of the IMRI, the properties of the cluster, and the position of the IMRI in the cluster. We find that $a$ is of the order $10^{-3}\text{--}1\,{\rm m\,s^{-2}}$ and of the order $10^{-5}\text{--}10^{-2}\,{\rm m\,s^{-2}}$ for a big cluster and a small cluster, respectively, depending on the position of the IMRI in the cluster. The acceleration induces either a time-dependent Doppler shift or an aberrational phase shift depending on the orientation of the acceleration relative to the line of sight~\cite{torres-orjuela_chen_2020}. In both cases, a phase shift $\delta\Phi$ is induced which is proportional to $a/c$ times the total duration of the encounters observed $t_{\rm tot}$. Such a phase shift can be detected if the SNR of the source fulfills $\rho\sim c/(at)$~\cite{lindblom_owen_2008,torres-orjuela_chen_2020}. Depending on the properties of the cluster and the position of the IMRI in the cluster, we can expect to observe either one encounter with low acceleration that lasts for the entire observation time or multiple short encounters with high accelerations. Using the lifetime of TianQin ($5\,{\rm yr}$)~\cite{tianqin_2016} as the maximal observation time, we can estimate that a SNR of around 1900 is required to detect one long encounter with a low acceleration of around $10^{-3}\,{\rm m\,s^{-2}}$ while a SNR of 1350 is necessary to detect multiple short encounters with accelerations of $10^{-1}\,{\rm m\,s^{-2}}$. For IMRIs, SNRs of several hundred thousand are possible (cf. Figure~\ref{fig:SNR_IMRIs0}); therefore, we conclude that the acceleration of an IMRI induced by its Brownian motion is detectable by TianQin.

\subsubsection{Dephasing in Gas}\label{subsubsec:imriagn}

IMRIs can naturally form in the AGN disks. 
On the one hand, IMBHs can be brought to the center of an AGN by either inspiraling globular clusters or galaxy mergers and then get captured by AGN accretion disks~\cite{Volonteri03, Mastrobuono14}. Moreover, IMBHs could also grow from the stellar-mass BHs embedded in the disk. These stellar-mass BHs can either be captured by the gas in the disk from the nuclear star cluster or evolve from the massive stars in the outskirt of the AGN disk~\cite{Syer91, Goodman2004}. The stellar-mass BHs in the disk can then grow by so-called ‘hierarchical mergers’~\cite{Yang2019_spin, Gerosa:2021mno} or through gas accretion. The IMBHs can grow up to $100-1000\,{\rm M_\odot}$ during these processes~\cite{McKernan2012, Bartos2017, Stone17}. On the other hand, the IMBHs in AGNs can further produce IMRIs. An IMBH can migrate inward due to its interaction with the surrounding gas~\cite{Gould00, Armitage02} and finally spiral into the central SMBH to form a heavy IMRI. Moreover, the hierarchical mergers will continue after the formation of an IMBH, so that the subsequent mergers between the IMBH and stellar BHs will produce light IMRIs. 
In particular, if an IMBH forms around a so-called migration trap in AGN accretion disk~\cite{Bellovary16, Secunda2019, Secunda2020}, it can capture other trapped stellar-mass BHs which enhances the light IMRI event rate~\cite{McKernan20}. 

\begin{figure}
    \centering
    \includegraphics[width=0.5\textwidth]{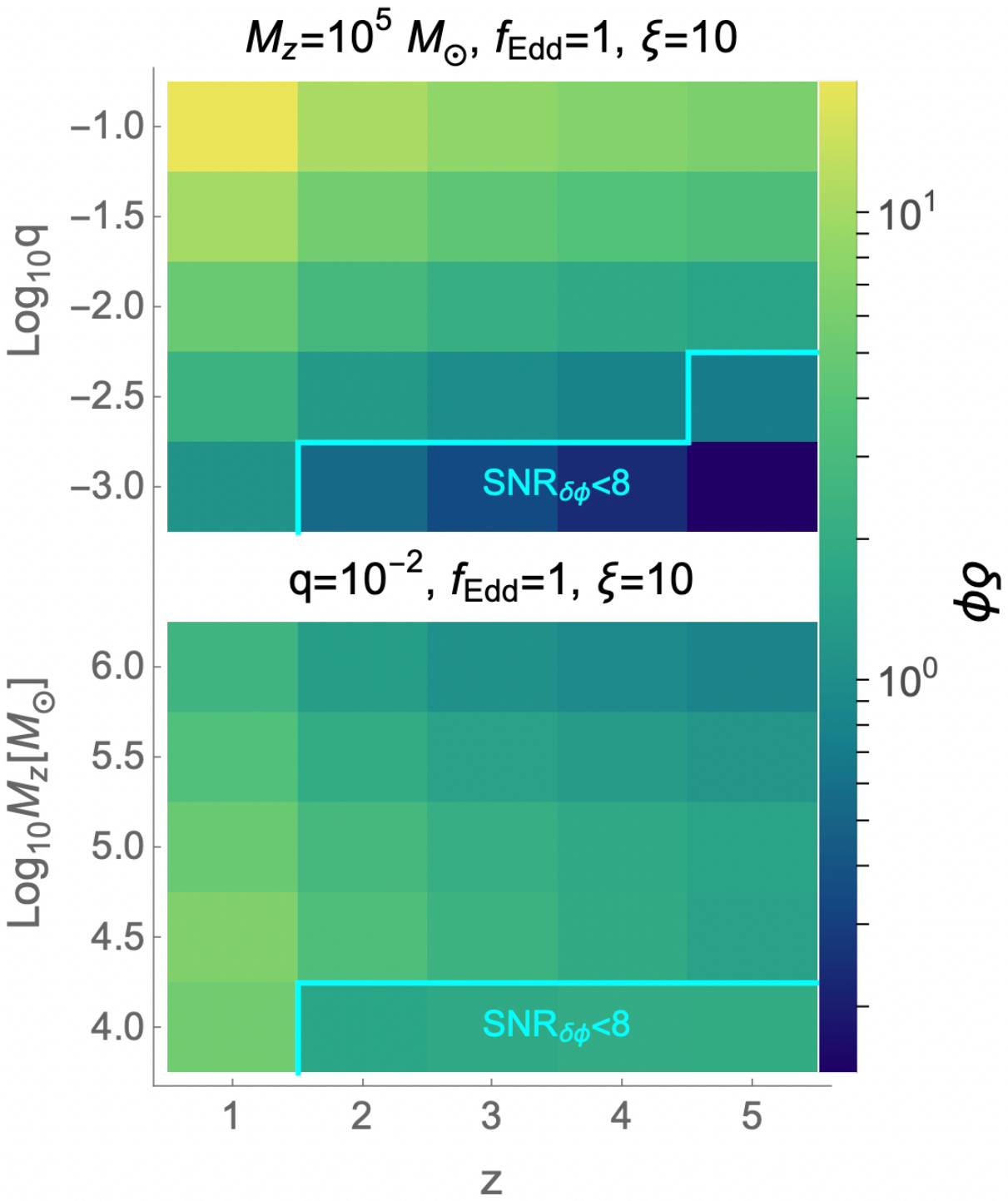}
    \caption{Dephasing of an IMRI embedded in a gaseous environment. In both panels, we vary the redshift $z$ from $1$ to $5$ and keep the Eddington ratio ${\rm f}_{\rm Edd}=1$ and strength of the torque $\xi=10$ constant. For the top panel, we fix the total mass $M_z=10^5~{\rm M}_\odot$ and increase the mass ratio $q$ from $10^{-3}$ to $10^{-1}$. In the bottom panel, we keep $q=10^{-2}$ and vary $M_z$ between $10^4~{\rm M}_\odot$ and $10^6~{\rm M}_\odot$. We add a cyan to separate regions of non-observable phase shifts (SNR$_{\delta\phi}<8$).}
    \label{fig:IMRIDephasing}
\end{figure}

If IMRIs are evolving in a thin accretion disk~\cite{ShakuraSunyaev1973}, then for a suitable parameter space, gas can leave measurable imprints on the GWs~\cite{Yunes2011,Kocsis2011,Barausse2014,Derdzinski2019,Derdzinski2021,Zwick2022,Garg2022,Garg2024a,Dittmann2023,Garg2024b,Garg2024c}. Depending on the details of the accretion flow and binary-disk parameters, gas can either slow down or quickens the binary inspiral via phase shifts, excite measurable eccentricities, or do both. Figure~\ref{fig:IMRIDephasing} shows the dephasing of such an IMRI embedded in the accretion disk of an AGN, where, for illustrative purposes,  we have considered only the gas-induced orbital evolution and neglected the growth of the BHs by gas accretion. The evolution is driven mainly by the so-called type-II migration~\cite{Garg2022}, and the orbital decay rate can be estimated with
\begin{align}
    \dot r_{\rm gas}=\frac{\xi {\rm f}_{\rm Edd}}{25~{\rm Myr}}\frac{(1+q)^2}{q}r
\end{align}
assuming typical radiative efficiency for the accretion disk, where $q$ denotes the mass ratio of the two BHs in the IMRI, $\xi$ is the strength of the gas torque which mainly depends upon the thermodynamics of the disk and the mass ratio, and as per current hydrodynamics simulations, should be between $-10$ to $10$~\cite{Garg2022}, ${\rm f}_{\rm Edd}$ is the Eddington ratio of the accretion disk, and $r$ is the binary separation in the detector-frame. With the reasonable assumption that five years before merger $\dot r_{\rm gas}$ is much smaller than the orbital decay rate due to GW radiation, the dephasing can be calculated from
\begin{equation}
    \delta\phi\approx2\pi\int^{r_f}_{r_i}{\rm d}r~f\frac{\dot r_{\rm gas}}{\dot r_{\rm GW}^2},
\end{equation}
where $f\equiv(1/\pi)\sqrt{GM_z/r^3}$ is the observed GW frequency. We find at maximum $\sim 20$ radians gas-induced phase shift, which corresponds to around $3$ GW cycles lost over $5$ years. 





To assess if $\delta\phi$ is observable by TianQin, we set the threshold of the SNR of the dephasing (${\rm SNR}_{\delta\phi}$) to be at least 8. We compute this as follows~\cite{Kocsis2011}:
\begin{equation}\label{Eq19}
    {\rm SNR}_{\delta\phi}=\sqrt{4\int^{{f}_{\text{f}}}_{{f}_{\text{i}}}{\rm d}{f}^\prime\frac{2h_{\rm{c}}^2({f}^\prime)(1-\cos{\delta\phi})}{S_{\rm{t}}({f}^\prime){f}^{\prime2}}},
\end{equation}
where $S_t(f)$ is the noise curve for TianQin~\cite{Mei2021}, and we have assumed that TianQin is switching on and off in an interval of three months during the five years of observation. The result suggests that dephasing is more easily detectable if the IMRI has lower mass ($\lesssim10^5M_\odot$), near-equal mass ratio ($q\sim10^{-1}$), and resides at relatively low redshifts ($z\sim1$).
Environmental dephasing in gas-rich disks is discussed in the context of evolution in AGN disk (Section~\ref{subsec:AGN_BBH}) and MBH accretion (Section~\ref{subsec:Coevolution}).


\subsubsection{Other environmental effects}

In addition to the astrophysical effects presented above, if an IMRI is close enough to the central SMBH, several other effects can distort the waveform of the IMRI signal. For example, the SMBH can induce some dynamical effects, such as the Zeipel-Lidov-Kozai effect~\cite{Liu18} and the evection resonance~\cite{Munoz22}, on the IMRIs and excite the orbital eccentricity. The strong gravity of the SMBH can also induce lensing and redshift effect on the signal~\cite{Kocsis13, Orazio20, Peng2021}. The orbital motion of the IMRI around the central SMBH can also induce dephasing in the waveform due to Doppler shift and aberrational phase shift~\cite{yunes_miller_2011,torres-orjuela_chen_2020}. These earlier studies have considered a LISA-type detector and showed that the imprints of various environmental effects on the waveform could be detectable. Given TianQin's better sensitivity in slightly higher frequencies compared to LISA, and the fact that IMRIs emit higher frequencies than EMRIs during their last few years of evolution, we generally expect a better chance of using TianQin to detect the environmental effects in IMRIs.

\subsubsection{Impact on Testing Gravity Theory}

GR is a pillar of modern physics and our standard framework for the description of gravitational interactions as well as for the spacetime structure. The past 10~years have seen tremendous progress in our ability to explore the strong field regime~\cite{2016PhRvL.116v1101A,2019PhRvL.123a1102A,2018PhRvL.120e1101C,2021ApJ...913...79T,2020PhRvL.125n1104P,2023CQGra..40p5007V} and space-borne GW observatories like TianQin will expand this ability by taking us to an era of precision tests of GR in the strong and dynamic regime~\cite{cardenas-avedano_sopuerta_2024}. Theories beyond GR may alter standard predictions by modifying the gravitational field of BHs, the emission of GWs, and the properties of the GWs themselves~\cite{yang_han_2019}. Space-borne GW observatories have usually superior constraining power to test gravity theories beyond GR with respect to present and future ground-based GW detectors. For example, from the detection of EMRIs, TianQin can constrain parameters in Einstein-dilaton-Gauss-Bonnet gravity and in non-commutative gravity around one order of magnitude more stringent than current constraints with LIGO-Virgo-KAGRA data~\cite{2024arXiv240112940H,tan_zhang_2024}.

In 4-dimensional GR, the spacetime geometry around uncharged BHs is described by the Kerr solution and is completely specified by two parameters: the mass $M$ and the spin angular momentum $J$ of the BH. Macroscopic deviations from the Kerr solution are possible if GR is not the correct theory of gravity~\cite{2009PhRvD..79h4043Y,2011PhRvL.106o1104K,jiang_dai_2022,destounis_angeloni_2023,liang_xu_2023,zhang_gong_2023,brito_shah_2023,zhang_guo_2023}, in the presence of macroscopic quantum gravity effects~\cite{2011arXiv1112.3359D,2017NatAs...1E..67G,tu_zhu_2023}, or in the presence of exotic matter fields.~\cite{2014PhRvL.112v1101H,2016CQGra..33o4001H,yue_cao_2019,hannuksela_ng_2020,becker_sagunski_2022,dai_gong_2022,zhang_fu_2024}. With TianQin, we will be able to test these hypotheses from the study of the GW signal of EMRIs and IMRIs. In the case we can accurately measure the mass, the spin angular momentum, and the mass quadrupole moment of a BH, we can perform a consistency test normally called the test of the no-hair theorem. From the detection of EMRIs, TianQin can measure the dimensionless quadrupole moment ($Q/M^3$) of the MBH with an accuracy of order $10^{-5}$~\cite{fan_hu_2020,zi_zhang_2021,Liu:2020ghq}. Moreover, TianQin can improve current constraints on the parameter of parametrized post-Einsteinian theories from ground-based detectors by orders of magnitude~\cite{2023PhRvD.108b4030S}. For example, TianQin can improve the constraints at the 2PN order by 3 orders of magnitude with IMRIs.

The potential of TianQin and other space-borne GW observatories for studying environmental effects and modified gravity (henceforth, ``beyond vacuum-GR'' effects) in sources involving MBHs is well documented in the GW literature. However, the overwhelming majority of these studies have been conducted in isolation relative to one another, i.e., they only examine the detection and inference prospects of a single beyond vacuum-GR effect at a time, and as a result tend to report incredibly precise constraints on such effects. In reality, the potential presence of multiple beyond vacuum-GR effects in a single GW signal would either require the simultaneous detection/inference of these effects or a reliable way to distinguish them using only their combined imprint on the signal~\cite{Kejriwal:2023djc}. Unfortunately, most beyond vacuum-GR effects are modeled as perturbative (small-amplitude) and
secular (long-timescale) corrections to the vacuum-GR signal that they modify, especially in the case of large-mass-ratio binaries such as IMRIs and EMRIs. This is sometimes due to various simplifying assumptions made in the models, e.g., averaging over orbits --- but often does truly reflect the nature of such effects. As a result, most beyond vacuum-GR effects modify the signal in ways that are far from independent, and the parameters that describe them are thus significantly correlated with one another (as well as the vacuum-GR parameters). 

These correlations are fundamentally problematic for the detection and inference of beyond vacuum-GR effects in IMRIs/EMRIs~\cite{Kejriwal:2023djc,shen_han_2023}. The issue of correlations among beyond vacuum-GR effects can be mitigated by using more detailed models that can make stronger predictions on the expected signals, or by focusing on effects with a more distinctive signature, e.g., transient resonances in IMRIs/EMRIs~\cite{PhysRevD.78.064028,Gupta:2022fbe}. Null tests that target the combined imprint of multiple putative effects on the signal might also be used, although these are easy to abuse~\cite{Chua:2020oxn}, and can lead to selection bias or cause practical problems in analysis/interpretation~\cite{Kejriwal:2023djc}. Ultimately, though, many of the isolated claims in the GW literature on the detectability of beyond vacuum-GR effects might turn out to be overly optimistic --- more joint analyses of important classes of effects will be needed to paint a fuller picture of which ones might more feasibly be detected with TianQin and other space-borne observatories.

\vspace{1cm}

\begin{tcolorbox}[colback=red!5!white,colframe=red!75!black]
\textbf{\textcolor{NavyBlue}{
\begin{it}
\textcolor{blue}{
EMRIs and IMRIs are expected to form in dense environments like star clusters and AGN disks. The environmental effects induced by these systems might bias GW detection -- e.g., in tests of GR or parameter extraction -- if not treated properly but open up the option to study dense astrophysical systems when accounting for them. Two major kinds of environmental effects have gained major attraction, a motion of the source and a dephasing induced by gas, but other effects like the Zeipel-Lidov-Kozai effect or eviction resonance might also play an important role. Due to TianQin having a better sensitivity than other space-based detectors at higher frequencies, it is particularly well equipped to detect environmental effects on light IMRIs. For example, it will be able to `see' the Brownian motion of a light IMRI in a globular cluster by detecting accelerations as small as $1\,{\rm m\,s^{-2}}$ or to measure the dephasing effect of gas for low-mass ($\lesssim10^5\,{\rm M_\odot}$), near-equal mass ratio ($q\sim10^{-1}$) IMRIs embedded in AGN disks.}
\end{it}
}}
\end{tcolorbox}

\subsection{Multi-messenger detections}\label{subsec:multi}

\contributors{Andrea Derdzinski, Xian Chen, Alvin Chua, Lixin Dai, Rongfeng Shen, Alejandro Torres-Orjuela, Martina Toscani, Lianggui Zhu}

Sources of GWs are expected to also emit EM radiation in the presence of matter. In this case, we talk about `multi-messenger sources' and the options opened by their detection are wide-ranging as well as promising. Therefore, we discuss a variety of sources emitting GWs and EM radiation detectable by TianQin like EMRIs and IMRIs in AGN, tidal disruption events (TDEs), and (possibly) quasi-periodic eruptions (QPEs).

\subsubsection{EMRIs and IMRIs in AGN}
\label{subsubsec:eimriem}

EMRI formation may occur efficiently in bright AGN systems where the accretion rate is near-Eddington~\cite{2021PhRvD.104f3007P, 2023MNRAS.521.4522D}; whereby it is speculated that this rate is higher around lower mass MBHs depending on the formation pathway~\cite{2023MNRAS.521.4522D}. When the EMRI is fully embedded in the disk, the possibility of EM counterparts is limited; albeit this scenario is not fully explored. Aside from the QPE scenarios described below, the possibility of ubiquitous EM counterparts for disk-embedded EMRIs is unlikely, given their low secondary mass and consequently low accretion luminosity compared to the central engine. To further complicate the picture, there exists a diversity of emission features present in AGN, the causes of which are difficult to disentangle. 

Heavy IMRIs, on the other hand, can perturb their host accretion disks substantially, leading to characteristic emission variability and spectral features that can be used to identify (or narrow down the list) of potential host galaxies. Aside from the potential connection between IMBH plunges and QPEs, the secondary's perturbation of an AGN disk via gravitational torques can produce time-varying spectral features, for example in the broad Fe-K$\alpha$ line detectable in the X-ray band~\cite{McKernan2014}. Sources may also exhibit a soft X-ray excess due to accretion onto the secondary BH. Following the merger, perturbations in the surrounding disk induced by the BH natal kick can also excite shocks and a post-merger flare on a timescale of weeks. Readers can find a detailed description of different emission signatures in \citet{2022LRR....25....3B} and \citet{lisa_2023}.

For multi-messenger detection the prospect of detecting time-domain emission signatures coincident (or following) GW detection, such that the host galaxy can be uniquely identified, is crucial. To assess the feasibility of finding such counterparts, it is important to discuss the expected localization of a detection. MBHs at low redshifts represent the ideal case for host galaxy identification. EMRIs can be detected up to a redshift of $z \approx 1.5 - 2$ with errors on the luminosity distance and sky localization of around 10\,\% and $10\,{\rm deg^2}$, respectively~\cite{fan_hu_2020}. Comparing these limits to the field of view of observatories that may catch an EM counterpart, it becomes clear that wide-field, high-cadence time-domain surveys in the optical band provide the best chance of detecting an EM counterpart, especially for EMRIs and IMRIs. As an example, the upcoming time-domain survey by the Vera Rubin Observatory will have a field of view of $\sim10\,{\rm deg^2}$~\cite{2009arXiv0912.0201L} while, e.g., the wide-field X-ray observatory Athena will have a relatively limited field of view of less than a square degree~\cite{nandra_et_al_2013}.

\begin{figure} \centering \includegraphics[width=0.48\textwidth]{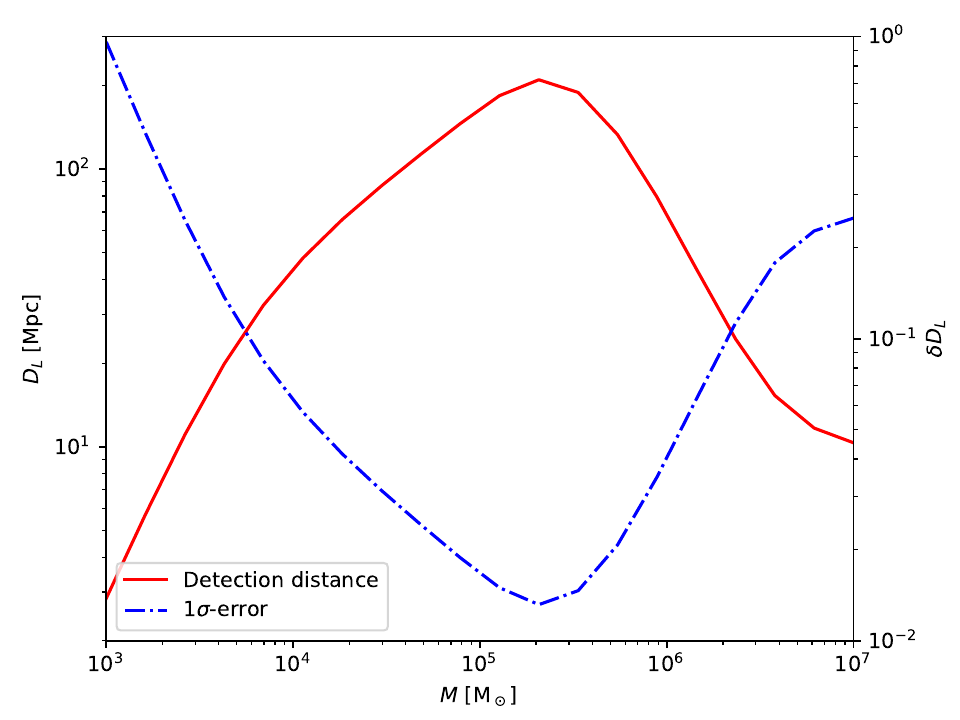}
\includegraphics[width=0.48\textwidth]{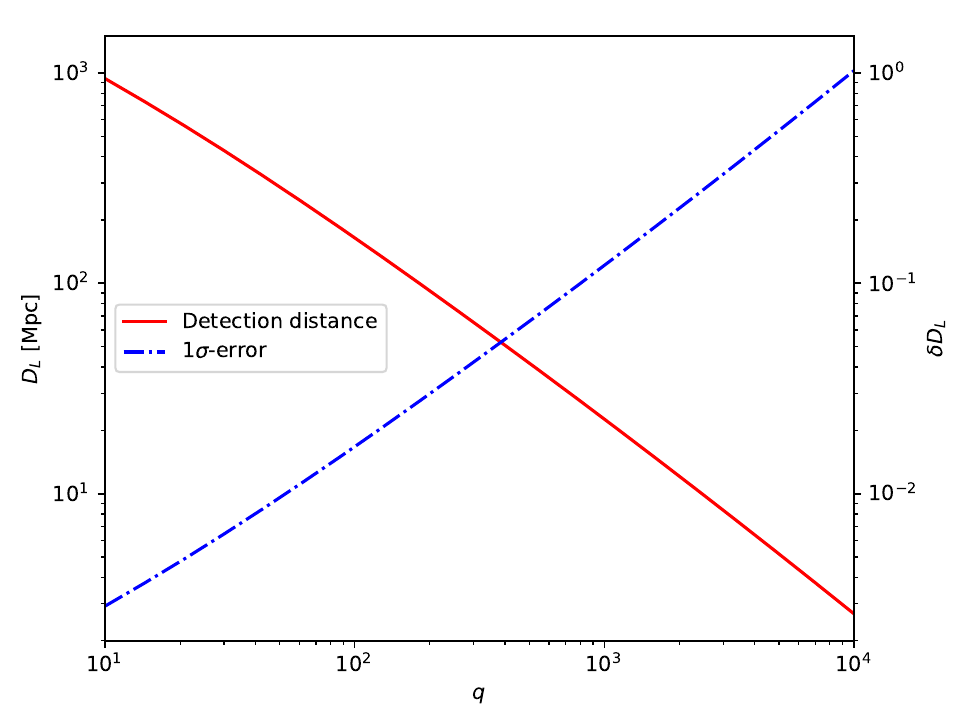}
\caption{ 
Left panel: The luminosity distance detection limit as a function of the primary BH mass, for a fixed mass ratio of $q=100$. The right axes show the relative error found assuming a fixed luminosity distance of 43 Mpc, or $z\sim 0.01$.  The values are obtained as an average of 100 sky localizations that are evenly distributed across the sky.
Right panel: The same quantities but as a function of mass ratio, for a fixed primary mass of $10^5 M_{\odot}$. Luminosity distance measurements are best for binaries with intermediate to equal mass ratios and primary masses around $\sim10^5 M_{\odot}$.
}
\label{fig:distancelim_imris}
\end{figure}

In Figure~\ref{fig:distancelim_imris}, we show the luminosity distance detection limit for an example IMRI source for different total masses and mass ratios. The measurement assumes a detection SNR threshold of 15 and shows how the luminosity distance measurement $D_L$ varies with binary mass (for a fixed mass ratio of $q=100$) and binary mass ratio (for a fixed primary mass of $10^5\,{\rm M_\odot}$). A relative error $\delta D_L$ is also shown, which is found assuming a fixed cosmological redshift of around 0.01. The values are obtained as an average of 100 sky localizations that are evenly distributed across the sky. Note that the waveforms adopted for this computation neglect higher modes beyond $\ell=10$, which become important for more extreme mass ratio binaries (higher $q$). The detection error for high mass ratio systems may be improved when higher modes are included in the model, although the trend of increasing localization error is still expected. We refer the reader to \citet{torres-orjuela_huang_2023} for a more detailed discussion on detectability limits and localization errors of EMRIs and IMRIs with TianQin. 

It is of interest to determine for which sources the localization of an EMRI or IMRI signal will be sufficient to narrow down the AGN population to a value for which host identification becomes realistic. It was recently shown that the localization of EMRIs with TianQin can reach to a volume that contains $\sim10^2$ galaxies for the highest SNR detections~\cite{zhu_fan_2024}. Thus if the EMRI demonstrates AGN-like characteristics (e.g. low eccentricity or environmental effects discussed in Section~\ref{subsec:enviro}), then the host galaxy can be narrowed down to a fraction of galaxies within the localization volume. Approximately a few percent of these galaxies contain AGNs, depending on the luminosity range and type considered, and barring observational biases that overlook lower mass, lower-luminosity sources~\cite{2013MNRAS.428..421S}. The precise number density depends on AGN classification, the MBH mass, and redshift, and the fraction drops significantly when considering only the brightest AGN sources. The local number density for near-Eddington AGN with 
SMBHs in the mass range $10^5$ -- $10^{7}\,{\rm M_\odot}$ ranges between $n_{\rm AGN} \sim 10^{-6} -- 10^{-5}\,{\rm Mpc^{-3}}$, respectively~\cite{2007ApJ...667..131G,2009ApJ...704.1743G}. This number becomes increasingly higher for lower luminosity systems, reaching values around $10^{-2}\,{\rm Mpc^{-3}}$ when considering the local population of Seyfert galaxies~\cite{2005AJ....129.1795H}. If we assume that AGN are responsible for the facilitation of E/IMRI events, then the number of galaxies can be significantly reduced within a localization error volume of a typical source, if we have an understanding of which class of AGN is most relevant. 

The number of possible hosts grows higher when expanding to other AGN types, including lower-luminosity sources which can still host accreting MBHs. Low-luminosity AGNs significantly outnumber higher-luminosity systems at low cosmological redshifts. Out of these, a significant number of galaxies (up to $30\,\%$) show low-ionization nuclear emission lines (so-called LINERs), which can be a result of gas accretion along with a compact, nuclear starburst~\cite{2004ApJ...605..105C,2017hsa9.conf..181M}. These nuclei are also proposed as hosts of gas-embedded IMBHs, given the expectation that an IMBH will perturb the accretion flow, reducing the accretion luminosity while also causing unique spectral features~\cite{McKernan2014}. If LINERs indeed host IMRIs, then the number of potential host galaxies in the detection limit grows significantly, making host identification difficult unless time-domain surveys are employed to catch coincident and characteristic emission signatures. In summary, the localization of high SNR sources presents a promising case for detecting EM counterparts and performing multi-messenger science. As discussed in Section~\ref{subsubsec:joint} below, the localization of a GW source can be improved significantly if TianQin and LISA conduct joint observations~\cite{Gong2021,torres-orjuela_huang_2023}. While direct studies for EMRIs and IMRIs are still scarce, previous works for MBHs suggest that the joint observation could pinpoint the location of the GW source in a sky area so small that only a couple of galaxies or AGNs can be found~\cite{zhu_chen2024}. Such a precise localization could help reveal the formation channel of EMRIs and IMRIs (e.g., in normal galaxies or in AGNs), as well as allow us to better determine the redshift so that the EMRI/IMRI can be used as a bright standard siren.

A recent study~\cite{Toscani:24aa} has explored the concept of lensed EMRIs,  whose GW signal has been strongly lensed. It is possible to extract these lensed signals directly from the detector data stream without assuming a specific lens model, thus allowing us to accurately constrain cosmological parameters. In fact, beyond determining the source distance through the GW measurement, we can also constrain the cosmological redshift as the EM emission from the host galaxy will also be strongly-lensed~\cite{Hannuksela:20aa, Wempe:22aa, Huang:2024zvk}.
Therefore, an identified lensed EMRI will yield percent constraints on the Hubble constant. Although the detectability of lensed EMRIs heavily relies on the assumption of the EMRI population, which remains highly speculative, they present a captivating new target for GW observatories.

\subsubsection{Tidal Disruption Events}
TDEs happen when a star gets too close to an MBH and gets destroyed by its tidal force~\cite{hills_1975, Carter:82aa, Rees88}. The stellar debris then orbit around the MBH and about half of the debris eventually accretes onto it. During this process, a bright flare can be produced, which is expected to rise over weeks to months and then decay over years following a characteristic pattern of $t^{-5/3}$~\cite{Evans89, Phinney:89aa, Guillochon13}. Calculations using the loss cone dynamics by mainly considering two-body scattering show that a TDE happens every $10^4-10^5$ years in a single galaxy~\cite{WangMerritt04}. The study of TDEs has caught a lot of attention recently for multiple reasons, including: 1) TDEs are unique, transient probes of quiescent MBHs; 2) TDEs can be used to detect IMBHs and binary SMBHs; 3) TDEs can disclose information of the stars in the innermost region of galaxies; 4) TDEs are ideal laboratories for studying BH disk formation, accretion and outflow physics, especially in the regime of super-Eddington accretion (see reviews~\cite{Rossi21, Bonnerot21, Dai21}).

About 100 TDEs have been detected up to now, with the majority seen through optical or soft X-ray surveys or telescopes\cite{Gezari21}. While it is widely agreed that the thermal X-ray emissions are produced from the accretion disk, there is still an ongoing debate about the production of the strong optical emission~\cite{Ulmer99, Roth20}. Some think that it is powered by the collision of the stellar debris, while others argue that it might be generated by the reprocessing of X-ray emissions in a thick envelope~\cite{Piran15, Dai18}. On top of this, some TDEs also produce radio, infrared, UV, or hard X-ray emissions. In particular, the events producing hard X-ray emissions are expected to launch relativistic jets with possible high-energy neutrinos associated~\cite{Bloom11, Cenko12, Stein:21aa}. 

As stars need to get relatively close to the MBHs to be tidally disrupted, it is naturally expected that TDEs produce GWs. Specifically, we expect the production of GWs while the star is approaching the pericenter and becomes distorted by tidal forces, causing changes in the stellar mass quadrupole moment~\cite{Stone:13aa,fan_zhong_2022,han_zhong_2020}. Additionally, GWs are expected during the disruption phase, due to the changes over time of the system (BH-star) mass quadrupole moment~\cite{Kobayashi:04aa, toscani_lodato_2022,garain_sarkar_2024}. The disruption phase is typically the dominant source of GWs, with peak amplitudes of around $10^{-22} - 10^{-21}$ (assuming a source distance of $\approx 20 \,{\rm Mpc}$) and frequencies ranging from millihertz to decihertz. The multi-messenger detection prospect of full TDEs (in which the star is totally destroyed and no remnant is left) has been considered~\cite{Pfister22} and it has been shown that TianQin and LISA will detect GWs from TDEs in the case of high-mass, young stars (stellar mass $\gtrsim60\,{\rm M_\odot}$) in the local universe ($z\lesssim 0.1$). 

Observing both GW and EM signals from a TDE offers remarkable opportunities for multimessenger studies~\cite{wevers_ryu_2023}. In particular, GWs could pinpoint the moment of stellar disruption at the initial close approach, a detail otherwise unobservable~\cite{Rossi21}. A measurement of the time delay between the GW signal and the subsequent EM emission would decisively help discriminate between different EM emission mechanisms~\cite{Toscani:23aa} and allow for cosmological constraints~\cite{Toscani:23aa, wong_2023}. Moreover, the actual number of detections can be used to constrain the stellar population near MBHs. Last but not least, the remnant of partial TDEs (in which the stellar envelope is destroyed but the stellar core remains intact) can possibly orbit around MBHs for multiple orbits and might also produce relatively strong GWs. Such topics are worthy of further exploration.

\subsubsection{Quasi-periodic eruptions}

QPEs are sources of soft X-rays that experience an increase in radiation by more than an order of magnitude relative to a quiescent level in a repeating pattern~\cite{miniutti_saxton_2019,giustini_miniutti_2020,arcodia_merloni_2021,chakraborty_kara_2021,arcodia_miniutti_2022,miniutti_giustini_2023a,miniutti_giustini_2023b}. The pattern consists of short and long periods of low emission that last for several hours followed by flares of less than an hour with peak luminosities of around $10^{43}\,{\rm erg\,s^{-1}}$, where the amplitude can differ for flares after short and long quiescent phases~\cite{miniutti_saxton_2019,giustini_miniutti_2020,arcodia_merloni_2021,miniutti_giustini_2023a}. QPEs have been observed to last for periods of at least a few years and are found in nearby small-mass galaxies, thus being associated with low-mass MBHs ($\approx 10^5-10^6\,{\rm M_\odot}$) in the center of those galaxies~\cite{giustini_miniutti_2020,arcodia_merloni_2021,chakraborty_kara_2021,arcodia_miniutti_2022,wevers_pasham_2022,miniutti_giustini_2023a,miniutti_giustini_2023b}. The host galaxies have been observed to be previously quiescent as well as active~\cite{arcodia_merloni_2021,giustini_miniutti_2020,wevers_pasham_2022} and there are hints that the formation of QPEs are related to preceding TDEs~\cite{chakraborty_kara_2021,miniutti_saxton_2019,miniutti_giustini_2023a,quintin_webb_2023}.

Although the possibility that QPEs are caused by the instability of accretion disks has not been safely excluded~\cite{Pan2022,Pan2023}, we focus here on models of QPEs that are linked to GW sources or their predecessors. Most of these models involve a (sub-)stellar mass object orbiting an MBH but differ in how the emitted flares are produced. The first class of models proposes that a WD inspiraling into an IMBH due to the emission of GWs is either partially tidally disrupted or its envelope is stripped away by tidal forces~\cite{chen_shen_2023,king_2020}. To avoid a runaway expansion of the WD which would make the QPE too short-lived to detect, it has also been proposed that the small body in a QPE is the core of a partially disrupted star~\cite{Zhao2022}. Another class of models suggests that QPEs could be formed by a main-sequence star in an EMRI system around an IMBH where the flares are produced when the star pierces through a gas disk around the IMBH~\cite{xian_zhang_2021}, although it has been pointed out that the formation rate of these sources is too low to explain the formation of (all) QPEs~\cite{metzger_stone_2022}. The X-ray flare is then produced by either the shock exerted on the gas or by a change of gas inflow onto the MBH~\cite{linial_metzger_2023,sukova_zajacek_2021}. Moreover, it has been suggested that the gas around the IMBH could have been deposited by a previous TDE event. Similar models discuss the plunge of stars or stellar-mass BHs in AGN disks to produce the flares of QPEs~\cite{franchini2023, Tagawa2023_QPE}. A third class of models involves two stars orbiting a low-mass MBH in similar orbits~\cite{metzger_stone_2022}. In this model, the two stars form independent EMRIs that evolve slowly due to the emission of GWs. X-ray flares are produced when the two stars pass near each other which leads to an increased Roche overflow onto the MBH.
QPEs as EM counterparts to IMRIs are related to broader multi-messenger strategies for MBHs (Section~\ref{subsec:multi-det}).

The period of QPEs is of the order of $10\,{\rm hr}$ thus corresponding to an orbital frequency of around $10^{-5}\,{\rm Hz}$ which is outside the sensitivity band of TianQin and other space-borne detectors. However, if the orbit has a high eccentricity the GW spectrum peaks at higher modes~\cite{chen_qiu_2022} leading to a signal detectable by TianQin. For the case of a WD that is partially tidally disrupted while inspiraling into an IMBH on a highly eccentric orbit ($e\gtrsim0.7$), it has been estimated that the signal could be detected within the Local Supercluster ($\approx30\,{\rm Mpc}$)~\cite{ye_chen_2024, chen_shen_2023}. The detection of QPEs by TianQin is not only restricted by the fact that they need to be highly eccentric to emit a significant amount of GWs in the TianQin sensitivity band but also because their lifetime is limited to only a few years since the stellar-mass object will get fully tidally disrupted long before plunging into the MBH~\cite{wang_yin_2022,linial_metzger_2023}. However, even in the case that a big fraction of QPEs cannot be detected directly by space-borne GW detectors they are still interesting sources for GW astronomy because they allow constraining the properties of the central MBH as well as of the orbit which can help understand the properties of otherwise elusive IMBHs and the formation of inspiraling systems around them. The properties that can be constrained include the mass and the spin of the IMBH as well as the semi-major axis, the eccentricity, the inclination, and the precession of the orbit of the stellar-mass object~\cite{king_2020,xian_zhang_2021,linial_metzger_2023}. In addition, QPEs as a population could contribute to the GW background with a significant SNR in the $\rm mHz$ band~\cite{chen_qiu_2022}. Finally, while the link between the less abrupt quasi-periodic oscillations (QPOs) and EMRIs is more tenuous, it has been suggested that QPEs and QPOs can originate from the same source under different configurations~\cite{king2023,linial_metzger_2023}, and that the QPO source RE J1034+396 might be detectable by space-based detectors in the 2030s if it is indeed caused by the inspiral of a $\sim 10 M_\odot$ BH into the central MBH~\cite{kejriwal2024}.

\vspace{1cm}

\begin{tcolorbox}[colback=red!5!white,colframe=red!75!black]
\textbf{\textcolor{NavyBlue}{
\begin{it}
GW sources that also emit EM radiation –- so-called multi-messenger sources –- are promising targets for studying a multitude of astrophysical and physical phenomena. Although typical EMRIs and IMRIs are not expected to emit EM radiation, this picture changes entirely when the GW source is embedded in gas or when the smaller component is replaced by a star. EMRIs and IMRIs in AGN disks are expected to emit light during and after the merger. For high-SNR sources in the close universe, sky localization might be accurate enough to identify an AGN as the host system, in particular, if time-domain surveys are employed to catch coincident and characteristic emission signatures. TDEs might also be promising multi-messenger sources -- in particular, when high-mass young stars are disrupted –- that will not only help us constrain the mass function of MBHs but also might significantly progress our understanding of the disruption process itself. Another type of (potential) multi-messenger source that might be related to the previous two, is QPEs which have gained much attraction recently but remain widely elusive. Detecting GWs from one of these systems would help exclude numerous models that do not consider close binaries as the underlying physical systems leading to QPEs while not detecting a GW signal would widely exclude close binaries as the correct models. However, even in the case of not detecting GWs, QPEs remain valuable sources for GW physics as they can help further constrain the mass function of MBHs.
\end{it}
}}
\end{tcolorbox}

\subsection{Contributions from GW observations with TianQin}\label{subsec:unique}

\contributors{Alejandro Torres-Orjuela, Hui-Min Fan, Xue-Ting Zhang}
\vspace{-0.5cm}

\subsubsection{Joint detection}\label{subsubsec:joint}

TianQin and LISA are planned to be launched at a similar time and share similar features~\cite{tianqin_2016,lisa_2017}. Therefore, they will detect the same sources but to different distances and with different accuracy~\cite{torres-orjuela_huang_2023,torres-orjuela_2024}. TianQin is more sensitive to lighter sources like light IMRIs while LISA is more sensitive to heavier sources such as heavy IMRIs and EMRIs. Nevertheless, we can expect improved detection distance and accuracy from joint detection. Here we discuss to scenarios for joint detection: (i) simultaneous joint detection if TianQin and LISA are online at the same time and (ii) consecutive joint detection if they are online one after the other.

If TianQin and LISA have a significant time overlap we get a higher SNR~\cite{isoyama_nakano_2018}, $\rho_{\rm joint}^2 = \rho_{\rm TianQin}^2 + \rho_{\rm LISA}^2$, which results in more accurate parameter estimation while further improvements can be expected from a more complete coverage of the sky or the off-times of the particular detectors~\cite{torres-orjuela_huang_2023}. For IMRIs, we use the Numerical Relativity surrogate model \texttt{BHPTNRSur1dq1e4}~\cite{islam_field_2022} which can only simulate the late inspiral and the merger phases, and thus the results presented here can be considered pessimistic. The waveforms for EMRIs are generated using an `augmented analytic kludge model' from the `EMRI waveform software suite'~\cite{chua_moore_2017}. The response of TianQin and LISA is modeled considering them as Michelson-type interferometers with an antenna response as in \citet{feng_wang_2019}.

\begin{figure}[htbp]
    \centering
    \subfloat[light IMRI\label{fig:limri_rad}]{
        \includegraphics[width=0.45\linewidth]{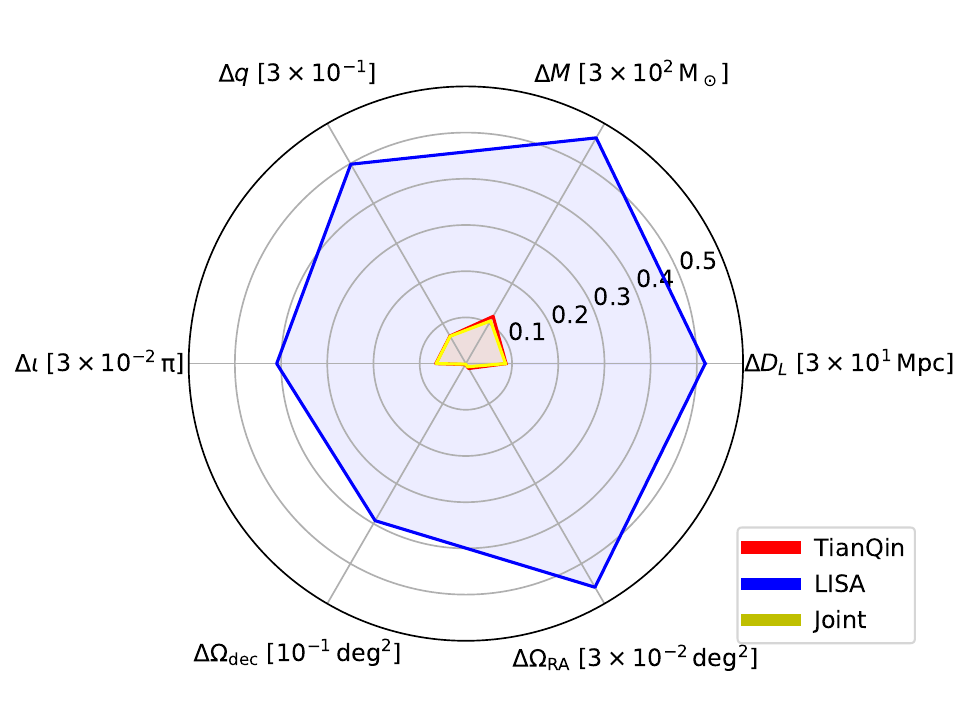}}
    \subfloat[heavy IMRI \label{fig:himri_rad}]{
        \includegraphics[width=0.45\linewidth]{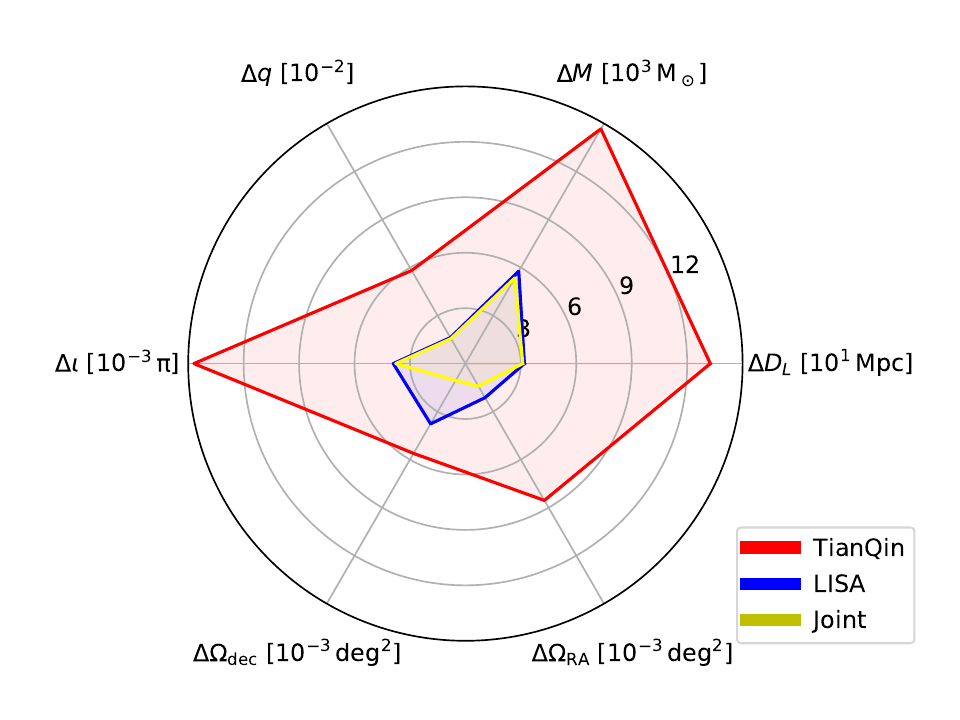}}
    \caption{The average detection error of TianQin (red), LISA (blue), and joint detection (yellow) for the luminosity distance $D_L$, the total mass in the observer frame $M$, the mass ratio $q$, the inclination $\iota$, and the sky localization $\Omega$ as a function of $\theta_{\rm bar}$ and $\phi_{\rm bar}$ for a light IMRI (left) and a heavy IMRI (right).}
    \label{fig:imri_rad}
\end{figure}

In Figure~\ref{fig:imri_rad}, we show the average detection error for IMRIs, where for light IMRIs (left) we consider the mass range $M\in[10^{2.75},10^5]\,{\rm M_\odot}$ and the luminosity distances $D_L\in[1,880]\,{\rm kpc}$ while for heavy IMRIs we consider the mass range $M\in[10^{4},10^{6.2}]\,{\rm M_\odot}$ and the distances $D_L\in[1,1600]\,{\rm Mpc}$. All other parameters are considered in the same ranges for both kinds of sources; for the mass ratio, we consider $q\in[1,1000]$, for the inclination $\iota\in[0,\pi/2]$, for the declination $\rm dec\in[0,+90^\circ]$, and for the right ascension $\rm RA\in[0^{\rm h},24^{\rm h}]$. The average errors of the parameters are calculated by computing the error for each value separately and then taking the arithmetic mean of the single errors. We see in Figure~\ref{fig:limri_rad} that the detection of light IMRIs by TianQin is significantly better than their detection by LISA and hence the joint detection yields similar results to TianQin. The detection error in TianQin is for all parameters smaller than LISA's by at least a factor of five, reaching on average errors for the luminosity distance of less than $3\,{\rm Mpc}$ and for the total mass of roughly $150\,{\rm M_\odot}$. Moreover, we find that despite the sky localization error $\Delta\Omega$ in TianQin being smaller than in LISA by one to two orders of magnitude, we still can have significant gains from joint detection reducing the error from on average $2.4\times10^{-3}$ for $\Delta\Omega_{\rm dec}$ in TianQin to $1.3\times10^{-3}$ in joint detection and for $\Delta\Omega_{\rm RA}$ from $1.2\times10^{-2}$ in TianQin to $7.9\times10^{-3}$ in joint detection. In Figure~\ref{fig:himri_rad}, we see that in contrast to light IMRIs for heavy IMRIs LISA's detection errors are smaller than TianQin's although the difference is less pronounced. We find that on average joint detection will measure $D_L$ with an accuracy of roughly $30\,{\rm Mpc}$ and the total mass with an error of around $5000\,{\rm M_\odot}$. The gain in the sky localization from joint detection is again most prominent reaching accuracies of on average $1-1.5\times10^{-3}\,{\rm deg^2}$ and thus multiple times smaller than in TianQin and LISA.

\begin{figure}[tpb] \centering \includegraphics[width=0.48\textwidth]{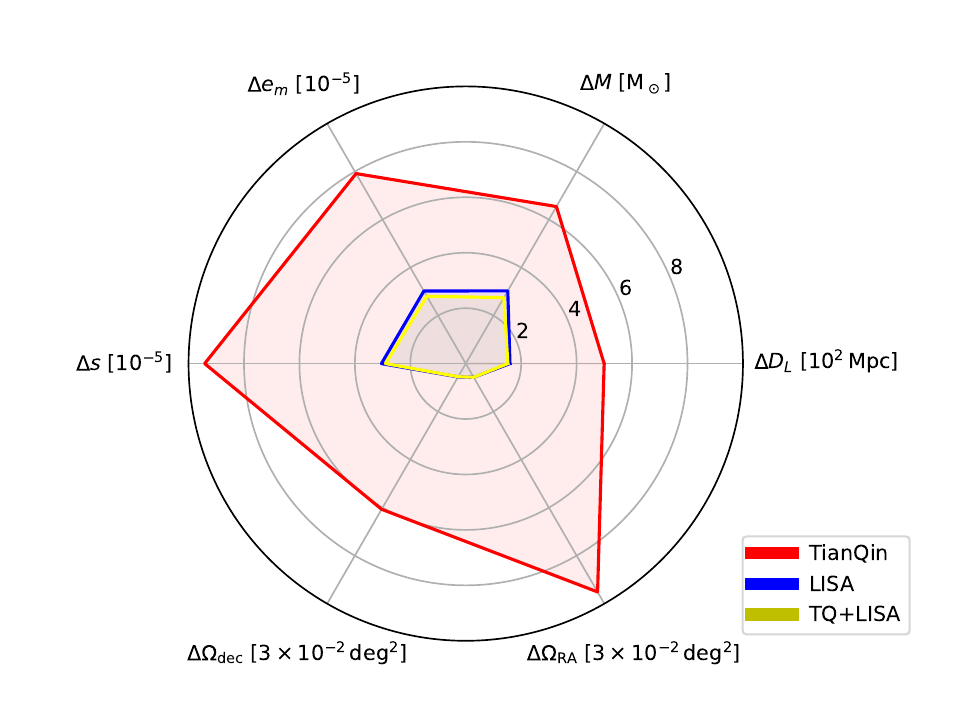}
\caption{
    The average detection error for the luminosity distance $D_L$, the total mass in the observer frame $M$, the mass ratio $q$, the inclination $\iota$, and the sky localization $\Omega$ as a function of $\theta_{\rm bar}$ and $\phi_{\rm bar}$ of an EMRI for TianQin (red), LISA (blue), and joint detection (yellow).
    }
\label{fig:emri_rad}
\end{figure}

Figure~\ref{fig:emri_rad} shows the average detection error of EMRIs. We consider the following range of parameters: for the luminosity distance $D_L\in[0.01,10]\,{\rm Gpc}$, for the mass of the SMBH $M\in[10^5,10^7]\,{\rm M_\odot}$, for the eccentricity at merger $e_m\in[2\times10^{-3},0.5]$, for the magnitude of the SMBH's spin $s\in[0.8,1]$, for the declination $\rm dec\in[-90^\circ,+90^\circ]$, and for the right ascension $\rm RA\in[0^{\rm h},24^{\rm h}]$. Moreover, we assume the small compact object to have a mass of $10\,{\rm M_\odot}$. We find, similar to heavy IMRIs, that LISA's detection error is smaller than TianQin's, where the accuracy differs by a factor of roughly two to twenty. The averaged errors we find for the joint detection are roughly $150\,{\rm Mpc}$ for $D_L$, $3\,{\rm M_\odot}$ for $M$, $3\times10^{-5}$ for $e_m$ and $s$, and $1.5\times10^{-3}\,{\rm deg^2}$ for the sky localization $\Omega$.  

It is also possible that the two missions will be online at different times. This case would allow a longer observation of sources which should result in a higher SNR as it scales with the square root of the observation time when assuming a constant sensitivity of the detector~\cite{sathyaprakash_schutz_2009}. However, neither TianQin nor LISA have constant sensitivity curves; on top of that, their sensitivities differ for most frequencies~\cite{tianqin_2016,lisa_2017,Mei2021, Colpi_et_al_2024}. Therefore, we study two possible cases of consecutive detection: (i) TianQin first collects data for $5\,{\rm yr}$ and then LISA collects data for $4\,{\rm yr}$ and (ii) LISA first collects data for $4\,{\rm yr}$ and then TianQin collects data for $5\,{\rm yr}$. This means in both cases the source data is collected for $9\,{\rm yr}$ but different phases of the source evolution are detected by the two detectors. To give an idea of the difference between the two detection scenarios, we show the SNR accumulated over the nine years for an EMRI with a central SMBH of the mass $M\approx1.2\times10^6\,{\rm M_\odot}$ (in the observer frame) and a spin magnitude $\chi=0.98$, a small compact object with an observer frame mass $m\approx12\,{\rm M_\odot}$, and an initial eccentricity $e_0\approx0.35$. Moreover, we set the source at a cosmological redshift of $z\approx0.2$, and a sky location $\rm dec\approx-37^\circ$ and $\rm RA\approx10.7^{\rm h}$. Note that we do not consider the case of an IMRI because there are currently no waveform models that can reliably simulate their signal from $9\,{\rm yr}$ before the merger until the merger.

\begin{figure}[tpb] \centering \includegraphics[width=0.48\textwidth]{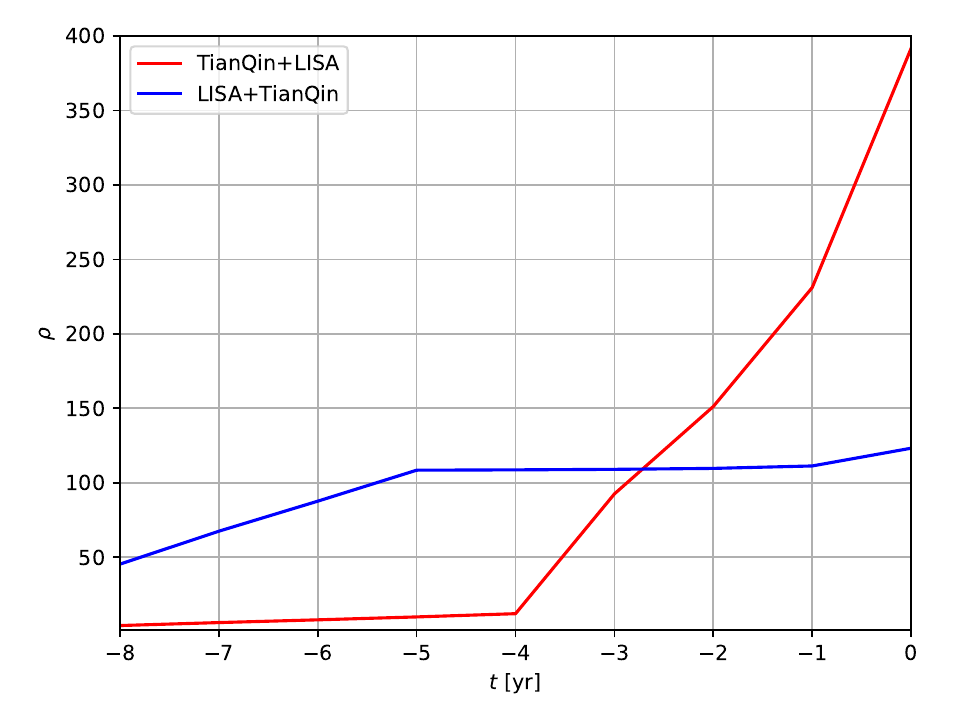}
\caption{
    The total SNR of an EMRI accumulated over $9\,{\rm yr}$ for two detection scenarios: (i) `TianQin+LISA': TianQin collects data for $5\,{\rm yr}$ and then LISA collects data for another $4\,{\rm yr}$, and (ii) `LISA+TianQin': LISA first collects data for $4\,{\rm yr}$ and then TianQin collects data for $5\,{\rm yr}$. 
    }
\label{fig:consecutive}
\end{figure}

We see in Figure~\ref{fig:consecutive}, that when TianQin is launched first (`TianQin+LISA') the SNR accumulated in the first five years is low only reaching around 12 after five years. The SNR accumulated in this period is low because the EMRI is still far from the merger, thus emitting GWs at relatively low frequencies where TianQin is less sensitive. However, in the last four years, the SNR increased rapidly reaching more than 230 one year before the merger and almost 400 at the time of the merger. In the scenario where LISA is launched first (`LISA+TianQin'), the SNR increases significantly in the first four years reaching $\rho\approx45$ after one year and $\rho\approx108$ after $4\,{\rm yr}$. In this detection scenario, the SNR almost does not increase between $5\,{\rm yr}$ and $1\,{\rm yr}$ before the merger but goes up to more than 120 in the last year. The increase in SNR leads to improved detection errors but the expected improvement for EMRIs is not very significant due to the stark contrast between TianQin's and LISA's ability to detect them. For heavy IMRIs, a more significant improvement is expected from consecutive detection as TianQin and LISA show more similar detection capacities for these sources. Nevertheless, the gain might be restricted because heavy IMRIs tend to last shorter in the band than EMRIs and are expected to accumulate a bigger portion of their SNR in the late inspiral and merger phase~\cite{sathyaprakash_schutz_2009}. In the case of light IMRIs, a similar trend to EMRIs for the collection of the SNR is expected but with inverted roles between TianQin and LISA.

\subsubsection{Studying IMRIs with TianQin}\label{sucsubsec:imriintq}

The detection of IMBHs has proved to be challenging but there is hope that GW detection will revolutionize this field. One of the most prominent sources involving IMBHs are IMRIs which TianQin will detect significantly better than LISA if their total masses is $\lesssim10^5\,{\rm M_\odot}$~\cite{torres-orjuela_huang_2023,torres-orjuela_2024}. We discussed extensively the detection of light IMRIs formed in globular clusters and nuclear star clusters by TianQin in Section~\ref{subsec:imri}, the measurement of their motion from the detection of higher modes and from a time-dependent phase shift for accelerated sources in Section~\ref{subsubsec:motion}, as well as their evolution inside AGN disks and the detection of the potential EM counterparts in Section~\ref{subsubsec:imriagn} and Section~\ref{subsubsec:eimriem}, respectively. Here, we summarize the main results of the detection of IMRIs.

We find that light IMRIs with total masses $\sim10^2-10^4\,{\rm M_\odot}$ will be detectable by TianQin out to distances of $5\,{\rm Mpc}$ while sources inside the Milky Way can have a SNR of several 10,000. For these high-SNR sources, we will be able to detect multiple higher modes which allow measuring their (constant) velocity, in particular, for sources with a high mass. Moreover, IMRIs in globular clusters are expected to undergo an (accelerated) Brownian motion due to the interaction with the stars in the system. In this case, even for IMRIs with relatively moderate SNRs of several ten, TianQin will be able to detect accelerations of an order of just $1-10\,{\rm m\,s^{-2}}$. If IMRIs are evolving in a thin accretion disk, gas can leave measurable imprints on the GWs. TianQin will be able to detect the dephasing induced by gas in these systems for a big portion of IMRIs, where the effect is particularly strong for systems with low total mass $M\lesssim10^5\,{\rm M_\odot}$ and near-equal mass ratio $q\sim10^{-1}$. We also expect to detect the EM counterpart for some of these sources produced by either perturbations of the gas during the inspiral phase of the IMRI -- if it is heavy enough and the two components are not too different -- or by the kick induced after the merger. If these systems are close enough, sky localization from GW detection by TianQin is expected to be accurate enough to identify the host system.

\vspace{1cm}

\begin{tcolorbox}[colback=red!5!white,colframe=red!75!black]
\textbf{\textcolor{NavyBlue}{
\begin{it}
TianQin will significantly expand the range of detectable sources as well as improve the estimation of their parameters. As extensively discussed in previous sections, TianQin is particularly well equipped to detect light IMRIs formed by an IMBH and a stellar-mass BH. These sources are barely detectable by other space-based GW detectors but are of high astrophysical interest as they allow us to study IMBHs in the low mass range as well as the environments where they form. Moreover, gains are particularly promising if TianQin conducts coordinated detections with other space-based detectors like LISA. Simultaneous joint detection will increase the SNR of EMRIs and IMRIs, leading to improved parameter extraction, particularly, for extrinsic parameters like the sky localization. The consecutive detection of GWs by TianQin and LISA (if they are launched several years apart) also allows us to obtain a higher SNR. However, the gains are not as prominent as in the case of simultaneous detection because it implies detecting sources long before the merger where the signal is relatively low and is also mainly expected for EMRIs where LISA's contribution is more significant than TianQin's. 
\end{it}
}}
\end{tcolorbox}

\clearpage
\section*{Acknowledgement}

Y.-M.H. is supported by Guangdong Major Project of Basic and Applied Basic Research (Grant No. 2019B030302001), the National Key Research and Development Program of China (No. 2020YFC2201400), and the Natural Science Foundation of China (Grants  No.  12173104, No. 12261131504). 
C.B. is supported by the Natural Science Foundation of China (Grant No. 12250610185 and No. 12261131497) and the Natural Science Foundation of Shanghai (Grant No. 22ZR1403400).
P.R.C. acknowledges support from the Swiss National Science Foundation under the Sinergia Grant CRSII5\_213497 (GW-Learn).
A.D. is supported by the National Science Foundation grant AST-2319441. 
K.I. is supported by the National Natural Science Foundation of China (12073003, 11991052, 11721303, 12150410307, and 11950410493), and the China Manned Space Project Nos. CMS-CSST-2021-A04 and CMS-CSST-2021-A06. 
H.T. is supported by the National Key R\&D Program of China (grant No. 2021YFC2203002). 
J.-d.Z. is supported by the Guangdong Basic and Applied Basic Research Foundation(Grant No. 2023A1515030116).
S.-J.H. is supported by the Postdoctoral Fellowship Program of CPSF (Grant No. GZC20242112). 
P.A.S. acknowledges the Foreign Scholar Research Program of Peking University for supporting his visit.
AA acknowledges support for this research from project No. 2021/43/P/ST9/03167 co-funded by the Polish National Science Center (NCN) and the European Union Framework Programme for Research and Innovation Horizon 2020 under the Marie Skłodowska-Curie grant agreement No. 945339. For the purpose of Open Access, the authors have applied for a CC-BY public copyright license to any Author Accepted Manuscript (AAM) version arising from this submission.
MG and AH acknowledge support by the Polish National Science Center (NCN) through the grant 2021/41/B/ST9/01191.
H.G. is supported by the NSFC Nos.12288102, 12125303, 12173081, the National Key R\&D Program of China (No. 2021YFA1600403), 
the Key Research Program of Frontier Sciences, CAS (No. ZDBS-LY-7005) and Yunnan Fundamental Research Projects (Nos. 202101AV070001).

\bibliography{refs}

\end{document}